\documentclass{article}
\usepackage{latexsym,amsmath,amssymb,amsfonts}   
\usepackage{pstricks,pst-node}
\usepackage{feynmf}
\usepackage{sty-frabetti-colombia2007}

\begin{document}

\begin{fmffile}{graphs}

\title{Renormalization Hopf algebras and combinatorial groups}
 
\author{
Alessandra Frabetti \\
Universit\'e de Lyon, Universit\'e Lyon 1, CNRS, \\  
UMR 5208 Institut Camille Jordan, \\ 
B\^atiment du Doyen Jean Braconnier, \\
43, blvd du 11 novembre 1918, F-69622 Villeurbanne Cedex, France. \\
{\tt frabetti@math.univ-lyon1.fr}}
\date{\today}
 
\maketitle


\begin{abstract}
These are the notes of five lectures given at the Summer School 
{\em Geometric and Topological Methods for Quantum Field Theory\/}, 
held in Villa de Leyva (Colombia), July 2--20, 2007. 
The lectures are meant for graduate or almost graduate students in 
physics or mathematics. They include references, many examples and 
some exercices. The content is the following. 

The first lecture is a short introduction to algebraic and 
proalgebraic groups, based on some examples of groups of matrices and 
groups of formal series, and their Hopf algebras of coordinate functions. 

The second lecture presents a very condensed review of classical 
and quantum field theory, from the Lagrangian formalism to the 
Euler-Lagrange equation and the Dyson-Schwinger equation for Green's
functions. It poses the main problem of solving some non-linear differential 
equations for interacting fields. 

In the third lecture we explain the perturbative solution of the 
previous equations, expanded on Feynman graphs, in the simplest case 
of the scalar $\phi^3$ theory. 

The forth lecture introduces the problem of divergent integrals appearing 
in quantum field theory, the renormalization procedure for the graphs, 
and how the renormalization affects the Lagrangian and the Green's 
functions given as perturbative series. 

The last lecture presents the Connes-Kreimer Hopf algebra of 
renormalization for the scalar theory and its associated proalgebraic 
group of formal series. 
\end{abstract}

\vfill 

{\small
\tableofcontents} 
\bigskip\bigskip 

\noindent
{\bf Aknowledgments.} 
These lectures are based on a course for Ph.D. students in mathematics, 
held at {\em Universit\'e Lyon 1\/} in spring 2006, by Alessandra Frabetti 
and Denis Perrot. Thanks Denis! 

During the Summer School {\em Geometric and Topological Methods for 
Quantum Field Theory\/}, many students made interesting questions and 
comments which greatly helped the writing of these notes. 
Thanks to all of them!


\section*{Lecture I - Groups and Hopf algebras}
\addcontentsline{toc}{section}{\bf Lecture I - Groups and Hopf algebras}
\label{lecture1}

In this lecture we review the classical duality between groups and 
Hopf algebras of certein types. Details can be found for instance in 
\cite{Hochschild}. 


\subsection{Algebras of representative functions}
\label{Hopf-def}

Let $G$ be a group, for instance a group of real or complex matrices, 
a topological or a Lie group. Let 
\begin{align*}
F(G)&= \{f:G\longrightarrow \C\ (\mbox{or $\R$}) \} 
\end{align*}
denote the set of functions on $G$, eventually continuous or differentiable. 
Then $F(G)$ has a lot of algebraic structures, that we describe in
details. 

\begin{point}{Product}
The natural vector space $F(G)$ is a unital associative and
commutative algebra over $\C$, with {\em product\/} $(f g)(x)=f(x) g(x)$, 
where $f,g\in F(G)$ and $x\in G$, and unit given by the constant 
function $1(x)=1$. 
\end{point}

\begin{point}{Coproduct}
For any $f\in F(G)$, the group law 
$G\times G\overset{\cdot}{\longrightarrow}G$ induces an element 
$\D f\in F(G\times G)$ defined by $\D f(x,y)=f(x\cdot y)$. 
Can we characterise the algebra 
$F(G\times G)= \{f:G\times G\longrightarrow \C\}$ starting from $F(G)$? 

Of course, we can consider the tensor product 
\begin{align*}
F(G)\otimes F(G) 
&= \left\{\sum_{\mbox{\small finite}} f_i\otimes g_i,\ f_i,g_i\in F(G)\right\},
\end{align*}
with componentwise product 
$(f_1\otimes g_1)(f_2\otimes g_2)= f_1 g_1 \otimes f_2 g_2$, 
but in general this algebra is a strict subalgebra of 
$F(G\times G)=\{\sum_{\mbox{\small infinite}}f_i\otimes g_i\}$ 
(it is equal for finite groups). 
For example, $f(x,y)=\exp(x+y)\in F(G)\otimes F(G)$, 
but $f(x,y)=\exp(xy)\notin F(G)\otimes F(G)$. 
Similarly, if $\delta(x,y)$ is the function equal to $1$ when $x=y$ 
and equal to $0$ when $x\neq y$, then $\delta\notin F(G)\otimes F(G)$. 
To avoid this problem we could use the {\em completed\/} or 
{\em topological\/} tensor product $\hat{\otimes}$ such that 
$F(G)\hat{\otimes} F(G)=F(G\times G)$. However this tensor product is 
difficult to handle, and for our purpuse we want to avoid it. 
In alternative, we can consider the subalgebras $R(G)$ of $F(G)$ such that 
$R(G)\otimes R(G)=R(G\times G)$. Such algebras are of course much easier to
describe then a completed tensor product. 
For our purpuse, we are interested in the case when one of these subalgebras 
is big enough to describe completely the group. That is, it does not
loose too much informations about the group with respect to $F(G)$. 
This condition will be specified later on.

Let us then suppose that there exists a subalgebra $R(G)\subset F(G)$ 
such that $R(G)\otimes R(G)=R(G\times G)$. Then, the group law 
$G\times G\overset{\cdot}{\longrightarrow}G$ induces a 
{\em coproduct\/} $\D:R(G)\longrightarrow R(G)\otimes R(G)$ defined by
$\D f(x,y)=f(x\cdot y)$. 
We denote it by $\D f=\sum_{\mbox{\small finite}} f_{(1)}\otimes f_{(2)}$. 
The coproduct has two main properties: 
\begin{enumerate}
\item 
$\D$ is a homomorphism of algebras, in fact 
\begin{align*}
\D (fg)(x,y) &= (fg)(x\cdot y) = f(x \cdot y) g(x\cdot y) 
= \D f(x,y)\D g(x,y), 
\end{align*}
that is $\D (fg) = \D (f) \D (g)$. This can also be expressed as 
$\sum (fg)_1 \otimes (fg)_2 = \sum f_1 g_1 \otimes f_2 g_2$. 

\item 
$\D$ is coassociative, that is $(\D\otimes\Id)\D = (\Id\otimes\D)\D$, 
because of the associativity $(x\cdot y)\cdot z=x\cdot(y\cdot z)$ 
of the group law in $G$. 
\end{enumerate}
\end{point}

\begin{point}{Counit}
The neutral element $e$ of the group $G$ induces a {\em counit\/} 
$\varepsilon:R(G)\longrightarrow \C$ defined by 
$\varepsilon(f)=f(e)$. The counit has two main properties: 
\begin{enumerate}
\item 
$\varepsilon$ is a homomorphism of algebras, in fact 
\begin{align*}
\varepsilon (fg) &= (fg)(e) = f(e) g(e) = \varepsilon(f) \varepsilon(g). 
\end{align*}

\item 
$\varepsilon$ satisfies the equality 
$\sum f_{(1)} \varepsilon(f_{(2)}) = \sum \varepsilon(f_{(1)}) f_{(2)}$, 
induced by the equality $x\cdot e=x=e\cdot x$ in $G$. 
\end{enumerate}
\end{point}

\begin{point}{Antipode}
\label{antipode-def}
The operation of inversion in $G$, that is $x\to x^{-1}$, induces 
the {\em antipode\/} $S:R(G)\longrightarrow R(G)$ defined by 
$S(f)(x)=f(x^{-1})$. The counit has four main properties: 
\begin{enumerate}
\item 
$S$ is a homomorphism of algebras, in fact 
\begin{align*}
S(fg)(x) &= (fg)(x^{-1}) = f(x^{-1}) g(x^{-1}) = S(f)(x) S(g)(x). 
\end{align*}

\item 
$S$ satisfies the 5-terms equality 
$m(S\otimes\Id)\D=u\varepsilon=m(\Id\otimes S)\D$, 
where $m:R(G)\otimes R(G)\longrightarrow R(G)$ denotes the product and
$u:\C\longrightarrow R(G)$ denotes the unit. 
This is induced by the equality $x\cdot x^{-1}=e=x^{-1}\cdot x$ in $G$. 

\item 
$S$ is anti-comultiplicative, that is 
$\D\circ S=(S\otimes S)\circ\tau\circ\D$, 
where $\tau(f\otimes g)=g\otimes f$ is the twist operator. This
property is induced by the equality $(x\cdot y)^{-1}=y^{-1}\cdot
x^{-1}$ in $G$. 

\item
$S$ is nilpotent, that is $S\circ S=\Id$, because of the identity 
$(x^{-1})^{-1}=x$ in $G$. 
\end{enumerate}
\end{point}

\begin{point}{Abelian groups}
Finally, $G$ is abelian, that is $x\cdot y=y\cdot x$ for all $x,y\in
G$, if and only if the coproduct is cocommutative, that is 
$\D=\D\circ\tau$, i.e. 
$\sum f_{(1)}\otimes f_{(2)}=\sum f_{(2)}\otimes f_{(1)}$. 
\end{point}

\begin{point}{Hopf algebras}
A unital, associative and commutative algebra $\H$ endowed with a 
coproduct $\D$, 
a counit $\varepsilon$ and an antipode $S$, satisfying all the 
properties listed above, is called a commutative Hopf algebra. 

In conclusion, we just showed that if $G$ is a (topological) group, 
and $R(G)$ is a subalgebra of (continuous) functions on $G$ such that 
$R(G)\otimes R(G)=R(G\times G)$, and sufficiently big to contain the
image of $\D$ and of $S$, then $R(G)$ is a commutative Hopf algebra. 
Moreover, $R(G)$ is cocommutative if and only if $G$ is abelian. 
\end{point}

\begin{point}{Representative functions}
We now turn to the existence of such a Hopf algebra $R(G)$. 
If $G$ is a finite group, then the largest such algebra is simply 
the linear dual $R(G)=F(G)=(\C G)^*$ of the group algebra. 

If $G$ is a topological group, then the condition 
$R(G)\otimes R(G)=R(G\times G)$ roughly forces $R(G)$ to be a
polynomial algebra, or a quotient of it. The generators are the 
{\em coordinate functions\/} on the group, but we do not always know
how to find them.  

For compact Lie groups, $R(G)$ always exists, and we can be more
precise. We say that a function $f:G\longrightarrow\C$ is 
{\em representative\/} if there exist a finite number of functions 
$f_1,...,f_k$ such that any translation of $f$ is a linear combination
of them. If we denote by $(L_xf)(y)=f(x\cdot y)$ the left translation 
of $f$ by $x\in G$, this means that $L_xf=\sum l_i(x) f_i$. 
Call $R(G)$ the set of all representative functions on $G$. 
Then, using representation theory, and in particular 
{\em Peter-Weyl Theorem\/}, one can show the following facts: 
\begin{enumerate}
\item 
$R(G)\otimes R(G)=R(G\times G)$; 
\item 
$R(G)$ is dense in the set of continuous functions; 
\item 
as an algebra, $R(G)$ is generated by the matrix elements of all the 
representations of $G$ of finite dimension; 
\item 
$R(G)$ is also generated by the matrix elements of one faithful 
representation of $G$, therefore it is finitely generated.  
\end{enumerate}
Moreover, for compact Lie groups, the algebra $R(G)$ has two additional
structures: 
\begin{enumerate}
\item 
because the group $G$ is a real manifold, and the functions have
complex values, $R(G)$ has an {\em involution\/}, that is a map 
$*:R(G)\longrightarrow R(G)$ such that $(f^*)^*=f$ and $(fg)^*=g^*f^*$; 
\item 
because $G$ is compact, $R(G)$ has a {\em Haar measure\/}, that is, 
a linear map $\mu:R(G)\longrightarrow\R$ such that $\mu(a a^*)> 0$ 
for all $a\neq 0$. 
\end{enumerate}
Similar results hold in general for groups of matrices, even if they
are complex manifolds, and even if they are not compact. In
particular, the algebra generated by the matrix elements of one faithful 
representation of $G$ satisfies the required properties. 

For other groups then those of matrices, a suitable algebra $R(G)$ can
exist, but there is no general procedure to find it. The best hint is
to look for a faithful representation, eventually with infinite
dimension. This may work also for groups which are not locally compact, as
shown in the examples (\ref{invertible-series}) and
(\ref{diffeomorphisms}), but in general not for groups of
diffeomorphisms on a manifold. 
\end{point}


\subsection{Examples}

\begin{point}{Complex affine plane}
Let $G=(\C^n,+)$ be the additive group of the complex affine plane. 
A complex group is supposed to be a holomorphic manifold. The
functions are also supposed to be holomorphic, that is they do not 
depend on the complex conjugate of the variables. 
The map 
\begin{align*}
\rho: (\C^n,+) & \longrightarrow GL_{n+1}(\C)=\Aut(\C^{n+1}) \\
(t_1,...,t_n) & \mapsto 
\left(\begin{array}{cccc} 
1&t_1&...&t_n\\0&1&...&0\\&&...&\\0&0&...&1 \end{array}\right)
\end{align*} 
is a faithful representation, in fact 
\begin{align*}
\rho\big((t_1,...,t_n)+(s_1,...,s_n)\big) 
&= \left(\begin{array}{cccc} 1&t_1+s_1&...&t_n+s_n
\\0&1&...&0\\&&...&\\0&0&...&1 \end{array}\right) \\ 
&= \left(\begin{array}{cccc} 1&t_1&...&t_n
\\0&1&...&0\\&&...&\\0&0&...&1 \end{array}\right) 
\left(\begin{array}{cccc} 1&s_1&...&s_n
\\0&1&...&0\\&&...&\\0&0&...&1 \end{array}\right) 
= \rho(t_1,...,t_n) \rho(s_1,...,s_n). 
\end{align*}
Therefore, there are $n$ local coordinates $x_i(t_1,...,t_n)=t_i$, 
for $i=1,...,n$, which are free of mutual relations.  
Hence the algebra of local coordinates on the affine line is the 
polynomial ring $R(\C^n,+)=\C[x_1,...,x_n]$. 
The Hopf structure is the following: 
\begin{itemize}
\item 
Coproduct: $\D x_i=x_i \otimes 1+1\otimes x_i$ and $\D 1= 1\otimes 1$. 
The group is abelian and the coproduct is indeed cocommutative. 
\item
Counit: $\varepsilon(x_i)=x(0)=0$, and $\varepsilon(1)=1$. 
\item
Antipode: $Sx_i=-x_i$ and $S1=1$. 
\end{itemize}
This Hopf algebra is usually called the {\em unshuffle Hopf
algebra\/}, because the coproduct on a generic monomial 
\begin{align*}
\D(x_{i_1}\cdots x_{i_l}) &= 
\sum_{p+q=l} \sum_{\sigma\in\Sigma_{p,q}} 
x_{\sigma(i_1)}\cdots x_{\sigma(i_p)} \otimes 
x_{\sigma(i_{p+1})}\cdots x_{\sigma(i_{p+q})} 
\end{align*}
makes use of the shuffle permutations $\sigma\in\Sigma_{p,q}$, 
that is the permutations of $\Sigma_{p+q}$ such that 
$\sigma(i_1)<\cdots <\sigma(i_p)$ and 
$\sigma(i_{p+1})<\cdots<\sigma(i_{p+q})$. 
\end{point}

\begin{point}{Real affine plane}
Let $G=(\R^n,+)$ be the additive group of the real affine plane. 
A real group is supposed to be a differentiable manifold. The
functions with values in $\C$ are the complexification of the 
functions with values in $\R$, that is, 
$R_{\C}(G)=R_{\R}(G)\otimes \C$. 
In principle, then, the functions depend also on the complex
conjugates, but the generators must be real: we expect that the 
algebra $R_{\C}(G)$ has an involution $*$. 
In fact, we have the following results: 
\begin{itemize}
\item
{\bf Real functions:} the map 
\begin{align*}
\rho: (\R^n,+) & \longrightarrow GL_{n+1}(\R)=\Aut(\R^{n+1}) \\
(t_1,...,t_n) & \mapsto 
\left(\begin{array}{cccc} 
1&t_1&...&t_n\\0&1&...&0\\&&...&\\0&0&...&1 \end{array}\right)
\end{align*} 
is a faithful representation. 
The local coordinates are $x_i(t_1,...,t_n)=t_i$, for $i=1,...,n$, 
and the algebra of real local coordinates is the polynomial ring 
$R_{\R}(\R^n,+)=\R[x_1,...,x_n]$. 
The Hopf structure is exactely as in the previous example. 
\item 
{\bf Complex functions:} complex faithful representation as before, 
but local coordinates $x_i(t_1,...,t_n)=t_i$ subject to an involution 
defined by $x_i^*(t_1,...,t_n)=\overline{t_i}$ and such that $x_i^*=x_i$. 
Then the algebra of complex local coordinates is the quotient 
\begin{align*}
R_{\C}(\R^n,+)
&=\frac{\C[x_1,x_1^*,...,x_n,x_n^*]}{\la x_i^*-x_i, i=1,...,n \ra},
\end{align*}
which is isomorphic to $\C[x_1,...,x_n]$ as an algebra, but not as 
an algebra with involution. 
Of course the Hopf structure is always the same. 
\end{itemize}
\end{point}

\begin{point}{Complex simple linear group}
The group 
\begin{align*}
SL(2,\C) 
&= \left\{ M=\left(\begin{array}{cc} m_{11}&m_{12}\\m_{21}&m_{22}
\end{array}\right)
\in M_2(\C),\ \det M = m_{11} m_{22} - m_{12}m_{21}=1 \right\}
\end{align*}
has a lot of finite-dimensional representations, and the smallest
faithful one is the identity 
\begin{align*}
\rho=\Id :\quad & SL(2,\C) \longrightarrow GL_2(\C) \\ 
& M \mapsto 
\left(\begin{array}{cc} m_{11}=a(M) &m_{12}=b(M)\\m_{21}=c(M)&m_{22}=d(M)
\end{array}\right). 
\end{align*}
Therefore there are 4 local coordinates $a,b,c,d:
SL(2,\C)\longrightarrow \C$, given by $a(M)=m_{11}$, etc, related by 
$\det M=1$. 
Hence the algebra of local coordinates of $SL(2,\C)$ is the quotient 
\begin{align*}
R(SL(2,\C))&=\frac{\C[a,b,c,d]}{\la ad-bc-1 \ra}.
\end{align*}
The Hopf structure is the following: 
\begin{itemize}
\item 
Coproduct: $\D f(M,N)= f(M N)$, therefore 
\begin{align*}
\D a=a\otimes a+b\otimes c \quad &\quad \D b=a\otimes b+b\otimes d\\
\D c=c\otimes a+d\otimes c \quad &\quad \D d=c\otimes b+d\otimes d
\end{align*}
To shorten the notation, we can write 
$\D \left(\begin{array}{cc} a&b\\c&d \end{array}\right) = 
\left(\begin{array}{cc} a&b\\c&d \end{array}\right) 
\otimes \left(\begin{array}{cc} a&b\\c&d \end{array}\right)$. 
\item
Counit: $\varepsilon(f)=f(1)$, hence
$\varepsilon \left(\begin{array}{cc} a&b\\c&d \end{array}\right) = 
\left(\begin{array}{cc} 1&0\\0&1 \end{array}\right)$. 
\item
Antipode: $Sf(M)=f(M^{-1})$, therefore
$S \left(\begin{array}{cc} a&b\\c&d \end{array}\right) = 
\left(\begin{array}{cc} d&-b\\-c&a \end{array}\right)$. 
\end{itemize}
\end{point}

\begin{point}{Complex general linear group}
For the group 
\begin{align*}
GL(2,\C) &= \left\{ M\in M_2(\C),\ \det M \neq 0 \right\}, 
\end{align*}
the identity $GL(2,\C)\longrightarrow GL(2,\C)\equiv \Aut(\C^2)$ is 
of course a faithful representation. We have then 4 local coordinates 
as for $SL(2,\C)$. However this time they satisfy the condition 
$\det M \neq 0$ which is not closed. To express this relation we use a
trick: since $\det M \neq 0$ if and only if there exists the inverse 
of $\det M$, we add a variable $t(M)=(\det M)^{-1}$. 
Therefore the algebra of local coordinates of $GL(2,\C)$ is the quotient 
\begin{align*}
R(GL(2,\C))&=\frac{\C[a,b,c,d,t]}{\la (ad-bc)t-1 \ra}.
\end{align*}
The Hopf structure is the same as that of $SL(2,\C)$ on the local 
coordinates $a,b,c,d$, and on the new variable $t$ is given by 
\begin{itemize}
\item 
Coproduct: since 
$\D t(M,N)= t(M N)=(\det(MN))^{-1}=(\det M)^{-1}(\det N)^{-1}=t(M)t(N)$, 
we have $\D t=t\otimes t$. 
\item
Counit: $\varepsilon(t)=t(1)=1$. 
\item
Antipode: $St(M)=t(M^{-1})=(\det(M^{-1}))^{-1}=\det M$, therefore 
$St= ad-bc$. 
\end{itemize}
\end{point}

\begin{point}{Simple unitary group}
The group 
\begin{align*}
SU(2)&= \left\{ M\in M_2(\C),\ \det M =1,\ M^{-1}=\overline{M}^t \right\}
\end{align*}
is a real group, infact it is one {\em real form\/} 
of $SL(2,\C)$, the other one being $SL(2,\R)$, and it is also the
maximal compact subgroup of $SL(2,\C)$. 
As a real manifold, $SU(2)$ is isomorphic to the 3-dimensional sphere
$S^3$, in fact
\begin{align*}
M=\left(\begin{array}{cc} a&b\\c&d\end{array}\right)\in SU(2) 
\quad\Longleftrightarrow\quad 
\begin{array}{c} 
ad-bc=1\\ \overline{a}=d\ ,\ \overline{b}=c 
\end{array} 
\quad\Longleftrightarrow\quad 
M=\left(\begin{array}{cc}
a&b\\-\overline{b}&\overline{a}\end{array}\right)
\quad\mbox{with $a\overline{a}+b\overline{b}=1$}. 
\end{align*}
If we set $a=x+iy$ and $b=u+iv$, with $x,y,u,v\in\R$, we then have 
\begin{align*}
a\overline{a}+b\overline{b}=1 
\quad\Longleftrightarrow\quad 
x^2+y^2+u^2+v^2=1 \quad\mbox{in $\R^4$}
\quad\Longleftrightarrow\quad 
(x,y,u,v)\in S^3. 
\end{align*}
We then expect that the algebra of complex functions on $SU(2)$ 
has an involution: 
\begin{align*}
R(SU(2))&=\frac{\C[a,b,c,d,a^*,b^*,c^*,d^*]}{\la a^*-d, b^*+c,
ad-bc-1\ra} 
\quad\cong\quad \frac{\C[a,b,a^*,b^*]}{\la aa^*+bb^*-1\ra} . 
\end{align*}

The Hopf structure is the same as that of $SL(2,\C)$, but expressed in
terms of the proper coordinate functions of $SU(2)$, that is: 
\begin{itemize}
\item 
Coproduct: $\D \left(\begin{array}{cc} a&b\\-b^*&a^* \end{array}\right) = 
\left(\begin{array}{cc} a&b\\-b^*&a^* \end{array}\right) 
\otimes \left(\begin{array}{cc} a&b\\-b^*&a^* \end{array}\right)$. 
\item
Counit: $\varepsilon \left(\begin{array}{cc} a&b\\-b^*&a^*
\end{array}\right) 
= \left(\begin{array}{cc} 1&0\\0&1 \end{array}\right)$. 
\item
Antipode: $S \left(\begin{array}{cc} a&b\\-b^*&a^* \end{array}\right) = 
\left(\begin{array}{cc} a^*&-b\\b^*&a \end{array}\right)$. 
\end{itemize}
\end{point}
\medskip

\begin{point}{Exercise: Heisenberg group} 
The Heisenberg group $H_3$ is the group of complex $3\times 3$ 
(upper) triangular matrices with all the diagonal elements equal to
$1$, that is 
\begin{align*}
H_3 &= \left\{ \left(\begin{array}{ccc} 1&a&b\\0&1&c\\0&0&1 \end{array}\right) 
\ \in GL(3,\C) \right\}. 
\end{align*}
Describe the Hopf algebra of complex representative (algebraic)
functions on $H_3$. 
\end{point}

\begin{point}{Exercise: Euclidean group} 
The group of rotations on the plane $\R^2$ is the special orthogonal
group 
\begin{align*}
SO(2,\R) &= 
\left\{ A \in GL(2,\R),\ \det A = 1,\ A^{-1}=A^t  \right\}. 
\end{align*}

The group of rotations acts on the group of translations
$T_2=(\R^2,+)$ as a product $Av$ of a matrix $A\in SO(2,\R)$ by a
vector $v\in\R^2$. 

The Euclidean group is the semi-direct product $E_2 = T_2 \rtimes SO(2,\R)$. 
That is, $E_2$ is the set of all the couples 
$(v,A) \in T_2 \times SO(2,\R)$, with the group law 
\begin{align*}
(v,A) \cdot (u,B) &:= (v+Au, AB). 
\end{align*}
\begin{enumerate}
\item 
Describe the Hopf algebra of real representative functions on $SO(2,\R)$. 
\item
Find a real faithful representation of $T_2$ of dimension 3. 
\item  
Describe the Hopf algebra of real representative functions on $E_2$. 
\end{enumerate}
\end{point}

\begin{point}{Group of invertible formal series}
\label{invertible-series}
The set 
\begin{align*}
\Gi(\C)&= \left\{ f(z)=\sum_{n=0}^\infty f_n\ z^n,\ f_n\in\C,\ f_0=1 \right\}
\end{align*}
of formal series in one variable, with constant term equal to $1$, 
is an Abelian group with  
\begin{itemize}
\item 
product: $\ds (f g)(z)= f(z)g(z) 
= \sum_{n=0}^\infty \left(\sum_{p+q=n} f_p\ g_q\right) z^n$; 
\item
unit: $1(z)=1$;
\item
inverse: by recursion, in fact \ 
$\ds(f f^{-1})(z)=\sum_{n=0}^\infty \left(\sum_{p+q=n} f_p\
(f^{-1})_q\right) z^n = 1$\ if and only if 
\begin{align*}
\mbox{for $n=0$} &\qquad f_0 (f^{-1})_0=1 \quad\Leftrightarrow\quad 
(f^{-1})_0=1 \quad\Leftrightarrow\quad f^{-1} \in\Gi(\C), \\ 
\mbox{for $n\geq 1$} &\qquad 
\sum_{p=0}^n f_p\ (f^{-1})_{n-p} =
f_0(f^{-1})_n+f_1(f^{-1})_{n-p}+\cdots
+f_{n-1}(f^{-1})_1+f_n(f^{-1})_0 \ =\ 0 \\ 
&\qquad \mbox{that is}\quad (f^{-1})_1=-f_1,\ (f^{-1})_2=f_1^2-f_2, ...
\end{align*}
\end{itemize}

This group has many finite-dimensional representations, of the form
\begin{align*}
\rho: \Gi(\C) & \longrightarrow GL_n(\C) \\
f(z)=\sum_{n=0}^\infty f_n\ z^n & \mapsto 
\left(\begin{array}{cccccc} 
1&f_1&f_2&f_3&...&f_{n-1}\\0&1&f_1&f_2&...&f_{n-2}\\0&0&1&f_1&...&f_{n-3}
\\&&&...&&\\0&0&&...&&1 \end{array}\right)
\end{align*} 
but they are never faithful! To have a faithful representation, 
we need to consider the map
\begin{align*}
\rho: \Gi(\C) & \longrightarrow GL_\infty(\C)
=\underset{\leftarrow}{\lim}\ GL_n(\C)\\
f(z) & \mapsto 
\left(\begin{array}{ccccc} 
1&f_1&f_2&f_3&...\\0&1&f_1&f_2&...\\0&0&1&f_1&...
\\&&...&&\\0&0&...&& \end{array}\right)
\end{align*} 
where $\underset{\leftarrow}{\lim}\ GL_n(\C)$ is the projective limit
of the groups $(GL_n(\C))_n$, that is, the limit of the groups such
that each $GL_n(\C)$ is identified with the quotient of $GL_{n+1}(\C)$
by its last column and row. 
Since $\underset{\leftarrow}{\lim}\ GL_n(\C)$ is not a group, it is 
necessary to restrict the image of the map $\rho$ to the triangular matrices 
$T_n(\C)$, whose projective limit $\underset{\leftarrow}{\lim}\ T_n(\C)$ 
indeed forms a group.\footnote{ 
Thanks to B.~Richter and R.~Holtkamp for pointing this to me.}

Therefore there are infinitely many local coordinates 
$x_n: \Gi(\C)\longrightarrow \C$, given by $x_n(f)=f_n$, which are 
free one from each other. 
Hence the algebra of local coordinates of $\Gi(\C)$ is the polynomial
ring 
\begin{align*}
R(\Gi(\C))&=\C[x_1,x_2,...,x_n,...]. 
\end{align*}
The Hopf structure is the following (with $x_0=1$): 
\begin{itemize}
\item 
Coproduct: $\D x_n= \sum_{k=0}^n x_k\otimes x_{n-k}$. 
\item
Counit: $\varepsilon(x_n)=\delta(n,0)$. 
\item
Antipode: recursively, from the 5-terms identity. In fact, for any
$n>0$ we have 
\begin{align*}
\varepsilon(x_n) 1= 0 &= \sum_{k=0}^n S(x_k)x_{n-k}
=S(1)x_n+S(x_1)x_{n-1}+S(x_2)x_{n-2}+\cdots+S(x_n)1
\end{align*}
and since $S(1)=1$ we obtain
$S(x_n) = -x_n-\sum_{k=1}^{n-1} S(x_k)x_{n-k}$. 
\end{itemize}
This Hopf algebra is isomorphic to the so-called {\em algebra of symmetric 
functions\/}, cf.~\cite{Macdonald}. 
\end{point}

\begin{point}{Group of formal diffeomorphisms}
\label{diffeomorphisms}
The set 
\begin{align*}
\Gd(\C)&=\left\{f(z)=\sum_{n=0}^\infty f_n\ z^{n+1},\ f_n\in\C,\ f_0=1\right\}
\end{align*}
of formal series in one variable, with zero constant term and linear
term equal to $1$, is a (non-Abelian) group with  
\begin{itemize}
\item 
product: given by the composition (or substitution)
\begin{align*}
(f\circ g)(z)& = f(g(z))=\sum_{n=0}^\infty f_n\ g(z)^n \\
&= z+(f_1+g_1)\ z^2+(f_2+2f_1g_1+g_2)\ z^3
+(f_3+3f_2g_1+2f_1g_2+f_1g_1^2+g_3)\ z^4 + \O(z^5). 
\end{align*}
\item
unit: $\id(z)=z$;
\item
inverse: given by the by the reciprocal series $f^{-1}$, such that 
$f\circ f^{-1}=\id=f^{-1}\circ f$, which can be found recursively, using for
instance {\em Lagrange Formula\/}, cf.~\cite{Stanley}. 
\end{itemize}

This group also has many finite-dimensional representations, which are
not faithful, and a faithful representation of infinite dimension:  
\begin{align*}
\rho: \Gd(\C) & \longrightarrow T_\infty(\C) 
=\underset{\leftarrow}{\lim}\ T_n(\C) \subset GL_\infty(\C)\\
f(z) & \mapsto 
\left(\begin{array}{cccccc} 
1&f_1&f_2&f_3&f_4&...\\0&1&2f_1&2f_2+f_1^2&2f_3+2f_1f_2&...
\\0&0&1&3f_1&3f_2+3f_1^2&...\\0&0&0&1&4f_1&...
\\&&&...&&\\0&0&&...&& \end{array}\right). 
\end{align*} 
Therefore there are infinitely many local coordinates 
$x_n: \Gd(\C)\longrightarrow \C$, given by $x_n(f)=f_n$, which are 
free one from each other. 
As in the previous example, the algebra of local coordinates of 
$\Gd(\C)$ is then the polynomial ring 
\begin{align*}
R(\Gd(\C))&=\C[x_1,x_2,...]. 
\end{align*}
The Hopf structure is the following (with $x_0=1$): 
\begin{itemize}
\item 
Coproduct: $\D x_n(f,g)= x_n(f\circ g)$, hence 
\begin{align*}
\D x_n = x_n\otimes 1+1\otimes x_n
+\sum_{m=1}^{n-1} x_m\otimes 
\underset{p_0,...,p_m\geq 0}{\sum_{p_0+p_1+\cdots +p_m=n-m}} 
x_{p_0}x_{p_1}\cdots x_{p_m}. 
\end{align*}
\item
Counit: $\varepsilon(x_n)=\delta(n,0)$. 
\item
Antipode: recursively, using
\begin{align*} 
S(x_n) = -x_n-\sum_{m=1}^{n-1} S(x_m)
\underset{p_0,...,p_m\geq 0}{\sum_{p_0+p_1+\cdots +p_m=n-m}} 
x_{p_0}x_{p_1}\cdots x_{p_m}. 
\end{align*}
\end{itemize}
This Hopf algebra is the so-called {\em Fa\`a di Bruno\/} Hopf
algebra, because the computations of the coefficients of the
Taylor expansion of the composition of two functions was firstly 
done by F.~Fa\`a di Bruno in \cite{FaadiBruno} (in 1855!). 
\end{point}


\subsection{Groups of characters and duality}

Let $\H$ be a commutative Hopf algebra over $\C$, with product $m$, 
unit $u$, coproduct $\D$, counit $\varepsilon$, antipode $S$ and
eventually an involution $*$. 

\begin{point}{Group of characters}
We call {\em character\/} of the Hopf algebra $\H$ a linear map 
$\alpha:\H\longrightarrow\C$ such that
\begin{enumerate}
\item 
$\alpha$ is a homomorphism of algebras, i.e. $\alpha(ab)=\alpha(a)\alpha(b)$;
\item 
$\alpha$ is unital, i.e. $\alpha(1)=1$. 
\end{enumerate}
Call $G_{\H}=\Hom_{Alg}(\H,\C)$ the set of characters of $\H$.
Given two characters $\alpha,\beta\in G_{\H}$, we call {\em convolution\/}
of $\alpha$ and $\beta$ the linear map
$\alpha\star\beta:\H\longrightarrow\C$ defined by 
$\alpha\star\beta = m_{\C}\circ(\alpha\otimes\beta)\circ\D$, that is, 
$\alpha\star\beta(a) = \sum \alpha(a_{(1)}) \beta(a_{(2)})$ for any
$a\in\H$. 
Applying the definitions, it is easy to prove the following properties: 
\begin{enumerate}
\item 
For any $\alpha,\beta\in G_{\H}$, the convolution $\alpha\star\beta$ 
is a unital algebra homomorphism, that is $\alpha\star\beta\in
G_{\H}$. 
\item 
The convolution product $G_{\H}\otimes G_{\H}\longrightarrow G_{\H}$ 
is associative. 
\item 
The counit $\varepsilon:\H\longrightarrow\C$ is the unit of the
convolution. 
\item 
For any $\alpha\in G_{\H}$, the homomorphism $\alpha^{-1}= \alpha\circ
S$ is the inverse of $\alpha$. 
\item 
The convolution product is commutative if and only if the coproduct is
cocommutative. 
\end{enumerate}
In other words, the set of characters $G_{\H}$ forms a group with the
convolution product. 
\end{point}

\begin{point}{Real subgroups}
If $\H$ is a commutative Hopf algebra endowed with an involution 
$*:\H\longrightarrow\H$ compatible with the Hopf structure, in the
sense that 
\begin{align*}
(ab)^* = b^* a^*, &\quad 1^* = 1 \\ 
\D(a^*)=(\D a)^*, &\quad \varepsilon(a^*) = \varepsilon(a), 
\qquad S(a^*)=(S a)^*, 
\end{align*}
then the subset 
\begin{align*}
G_{\H}^* &= \Hom_{*Alg}(\H,\C) 
= \Big\{\alpha\in G_{\H}, \alpha(a^*)=\overline{\alpha(a)} \Big\} 
\end{align*}
is a (real) subgroup of $G_{\H}$. 
\end{point}

\begin{point}{Compact subgroups}
If, furthermore, $\H$ is a commutative *Hopf algebra, finitely generated
and endowed with a Haar measure compatible with the Hopf structure,
that is, a linear map 
$\mu:\H\longrightarrow\R$ such that  
\begin{align*}
&(\mu\otimes\Id)\D = (\Id\otimes\mu)\D = u\circ\mu, \\
&\mu(a a^*)> 0 \qquad\mbox{for all $a\neq 0$},
\end{align*}
then $G_{\H}^*$ is a compact Lie group. 
\end{point}

\begin{point}{Comparision of $SL(2,\C)$, $SL(2,\R)$ and $SU(2)$}
Consider the commutative algebra 
$\ds\H=\frac{\C[a,b,c,d]}{\la ad-bc-1 \ra}$. 
If on $\H$ we consider the Hopf structure
\begin{align*}
\D \left(\begin{array}{cc} a&b\\c&d \end{array}\right) &= 
\left(\begin{array}{cc} a&b\\c&d \end{array}\right) 
\otimes \left(\begin{array}{cc} a&b\\c&d \end{array}\right) \\
\varepsilon \left(\begin{array}{cc} a&b\\c&d \end{array}\right) &= 
\left(\begin{array}{cc} 1&0\\0&1 \end{array}\right) \\
S \left(\begin{array}{cc} a&b\\c&d \end{array}\right) &= 
\left(\begin{array}{cc} d&-b\\-c&a \end{array}\right), 
\end{align*}
then $G_{\H}=SL(2,\C)$. 
If in addition we consider the involution 
\begin{align*}
\left(\begin{array}{cc} a&b\\c&d \end{array}\right)^* &= 
\left(\begin{array}{cc} a&b\\c&d \end{array}\right), 
\end{align*}
then $G_{\H}^*=SL(2,\R)$. 
If, instead, we consider the involution 
\begin{align*}
\left(\begin{array}{cc} a&b\\c&d \end{array}\right)^* &= 
\left(\begin{array}{cc} d&-c\\-b&a \end{array}\right), 
\end{align*}
then $G_{\H}^*=SU(2)$. 
\end{point}

\begin{point}{Duality}
We have seen first how to associate a Hopf algebra to a group, through
a functor $R$, and then how to associate a group to a Hopf algebra, through
a functor $G$. In general, these two functors are {\em adjoint\/} one to each
other, that is
\begin{align*}
\Hom_{Groups}(G,G_{\H}) \quad\cong\quad \Hom_{Alg}(\H,R(G)).  
\end{align*}
Sometimes, these two functors are dual one to each other. In
particular, we have the following results: 
\begin{itemize}
\item
Given a complex group $G$, and its Hopf algebra $R(G)$ of representative
functions, the map 
\begin{align*}
\Phi: G & \longrightarrow G_{R(G)}= \Hom_{Alg}(R(G),\C) \\ 
x & \mapsto \Phi_x:\R(G)\rightarrow \C, \Phi_x(f)=f(x)
\end{align*}
defines an isomorphism of groups to the characters group of $R(G)$. 
This result must be refined to the group $G_{R(G)}^*$ if $G$ is real. 
It is known as {\em Tannaka duality\/} for compact Lie groups. 

\item
Viceversa, given a commutative Hopf algebra over $\C$, the complex
group $G$ can be defined as the group of characters of $\H$, that is, 
by stating that its coordinate functions are given by $\H$.
If the Hopf algebra $\H$ has an involution and a Haar measure, and it
is finitely generated, then the map 
\begin{align*}
\Psi: \H & \longrightarrow R(G^*(\H)) \\ 
a & \mapsto \Psi_a:\Hom_{*Alg}(\H,\C)\rightarrow \C, \Psi_a(\alpha)=\alpha(a)
\end{align*}
defines an isomorphism of Hopf algebras. The underlying group is
compact, and this result is known as the {\em Krein duality\/}. 
\end{itemize}
\end{point}

\begin{point}{Algebraic and proalgebraic groups}
\label{proalgebraic group}
As we saw in the most of the examples, the group structure of many
groups does not depend on the field where the coefficients take
value. This is the case of matrix groups, but also of the groups of
formal series. Apart from the coefficients, such groups have in common
the form of their coordinate ring, that is the Hopf algebra $\H$. 
They are better described as follows. 

Given a commutative Hopf algebra $\H$ which is finitely generated, 
we call {\em algebraic group\/} associated to $\H$ the functor 
\begin{align*}
G_{\H}: \quad & \{\mbox{Commutative, associative algebras}\} 
\longrightarrow \{\mbox{Groups}\} \\ 
& A \mapsto G_{\H}(A) = \Hom_{Alg}(\H,A),
\end{align*}
where $G_{\H}(A)$ is a group with the convolution product. 
If $\H$ is not finitely generated, we call {\em proalgebraic group\/} 
the same functor. 

In particular, all the matrix groups $SL_n$, $GL_n$, etc., can have matrix 
coefficients in any commutative algebra $A$, not only $\C$ or $\R$, and 
therefore are algebraic groups. Similarly, the groups of formal series 
$\Gi$, $\Gd$, with coefficients in any commutative algebra $A$, 
are proalgebraic groups.  
\end{point}


\section*{Lecture II - Review on field theory}
\addcontentsline{toc}{section}{\bf Lecture II - Review on field theory}
\label{lecture2}

\subsection{Review of classical field theory} 
\label{ReviewCFT}

In this section we briefly review the standard Lagrangian tools 
applied to fields, and the main examples of solutions of the 
Euler-Lagrange equations.  

\begin{point}{Space-time} 
The space-time coordinates are points in the Minkowski space $\R^{1,3}$, 
that is, the space endowed with the flat diagonal metric 
$g=(1,-1,-1,-1)$. A transformation, called {\em Wick's rotation\/}, 
allows to reformulate the problems on the Euclidean space $\R^4$. 
For more generality, we then consider an Eucledian space $\R^D$ of 
dimension $D$, and we denote the space-time coordinates by $x=(x^\mu)$, 
with $\mu=0,1,...,D-1$.
\end{point}

\begin{point}{Classical fields} 
A {\em field\/} is a section of a bundle on a base space. If the base space 
is flat, as in the case we consider here, a field is just a vector-valued 
function. 
By {\em classical field\/}, we mean a real function 
$\phi:\R^D\longrightarrow \R$ of class $C^\infty$, with compact support 
and rapidly decreasing. 
To be precise, we can take the function $\phi$ in the Schwartz space 
$\S(\R^D)$, that is, $\phi$ is a $C^\infty$ function such that all 
its derivatives $\partial_\mu^n \phi$ converge rapidly to zero 
for $|x|\to\infty$.

The {\em observables\/} of the system described by a field $\phi$, that is, 
the observable quantities, are real functionals $F$ of the field $\phi$, 
and what can be measured of these observables are the values $F(\phi)\in\R$. 
To determine all the observables it is enough to know the field itself. 

When the field $\phi:\R^D\longrightarrow \C$ has complex (unreal) values, 
or vector values $\C^4$, it is called a {\em wave function\/}.  
In this case, what can be measured is not the value $\phi(x)$ itself, 
for any $x\in\R^D$, but rather the real value $|\phi(x)|^2$ which describes 
the probability to find the particle in the position $x$. 
\end{point}

\begin{point}{Euler-Lagrange equation}
A classical field is determined as the solution of a partial 
differential equation, called the {\em field equation\/}, which 
encodes its evolution. To any system is associated a {\em Lagrangian 
density\/}, that is a real function $\L:\R^D\longrightarrow\R$, 
$x\mapsto \L(x,\phi(x),\partial\phi(x))$, 
where $\partial\phi$ denotes the gradient of $\phi$. 
By {\em Noether's theorem\/}, the dynamics of the field $\phi$ is such that 
the {\em symmetries\/} of the field (i.e. the transformations which leave 
the Lagrangian invariant) are conserved. 
This conservation conditions are turned into a field equation by means
of the {\em action\/} of the field $\phi$: it is the functional $S$ of
$\phi$ given by 
\begin{align*}
\phi\mapsto 
S[\phi]=\int_{\R^D} d^Dx\ \L(x,\phi(x),\partial\phi(x)).
\end{align*}
The action $S$ is {\em stationary\/} in $\phi\in\S(\R^D)$ 
if for any other function $\delta\phi\in\S(\R^D)$ we have 
$\frac{d}{dt} S[\phi+t \delta\phi]_{|t=0}=0$. 
Then, {\em Hamilton's principle of least (or stationary) action\/} states 
that a field $\phi$ satisfies the field equation if and only if the 
action $S$ is stationary in $\phi$. 
In terms of the Lagrangian, the field equation results into the so-called 
{\em Euler-Lagrange equation\/}
\begin{align}
\label{Euler-Lagrange equation}
\left[\frac{\partial\L}{\partial\phi} - 
\sum_{\mu} \partial_{\mu}
\left(\frac{\partial\L}{\partial(\partial_{\mu}\phi)}\right)\right]
(x,\phi(x)) = 0. 
\end{align}
This is the equation that we have to solve to find the classical field $\phi$. 
In general, it is a non-homogenous and non-linear partial differential 
equation, where the non-homogeneous terms appear if the system is not 
isolated, and the non-linear terms appear if the field is self-interacting. 

For example, a field with Lagrangian density 
\begin{align}
\label{Klein-Gordon-Lagrangian}
\L(x,\phi(x),\partial\phi(x)) 
&= \frac{1}{2}\left(|\partial_\mu \phi(x)|^2+m^2\phi(x)^2\right)
- J(x)\phi(x) - \frac{\lambda}{3!}\phi(x)^3 - \frac{\mu}{4!}\phi(x)^4
\end{align} 
is subject to the Euler-Lagrange equation 
\begin{align}
\label{Klein-Gordon-equation}
(-\Delta+m^2)\phi(x) 
&= J(x)+\frac{\lambda}{2}\phi(x)^2+\frac{\mu}{3!}\phi(x)^3, 
\end{align}
where we denote 
$\Delta\phi(x)=\sum_{\mu}\partial_{\mu}\left(\partial_{\mu}\phi(x)\right)$. 
This equation is called the {\em Klein-Gordon equation\/}, because the operator 
$-\Delta+m^2$ is called the {\em Klein-Gordon operator\/}. 
\end{point}

\begin{point}{Free and interacting Lagrangian}
A generic relativistic particle with mass $m$, described by a field $\phi$, 
can have a Lagrangian density of the form 
\begin{align}
\label{general-Lagrangian}
\L(x,\phi(x),\partial\phi(x)) &= \frac{1}{2}\ \phi^t(x) A \phi(x)
- J(x)\phi(x) - \frac{\lambda}{3!}\phi(x)^3 - \frac{\mu}{4!}\phi(x)^4,  
\end{align} 
where $A$ is a differential operator such as the Dirac operator or
the Laplacian, typically summed up with the operator of multiplication
by the mass or its square. 
The term $\frac{1}{2}\phi^t A \phi$ (quadratic in $\phi$) is the kinetic 
term. It is also called {\em free Lagrangian density\/}, and denoted 
by $\L_0$. 

The field $J$ is an external field, which may represent a source for the 
field $\phi$. If $J=0$, the system described by $\phi$ is {\em isolated\/}, 
that is, it is placed in the {\em vacuum\/}. The term of the Lagrangian 
containing $J$ (linear in $\phi$) is the same for and field theory. 

The parameters $\lambda$, $\mu$ are called {\em coupling constants\/}, 
because they express the self-interactions of the field. They are usually 
measurable parameters such as the electric charge or the flavour, but can 
also be unphysical parameters added for convenience. 
The sum of the terms which are non-quadratic in $\phi$ (and non-linear) 
is called {\em interacting Lagrangian density\/}, and denoted by $\L_{int}$. 
\end{point}

\begin{point}{Free fields}
A free field, that we shall denote by $\phi_0$, has the dynamics of a free 
Lagrangian $\L(\phi_0)=\frac{1}{2}\ \phi_0^t A \phi_0 - J\ \phi_0$. 
The Euler-Lagrange equation is easily written in the form 
\begin{align}
\label{Euler-Lagrange-free}
A \phi_0(x) &= J(x). 
\end{align}
The general solution of this equation is well known to be the sum 
$\phi_0^g+\phi_0^p$ of the general solution of the homogeneous equation 
$A \phi_0^g(x) = 0$, and a particular solution $\phi_0^p(x)$ of the 
non-homogeneous one. 
In the Minkowski space-time the function $\phi_0^g$ is a wave (superposition 
of plane waves), in the Euclidean space-time the formal solution $\phi_0^g$ 
is not a Schwartz function and we do not consider it. 
Therefore the function $\phi_0$ is the convolution 
\begin{align*}
\phi_0(x) &= \int d^Dy\ G_0(x-y)\ J(y),  
\end{align*}
where $G_0(x)$ is the {\em Green's function\/} of the operator $A$, 
that is, the distribution such that $A G_0(x)=\delta(x)$. 
The physical interpretation of the convolution is that from each point $y$ 
of its support, the source $J$ affects the field $\phi$ at the position $x$ 
through the action of $G_0(x-y)$, which is then regarded as the 
{\em field propagator\/}. 

For instance, if $A=-\Delta+m^2$ is the Klein-Gordon operator, 
the Green's function $G_0$ is the distribution defined by the Fourier 
transformation 
\begin{align}
\label{KG-propagator}
G_0(x-y) &= \int_{\R^D} \frac{d^Dp}{(2\pi)^D}\ 
\frac{1}{p^2+m^2}\ e^{-i\ p\cdot (x-y)}. 
\end{align}
\end{point}

\begin{point}{Self-interacting fields}
A field $\phi$ with Lagrangian density of the form (\ref{general-Lagrangian})
satisfies the Euler-Lagrange equation 
\begin{align*}
A\phi(x) &= J(x)+\frac{\lambda}{2}\ \phi(x)^2+\frac{\mu}{3!}\ \phi(x)^3. 
\end{align*}
This differential equation is non-linear, and in general can not be solved 
exactely. If the coupling constants $\lambda$ and $\mu$ are suitably small, 
we solve it {\em perturbatively\/}, that is, we regard the interacting terms 
as perturbations of the free ones. In fact, the Euler-Lagrange equation 
can be expressed as a recursive equation 
\begin{align*}
\phi(x) &= \int_{\R^D} d^Dy\ G_0(x-y)\ 
\left[J(y)+\frac{\lambda}{2}\ \phi(y)^2+\frac{\mu}{3!}\ \phi(y)^3 \right], 
\end{align*}
where $G_0$ is the Green's function of $A$. This equation can then be 
solved as a formal series in the powers of $\lambda$ and $\mu$. 

For instance, let us consider the simpliest Lagrangian 
(\ref{general-Lagrangian}) with $\mu=0$, whose Euler-Lagrange equation is 
\begin{align}
\label{Euler-Lagrange-equation}
\phi(x) &= \int_{\R^D} d^Dy\ G_0(x-y)\ 
\left[J(y)+\frac{\lambda}{2}\phi(y)^2 \right]. 
\end{align}
If on the right hand-side of Eq.~(\ref{Euler-Lagrange-equation}) 
we replace $\phi(y)$ by its value, and we repeat the substitutions 
recursively, we obtain the following perturbative solution: 
\begin{align}
\label{Euler-Lagrange-solution}
\phi(x) &= \int d^Dy\ G_0(x-y)\ J(y) \\ 
&\nonumber 
+ \frac{\lambda}{2} \int d^Dy\ d^Dz\ d^Du\  
G_0(x-y)\ G_0(y-z)\ G_0(y-u)\ J(z)\ J(u) \\
&\nonumber 
+ \frac{2\lambda^2}{4} \int d^Dy\ d^Dz\ d^Du\ d^Dv\ d^Dw\ 
G_0(x-y)\ G_0(y-z)\ G_0(y-u)\ G_0(z-v)\ G_0(z-w)\\ 
&\nonumber 
 \hspace{3cm} \times J(z)\ J(u)\ J(v)\ J(w) \\
&\nonumber 
+ \frac{\lambda^3}{8} \int d^Dy\ d^Dz\ d^Du\ d^Dv\ d^Dw\ d^Ds\ d^Dt\   
G_0(x-y)\ G_0(y-z)\ G_0(y-u)\ G_0(z-v)\ G_0(z-w)\\ 
&\nonumber 
 \hspace{3cm} \times G_0(u-s)\ G_0(u-t)\ J(z)\ J(u)\ J(v)\ J(w)\
J(s)\ J(t)\ 
+ \O(\lambda^4)
\end{align}
which describes the self-interacting field in presence of an external
field $J$. 
\end{point}

\begin{point}{Conclusion}
\label{Conclusion-ReviewCFT}
To summerize, a typical classical field $\phi$ with Lagrangian
density of the form 
\begin{align*}
\L(\phi) &= \frac{1}{2}\ \phi^t A \phi-J(x)\ \phi(x)
-\frac{\lambda}{3!}\ \phi(x)^3, 
\end{align*}
can be described perturbatively as a formal series 
\begin{align*}
\phi(x) &= \sum_{n=0}^\infty \lambda^n\ \phi_n(x)
\end{align*}
in the powers of the coupling constant $\lambda$. 
Each coefficient $\phi_n(x)$ is a finite sum of integrals
involving only the field propagator and the source. 
We describe these coefficients in Lecture~III, 
using Feynman graphs. 
\end{point}


\subsection{Review of quantum field theory}
\label{ReviewQFT}

In this section we briefly review the standard tools to describe 
quantum fields. 

\begin{point}{Minkowski versus Euclidean approach}
In the Minkowski space-time coordinates, the quantization procedure 
is the so-called {\em canonical quantization\/}, based on the principle 
that the observables of a quantum system are self-adjoint operators acting 
on a Hilbert space whose elements are the {\em states\/} in which the 
system can be found. The probability that the measurerement of an observable 
$F$ is the value carried by a state $v$ is given by the expectation value 
$\la v|F|v \ra\in\R$. In this procedure, the quantum fields are {\em field 
operators\/}, which must be defined together with the Hilbert space of states 
on which they act. 

A standard way to deal with quantum fields is to Wick rotate the time, 
through the transformation $t\mapsto -it$, and therefore transform the 
Minkowski space-time into a Euclidean space. The quantum fields are then 
treated as {\em statistical fields\/}, that is, classical fields or wave 
functions $\phi$ which fluctuate around their expectation values. 
The result is equivalent to that of the Minkowski approach, and this 
quantization procedure is the so-called {\em path integral quantization\/}. 
\end{point}

\begin{point}{Green's functions through path integrals}
The first interesting expectation value is the mean value $\la\phi(x)\ra$ 
of the field $\phi$ at the point $x$. More generally, we wish 
to compute the {\em Green's functions\/} 
$\la \phi(x_1)\cdots\phi(x_k) \ra$, which represent the probability that 
the quantum field $\phi$ moves from the point $x_k$ to $x_{k-1}$ and so on, 
and reaches $x_1$. 

A quantum field does not properly satisfy the principle of stationary action, 
but can be interpretated as a fluctuation around the classical solution 
of the Euler-Lagrange equation. On the Euclidean space, the probability to 
observe the quantum field at the value $\phi$ is proportional to 
$\exp\left(-\frac{S[\phi]}{\hbar}\right)$
\footnote{On the Minkowski space this value is 
$\exp\left(i\frac{S[\phi]}{\hbar}\right)$.},  
where $\hbar=\frac{h}{2\pi}$ is the reduced Planck's constant. 
When $\hbar\to 0$ (classical limit), we recover a maximal probability 
to find the field $\phi$ at the minimum of the action, that is, 
to recover the classical solution of Euler-Lagrange equation. 
The Green's functions can then be computed as the {\em path integrals\/} 
\begin{align*}
\la \phi(x_1)\cdots\phi(x_k) \ra 
&= \frac{\int d\phi\ \phi(x_1)\cdots\phi(x_k)\ 
e^{-\frac{S[\phi]}{\hbar}}|_{J=0}}
{\int d\phi\ e^{-\frac{S[\phi]}{\hbar}}|_{J=0}}. 
\end{align*}

This approach presents a major problem: on the infinite dimensional set of 
classical fields, that we may fix as the Schwartz space $\S(\R^D)$, 
for $D>1$ there is no measure $d\phi$ suitable to give a meaning to such 
an integral. (For $D=1$ the problem is solved on continuous functions by 
the Wiener's measure.) 
However, assuming that we can give a meaning to the path integrals, 
this formulation allows to recover the classical values, 
for instance $\la\phi(x)\ra\sim\phi(x)$, when $\hbar\to 0$. 
\end{point}

\begin{point}{Free fields}
The quantization of a classical free field $\phi_0$ is easy. 
In fact, the action $S_0[\phi_0]= \frac{1}{2} \int d^Dx\ \phi_0(x)A\phi_0(x)$ 
is quadratic in $\phi_0$ and gives rise to a Gaussian measure,  
$\exp{\left(-\frac{S[\phi_0]}{\hbar}\right)} d\phi_0$. 
If the field is isolated, the Green's functions are then easily computed: 
\begin{itemize}
\item 
the mean value $\la\phi_0(x)\ra$ is zero; 
\item 
the 2 points Green's function $\la\phi_0(x)\phi_0(y)\ra$ coincides 
with the Green's function $G_0(x-y)$; 
\item
all the Green's functions on an odd number of points are zero; 
\item
the Green's functions on an even number of points are products 
of Green's functions exhausting all the points. 
\end{itemize}

If the field is not isolated, instead, as well as when the field is 
self-interacting, the computation of the Green's functions are more 
involved. 
\end{point}

\begin{point}{Dyson-Schwinger equation}
In general, the Green's functions satisfy an integro-differential equation 
which generalises the Euler-Lagrange equation, written in the form 
$\frac{\partial S[\phi]}{\partial\phi(x)}=0$. 
To obtain this equation, in analogy with the analysis that one would perform 
on a finite dimensional set of paths, one can proceed by introducing a 
generating functional for Green's functions. 
The self-standing of the results is considered sufficient to accept the 
intermediate meaningless steps. 

To do it, let us regard the action as a function also of the classical 
source field $J$, that is $S[\phi]=S[\phi,J]$.  
Then we define the {\em partition function\/} 
\begin{align*}
Z[J] &= \int d\phi\ e^{-\frac{S[\phi]}{\hbar}}, 
\end{align*}
and we impose the normalization condition $Z[J]|_{J=0}=1$. 
It is then easy to verify that the Green's functions can be derived 
from the partition function, as
\begin{align*}
\la \phi(x_1)\cdots\phi(x_k) \ra &= \frac{\hbar^k}{Z[J]}\ 
\frac{\delta^k Z[J]}{\delta J(x_1)\cdots\delta J(x_k)}|_{J=0},  
\end{align*}
where $\frac{\delta}{\delta J(x)}$ is the functional derivative. 
The Dyson-Schwinger equation for Green's functions, then, can be 
deduced from a functional equation which constrains the partition function: 
\begin{align*}
\frac{\delta S}{\delta\phi(x)}
\left[\hbar\frac{\delta}{\delta J}\right]\ Z[J]=0. 
\end{align*}
The notation used on the left hand-side means that in the functional 
$\frac{\delta S}{\delta\phi(x)}$ of $\phi$, we substitute the 
variable $\phi$ with the operator $\hbar\frac{\delta}{\delta J}$. 
Since $S[\phi]$ is a poynomial, we obtain an operator which contains some 
repeted derivations with respect to $J$, and which can then act on $Z[J]$. 
\end{point}

\begin{point}{Connected Green's functions}
If, starting from the partition function, we define the {\em free energy\/} 
\begin{align*}
W[J] &= \hbar \log Z[J], \quad\mbox{i.e.}\quad Z[J]=e^{\frac{W[J]}{\hbar}}, 
\end{align*}
with normalization condition $W[J]|_{J=0}=0$, we see that the Green's 
functions are sums of recursive terms (products of Green's functions 
on a smaller number of points), and additional terms which involve the 
derivatives of the free energy:  
\begin{align*}
\la\phi(x)\ra 
&= \frac{\hbar}{Z[J]}\ \frac{\delta Z[J]}{\delta J(x)}|_{J=0} 
= \frac{\delta W[J]}{\delta J(x)}|_{J=0}, \\ 
\la\phi(x)\phi(y)\ra &= \la\phi(x)\ra\ \la\phi(y)\ra 
+ \hbar \frac{\delta^2 W[J]}{\delta J(x)\delta J(y)}|_{J=0}, ...
\end{align*}
These additional terms 
\begin{align*}
G(x_1,...,x_k) &= 
\frac{\delta^k W[J]}{\delta J(x_1)\cdots\delta J(x_k)}|_{J=0}
\end{align*}
are called {\em connected Green's functions\/}, for reasons which will 
be clear after we introduced the Feynman diagrams. 
Of course, knowing the connected Green's functions $G(x_1,...,x_k)$ 
is enough to recover the {\em full\/} Green's functions 
$\la\phi(x_1)\cdots\phi(x_k)\ra$, through the relations: 
\begin{align}
\label{full-connected-green-functions}
\la\phi(x)\ra &= G(x), \\ \nonumber
\la\phi(x)\phi(y)\ra &= G(x)\ G(y)+\hbar\ G(x,y), \\ \nonumber
\la\phi(x)\phi(y)\phi(z)\ra &= G(x)\ G(y)\ G(z)
+\hbar\ \left[G(x)\ G(y,z)+G(y)\ G(x,z)+G(z)\ G(x,y)\right] 
+\hbar^2\ G(x,y,z), \\ \nonumber
\la\phi(x)\phi(y)\phi(z)\phi(u)\ra &= G(x)\ G(y)\ G(z)\ G(u) \\ \nonumber 
&\hspace{1cm} +\hbar\ \left[G(x)\ G(y)\ G(z,u)+\mbox{terms}\right] 
\\ \nonumber 
&\hspace{1cm} +\hbar^2\ \left[G(x,y)\ G(z,u)+\mbox{terms}+ 
G(x)\ G(y,z,u)+\mbox{terms}\right] \\ \nonumber 
&\hspace{1cm} +\hbar^3\ G(x,y,z,u)
\end{align}
and so on, where by ``terms'' we mean the same products
evaluated on suitable permutations of the points $(x,y,z,u)$. 
\end{point}

\begin{point}{Self-interacting fields}
The Dyson-Schwinger equation can be expressed in terms of the connected 
Green's functions. To be precise, we consider the typical quantum field 
with classical action 
\begin{align*}
S[\phi] &= \frac{1}{2} \phi^t A \phi - J^t \phi 
-\frac{\lambda}{3!}\int d^D x\ \phi(x)^3, 
\end{align*}
and we denote by $G_0=A^{-1}$ the resolvent of the operator $A$. 
Then, the Dyson-Schwinger equation for the 1-point Green's
function of a field in an external field $J$ is 
\begin{align}
\label{Dyson-Schwinger-equation-W}
\la\phi(x)\ra_J = \frac{\delta W[J]}{\delta J(x)} &= 
\int d^D u\ G_0(x-u) \left[J(u) +\frac{\lambda}{2} 
\left[\left(\frac{\delta W[J]}{\delta J(u)}\right)^2 
+ \hbar \frac{\delta^2 W[J]}{\delta J(u)^2}\right]\right].  
\end{align}

If we evaluate Eq.~(\ref{Dyson-Schwinger-equation-W}) at $J=0$, we obtain the 
Dyson-Schwinger equation for the 1-point Green's function of an
isolated field: 
\begin{align}
\label{Dyson-Schwinger-equation-1}
\la\phi(x)\ra=G(x) &= \frac{\lambda}{2} 
\int d^D u\ G_0(x-u) \left[G(u)^2 + \hbar\ G(u,u)\right]. 
\end{align}
If we derive Eq.~(\ref{Dyson-Schwinger-equation-W}) by 
$\frac{\delta}{\delta J(y)}$, and evaluate at $J=0$, we obtain 
the Dyson-Schwinger equation for the 2-points connected Green's 
function: 
\begin{align}
\label{Dyson-Schwinger-equation-2}
G(x,y) &= G_0(x-y) + \frac{\lambda}{2} \int d^D u\ G_0(x-u) 
\left[2\ G(u)\ G(u,y) + \hbar\ G(u,u,y)\right],  
\end{align}
which involves the 3-points Green's function. 
Repeating the derivation, we get the Dyson-Schwinger equation for the 
n-points connected Green's function. 

As for classical interacting fields, these equations can be solved 
perturbatively. For instance, the solution of 
Eq.~(\ref{Dyson-Schwinger-equation-W}), that is the mean value of a
field $\phi$ in an external field $J$, is: 
\begin{align}
\label{Dyson-Schwinger-solution-W}
\la\phi(x)\ra_J &= \int d^D u\ G_0(x-u)J(u) \\ 
&\nonumber 
+ \frac{\lambda}{2} \int d^Dy\ d^Dz\ d^Du\  
G_0(x-y)\ G_0(y-z)\ G_0(y-u)\ J(z)\ J(u) \\
&\nonumber 
+ \frac{2\lambda^2}{4} \int d^Dy\ d^Dz\ d^Du\ d^Dv\ d^Dw\ 
G_0(x-y)\ G_0(y-z)\ G_0(y-u)\ G_0(z-v)\ G_0(z-w)\\ 
&\nonumber 
 \hspace{3cm} \times J(z)\ J(u)\ J(v)\ J(w) \\
&\nonumber 
+ \hbar \frac{\lambda}{2} \int d^D y\ G_0(x-y)\ G_0(y-y)\\  
&\nonumber 
+ \hbar \frac{\lambda^2}{2} \int d^Dy\ d^Dz\ d^D u\ G_0(x-y)\
G_0(y-z)^2\ G_0(z-u)\ J(u) 
+ \O(\lambda^3). 
\end{align}
Of course, the mean value of the isolated field, that is the solution
of Eq.~(\ref{Dyson-Schwinger-equation-1}), is then obtained by
setting $J=0$: 
\begin{align}
\label{Dyson-Schwinger-solution-1}
G(x) &= \hbar \frac{\lambda}{2} \int d^D y\ G_0(x-y)\ G_0(y-y) 
+ \O(\lambda^3). 
\end{align}
\end{point}

\begin{point}{Exercise: 2-points connected Green's function} 
\label{G(x,y)-value}
Compute the first perturbative terms of the solution of 
Eq.~(\ref{Dyson-Schwinger-equation-2}), which represents the
Green's function $G(x,y)$ for an isolated field ($J=0$).  
\end{point}

\begin{point}{Conclusion}
\label{Conclusion-ReviewQFT}
For a typical quantum field $\phi$ with classical Lagrangian density 
of the form 
\begin{align*}
\L(\phi) &= \frac{1}{2}\ \phi^t A \phi
-\frac{\lambda}{3!}\ \phi(x)^3, 
\end{align*}
\begin{itemize}
\item
the full $k$-points Green's function $\la\phi(x_1)\cdots\phi(x_k)\ra$ 
is the sum of the products of the connected Green's functions 
exhausting the $k$ external points; 
\item 
the connected $k$-points Green's function can be described 
perturbatively as a formal series 
\begin{align*}
G(x_1,...,x_k) &= \sum_{n=0}^\infty \lambda^n\ G_n(x_1,...,x_k)
\end{align*}
in the powers of the coupling constant $\lambda$; 
\item 
the constant coefficient $G_0(x_1,...,x_k)$ is the Green's function 
of the free field; 
\item 
each higher order coefficient $G_n(x_1,...,x_k)$ is a finite sum of integrals 
involving only the free propagator. 
\end{itemize}
We describe the sums appearing in $G_n(x_1,...,x_k)$ in 
Lecture~III using Feynman graphs. 
\end{point}


\section*{Lecture III - Formal series expanded over Feynman graphs}
\addcontentsline{toc}{section}{\bf Lecture III - Formal series expanded 
over Feynman graphs}
\label{lecture3}

In this lecture we consider a quantum field $\phi$ with classical 
Lagrangian density of the form 
\begin{align*}
\L(\phi) &= \frac{1}{2}\ \phi^t A \phi-J(x)\ \phi(x)
-\frac{\lambda}{3!}\ \phi(x)^3,   
\end{align*}
where $A$ is a differential operator, typically the Klein-Gordon
operator. We denote by $G_0$ the Green's function of $A$. 
We saw in Section~\ref{ReviewQFT} that the Green's functions of this 
field are completely determined by the connected Green's functions, and 
that these can only be described as formal series in the powers of the 
coupling constant, 
\begin{align*}
G(x_1,...,x_k) &= \sum_{n=0}^\infty \lambda^n\ G_n(x_1,...,x_k).   
\end{align*}
In this section we describe the coefficients $G_n(x_1,...,x_k)$ using Feynman
diagrams. We begin by describing the coefficients of the perturbative solution 
$\phi(x)= \sum \lambda^n\ \phi_n(x)$ for the classical field.


\subsection{Interacting classical fields}
\label{series-classical}

\begin{point}{Feynman notations}
We adopt the following Feynman's notations for the field $\phi$: 
\begin{itemize}
\item
field\ $\phi(x)$ = 
\begin{picture}(16,4)(-6,0)
  \parbox{0cm}{\begin{center}
  \begin{fmfgraph*}(6,2)
  \setval
  \fmfforce{0w,1h}{v1}
  \fmfforce{1w,1h}{v2}
  \fmf{plain}{v1,v2}
  \fmfdot{v1}
  \fmflabel{$x$}{v1}
  \fmfblob{.5w}{v2}
  \end{fmfgraph*}
  \end{center}}
\end{picture};  
\item 
source\  $J(y)$ = 
\begin{picture}(14,4)(-2,0)
  \parbox{0cm}{\begin{center}
  \begin{fmfgraph*}(6,2)
  \setval
  \fmfforce{0w,1h}{v1}
  \fmfforce{1w,1h}{v2}
  \fmf{plain}{v1,v2}
  \fmfv{decor.shape=cross,decor.size=.4w}{v2}
  \fmflabel{$y$}{v2}
  \end{fmfgraph*}
  \end{center}}
\end{picture};  
\item
propagator\ $G_0(x-y)$ = 
\begin{picture}(16,4)(-6,0)
  \parbox{0cm}{\begin{center}
  \begin{fmfgraph*}(6,2)
  \setval
  \fmfforce{0w,1h}{v1}
  \fmfforce{1w,1h}{v2}
  \fmf{plain}{v1,v2}
  \fmfdot{v1,v2}
  \fmflabel{$x$}{v1}
  \fmflabel{$y$}{v2}
  \end{fmfgraph*}
  \end{center}}
\end{picture}.  
\end{itemize}
For each graphical object resulting from Feynman's notation, we call 
{\em amplitude\/} its analytical value. 
\end{point}

\begin{point}{Euler-Lagrange equation} 
The Euler-Lagrange equation (\ref{Euler-Lagrange-equation}) is 
represented by the following diagrammatic equation: 
\begin{align}
\label{Euler-Lagrange-diagrams}
\begin{picture}(16,4)(-6,0)
  \parbox{0cm}{\begin{center}
  \begin{fmfgraph*}(6,2)
  \setval
  \fmfforce{0w,1h}{v1}
  \fmfforce{1w,1h}{v2}
  \fmf{plain}{v1,v2}
  \fmfdot{v1}
  \fmflabel{$x$}{v1}
  \fmfblob{.5w}{v2}
  \end{fmfgraph*}
  \end{center}}
\end{picture} 
&= 
\begin{picture}(16,4)(-6,0)
  \parbox{0cm}{\begin{center}
  \begin{fmfgraph*}(6,2)
  \setval
  \fmfforce{0w,1h}{v1}
  \fmfforce{1w,1h}{v2}
  \fmf{plain}{v1,v2}
  \fmfdot{v1}
  \fmfv{decor.shape=cross,decor.size=.4w}{v2}
  \fmflabel{$x$}{v1}
  \fmflabel{$y$}{v2}
  \end{fmfgraph*}
  \end{center}}
\end{picture} 
 + \frac{\lambda}{2}  
\begin{picture}(22,4)(-6,0)
  \parbox{0cm}{\begin{center}
  \begin{fmfgraph*}(6,2)
  \setval
  \fmfforce{0w,1h}{v1}
  \fmfforce{1w,1h}{v2}
  \fmfforce{2w,2.5h}{v3}
  \fmfforce{2w,-.5h}{v4}
  \fmf{plain}{v1,v2}
  \fmf{plain}{v2,v3}
  \fmf{plain}{v2,v4}
  \fmfdot{v1,v2}
  \fmflabel{$x$}{v1}
  \fmflabel{$y$}{v2}
  \fmfblob{.5w}{v3}
  \fmfblob{.5w}{v4}
  \end{fmfgraph*}
  \end{center}}
\end{picture}. 
\end{align}
\end{point}

\begin{point}{Perturbative expansion on trees} 
Inserting the value of 
$\begin{picture}(16,4)(-6,0)
  \parbox{0cm}{\begin{center}
  \begin{fmfgraph*}(6,2)
  \setval
  \fmfforce{0w,1h}{v1}
  \fmfforce{1w,1h}{v2}
  \fmf{plain}{v1,v2}
  \fmfdot{v1}
  \fmflabel{$y$}{v1}
  \fmfblob{.5w}{v2}
  \end{fmfgraph*}
  \end{center}}
\end{picture}$
on the right hand-side of Eq.~\ref{Euler-Lagrange-diagrams}, and repeating 
the insertion until all the black boxes have disappeared on the right 
hand-side, we obtain a perturbative solution given by a formal series 
expanded on {\em trees\/}, which are graphs without loops in the space: 
\begin{align}
\label{Euler-Lagrange-diagram-solution}
&\nonumber \\
\begin{picture}(16,4)(-6,0)
  \parbox{0cm}{\begin{center}
  \begin{fmfgraph*}(6,2)
  \setval
  \fmfforce{0w,1h}{v1}
  \fmfforce{1w,1h}{v2}
  \fmf{plain}{v1,v2}
  \fmfdot{v1}
  \fmflabel{$x$}{v1}
  \fmfblob{.5w}{v2}
  \end{fmfgraph*}
  \end{center}}
\end{picture} 
&= 
\begin{picture}(14,4)(-6,0)
  \parbox{0cm}{\begin{center}
  \begin{fmfgraph*}(6,2)
  \setval
  \fmfforce{0w,1h}{v1}
  \fmfforce{1w,1h}{v2}
  \fmf{plain}{v1,v2}
  \fmfdot{v1}
  \fmfv{decor.shape=cross,decor.size=.4w}{v2}
  \fmflabel{$x$}{v1}
  \end{fmfgraph*}
  \end{center}}
\end{picture} 
+ \frac{\lambda}{2}  
\begin{picture}(20,4)(-6,0)
  \parbox{0cm}{\begin{center}
  \begin{fmfgraph*}(6,2)
  \setval
  \fmfforce{0w,1h}{v1}
  \fmfforce{1w,1h}{v2}
  \fmfforce{2w,2.5h}{v3}
  \fmfforce{2w,-.5h}{v4}
  \fmf{plain}{v1,v2}
  \fmf{plain}{v2,v3}
  \fmf{plain}{v2,v4}
  \fmfdot{v1}
  \fmflabel{$x$}{v1}
  \fmfv{decor.shape=cross,decor.size=.4w}{v3,v4}
  \end{fmfgraph*}
  \end{center}}
\end{picture}
 + \frac{\lambda^2}{2} 
\begin{picture}(26,4)(-6,0)
  \parbox{0cm}{\begin{center}
  \begin{fmfgraph*}(6,2)
  \setval
  \fmfforce{0w,1h}{v1}
  \fmfforce{1w,1h}{v2}
  \fmfforce{2w,2.5h}{v3}
  \fmfforce{2w,-.5h}{v4}
  \fmfforce{3w,4h}{v5}
  \fmfforce{3w,1h}{v6}
  \fmf{plain}{v1,v2,v3,v5}
  \fmf{plain}{v2,v4}
  \fmf{plain}{v3,v6}
  \fmfdot{v1}
  \fmflabel{$x$}{v1}
  \fmfv{decor.shape=cross,decor.size=.4w}{v4,v5,v6}
  \end{fmfgraph*}
  \end{center}}
\end{picture}
 + \frac{\lambda^3}{8} 
 \begin{picture}(28,4)(-6,0)
  \parbox{0cm}{\begin{center}
  \begin{fmfgraph*}(6,2)
  \setval
  \fmfforce{0w,1h}{v1}
  \fmfforce{1w,1h}{v2}
  \fmfforce{2w,3h}{v3}
  \fmfforce{2w,-1h}{v4}
  \fmfforce{3w,5h}{v5}
  \fmfforce{3w,2h}{v6}
  \fmfforce{3w,0h}{v7}
  \fmfforce{3w,-3h}{v8}
  \fmf{plain}{v1,v2,v3,v5}
  \fmf{plain}{v3,v6}
  \fmf{plain}{v2,v4,v7}
  \fmf{plain}{v4,v8}
  \fmfdot{v1}
  \fmflabel{$x$}{v1}
  \fmfv{decor.shape=cross,decor.size=.4w}{v5,v6,v7,v8}
  \end{fmfgraph*}
  \end{center}}
\end{picture}
+ \ldots  \\
&\nonumber
\end{align}

The coefficient of each tree $t$ contains a factor $\lambda^{V(t)}$
where $V(t)$ is the number of internal vertices of the tree, 
and at the denominator the symmetry factor $\Sym(t)$ of the tree, 
that is the number of permutations of the external crosses (the
sources) which leave the tree invariant. 

If we compare the diagrammatic solution
(\ref{Euler-Lagrange-diagram-solution}) with the explicit solution 
(\ref{Euler-Lagrange-solution}), we can write explicitely the value 
$\phi_t(x)$ of each tree $t$, for instance: 
\begin{align*}
t=
\begin{picture}(10,4)(-4,0)
  \parbox{0cm}{\begin{center}
  \begin{fmfgraph*}(6,2)
  \setval
  \fmfforce{0w,1h}{v1}
  \fmfforce{1w,1h}{v2}
  \fmf{plain}{v1,v2}
  \fmfdot{v1}
  \fmfv{decor.shape=cross,decor.size=.4w}{v2}
  \end{fmfgraph*}
  \end{center}}
\end{picture} 
&\quad\Longrightarrow\quad 
\phi_t(x)\ =\ \int d^Dy\ G_0(x-y)\ J(y) \quad, \\ 
t=  
\begin{picture}(12,4)(-2,0)
  \parbox{0cm}{\begin{center}
  \begin{fmfgraph*}(6,2)
  \setval
  \fmfforce{0w,1h}{v1}
  \fmfforce{1w,1h}{v2}
  \fmfforce{2w,2.5h}{v3}
  \fmfforce{2w,-.5h}{v4}
  \fmf{plain}{v1,v2}
  \fmf{plain}{v2,v3}
  \fmf{plain}{v2,v4}
  \fmfdot{v1}
  \fmfv{decor.shape=cross,decor.size=.4w}{v3,v4}
  \end{fmfgraph*}
  \end{center}}
\end{picture}
&\quad\Longrightarrow\quad 
\phi_t(x)\ =\ \int d^Dy\ d^Dz\ d^Du\  
G_0(x-y)\ G_0(y-z)\ G_0(y-u)\ J(z)\ J(u) \quad. 
\end{align*}

Finally note that the valence of the internal vertices of the trees 
depends directly on the interacting term of the Lagrangian. 
In the above example this term was $-\frac{\lambda}{3!}\ \phi^3$. 
If the Lagrangian contains the interacting term $-\frac{\mu}{4!}\ \phi^4$, 
the internal vertices of the trees turn out to have valence $4$, that is, 
the trees are of the form 
\begin{align*}
&\frac{\mu}{3!} 
\begin{picture}(20,4)(-6,0)
  \parbox{0cm}{\begin{center}
  \begin{fmfgraph*}(6,2)
  \setval
  \fmfforce{0w,1h}{v1}
  \fmfforce{1w,1h}{v2}
  \fmfforce{1.8w,3h}{v3}
  \fmfforce{2w,1h}{v4}
  \fmfforce{1.8w,-1h}{v5}
  \fmf{plain}{v1,v2,v3}
  \fmf{plain}{v2,v4}
  \fmf{plain}{v2,v5}
  \fmfdot{v1}
  \fmflabel{$x$}{v1}
  \fmfv{decor.shape=cross,decor.size=.4w}{v3,v4,v5}
  \end{fmfgraph*}
  \end{center}}
\end{picture}  .
\end{align*} 
\end{point}

\begin{point}{Feynman rules}
We can therefore conclude that the field 
$\phi(x)=\sum_n \lambda^n\ \phi_n(x)$ 
has perturbative coefficients $\phi_n(x)$ given by the finite sum of 
the {\em amplitude\/} $\phi_t(x)$ of all the trees $t$ with $n$ internal 
vertices, constructed according to the following {\em Feynman's rules\/}: 
\begin{itemize}
\item
consider all the trees with internal vertices of valence 3, 
and external vertices of valence 1; 
\item 
fix one external vertex called the {\em root\/} (therefore the trees 
are called {\em rooted\/}), and call the other external vertices the 
{\em leaves\/}; 
\item 
label the root by $x$; 
\item
label the internal vertices and the leaves by free variables $y,z,u,v,...$;
\item 
assign a weigth $G_0(y-z)$ to each edge joining the vertices $y$ and
$z$;
\item 
assign a weigth $\lambda$ to each internal vertex 
\begin{picture}(12,4)(-2,0)
  \parbox{0cm}{\begin{center}
  \begin{fmfgraph*}(4,2)
  \setval
  \fmfforce{0w,1h}{v1}
  \fmfforce{1w,1h}{v2}
  \fmfforce{2w,2.5h}{v3}
  \fmfforce{2w,-.5h}{v4}
  \fmf{plain}{v1,v2}
  \fmf{plain}{v2,v3}
  \fmf{plain}{v2,v4}
  \end{fmfgraph*}
  \end{center}}
\end{picture};
\item 
assign a weigth $J(y)$ to each leaf; 
\item 
to obtain $\phi_t(x)$ for a given tree $t$, 
multiply all the weigths and integrate over the free variables;
\item
divide by the symmetry factor $\Sym(t)$ of the tree.
\end{itemize}
\end{point}

\begin{point}{Conclusion}
A typical classical field $\phi$ with Lagrangian density of the form 
\begin{align*}
\L(\phi) &= \frac{1}{2}\ \phi^t A \phi-J(x)\ \phi(x)
-\frac{\lambda}{3!}\ \phi(x)^3, 
\end{align*}
can be described as a formal series in the coupling constant $\lambda$,  
\begin{align*}
\phi(x) &= \sum_{n=0}^\infty \lambda^n\ \phi_n(x),  
\end{align*}
where each coefficient $\phi_n(x)$ is a finite sum 
\begin{align*}
\phi_n(x) &= \sum_{V(t)=n} \frac{1}{\Sym(t)}\ \phi_t(x)
\end{align*}
of amplitudes $\frac{1}{\Sym(t)}\ \phi_t(x)$ associated to each tree $t$ 
with $n$ internal vertices of valence 3. 
Note that, in these lectures, the amplitude of a tree is considered 
modulo the factor $\frac{1}{\Sym(t)}$. 
\end{point}


\subsection{Interacting quantum fields}

\begin{point}{Feynman notations}
\label{FeynmanNotationQuantum}
We adopt the following Feynman's notations: 
\begin{itemize}
\item
$k$-points full Green's function $\la \phi(x_1)\cdots\phi(x_k) \ra$ = 
\begin{picture}(20,10)(-6,0)
  \parbox{0cm}{\begin{center}
  \begin{fmfgraph*}(10,10)
  \setval
  \fmfsurroundn{v}{7}
  \fmfforce{c}{c}
  \fmfv{decor.shape=circle,decor.size=.33w,decor.filled=full}{c}
  \fmf{plain}{v1,c,v2}
  \fmf{plain}{v3,c,v4}
  \fmf{plain}{v5,c,v6}
  \fmf{plain}{v7,c}
  \fmfdotn{v}{7}
  \fmflabel{$x_1$}{v4}
  \fmflabel{$x_2$}{v3}
  \fmflabel{$x_k$}{v5}
  \end{fmfgraph*}
  \end{center}}
\end{picture}; 
\bigskip 

\item 
$k$-points connected Green's function $G(x_1,...,x_k)$ = 
\begin{picture}(20,10)(-6,0)
  \parbox{0cm}{\begin{center}
  \begin{fmfgraph*}(10,10)
  \setval
  \fmfsurroundn{v}{7}
  \fmfforce{c}{c}
  \fmfv{decor.shape=circle,decor.size=.33w,decor.filled=shaded}{c}
  \fmf{plain}{v1,c,v2}
  \fmf{plain}{v3,c,v4}
  \fmf{plain}{v5,c,v6}
  \fmf{plain}{v7,c}
  \fmfdotn{v}{7}
  \fmflabel{$x_1$}{v4}
  \fmflabel{$x_2$}{v3}
  \fmflabel{$x_k$}{v5}
  \end{fmfgraph*}
  \end{center}}
\end{picture}; 

\item 
source $J(y)$ = 
\begin{picture}(12,4)(-2,0)
  \parbox{0cm}{\begin{center}
  \begin{fmfgraph*}(6,2)
  \setval
  \fmfforce{0w,1h}{v1}
  \fmfforce{1w,1h}{v2}
  \fmf{plain}{v1,v2}
  \fmfv{decor.shape=cross,decor.size=.4w}{v2}
  \fmflabel{$y$}{v2}
  \end{fmfgraph*}
  \end{center}}
\end{picture}; 

\item
propagator $G_0(x-y)$ = 
\begin{picture}(16,4)(-6,0)
  \parbox{0cm}{\begin{center}
  \begin{fmfgraph*}(6,2)
  \setval
  \fmfforce{0w,1h}{v1}
  \fmfforce{1w,1h}{v2}
  \fmf{plain}{v1,v2}
  \fmfdot{v1,v2}
  \fmflabel{$x$}{v1}
  \fmflabel{$y$}{v2}
  \end{fmfgraph*}
  \end{center}}
\end{picture}. 
\end{itemize}
\end{point}

\begin{point}{Exercise: Diagrammatic expression of the full
Green's functions}
\label{full-connected-diagrams}
Using Eqs.~(\ref{full-connected-green-functions}), 
draw the diagrammatic expression of the full
Green's functions in terms of the connected ones. 
\end{point}

\begin{point}{Dyson-Schwinger equations}
The Dyson-Schwinger equation for the 1-point connected
Green's function of a field in presence of an external field $J$ 
(cf. Eq.~(\ref{Dyson-Schwinger-equation-W})), is the following:  
\begin{align}
\label{Dyson-Schwinger-W}
\begin{picture}(16,4)(-6,0)
  \parbox{0cm}{\begin{center}
  \begin{fmfgraph*}(6,2)
  \setval
  \fmfforce{0w,1h}{v1}
  \fmfforce{1w,1h}{v2}
  \fmf{plain}{v1,v2}
  \fmfdot{v1}
  \fmflabel{$x$}{v1}
  \fmfblob{.5w}{v2}
  \end{fmfgraph*}
  \end{center}}
\end{picture} 
&= 
\begin{picture}(16,4)(-6,0)
  \parbox{0cm}{\begin{center}
  \begin{fmfgraph*}(6,2)
  \setval
  \fmfforce{0w,1h}{v1}
  \fmfforce{1w,1h}{v2}
  \fmf{plain}{v1,v2}
  \fmfdot{v1}
  \fmfv{decor.shape=cross,decor.size=.4w}{v2}
  \fmflabel{$x$}{v1}
  \end{fmfgraph*}
  \end{center}}
\end{picture} 
+ \frac{\lambda}{2}  
\begin{picture}(20,4)(-6,0)
  \parbox{0cm}{\begin{center}
  \begin{fmfgraph*}(6,2)
  \setval
  \fmfforce{0w,1h}{v1}
  \fmfforce{1w,1h}{v2}
  \fmfforce{2w,2.5h}{v3}
  \fmfforce{2w,-.5h}{v4}
  \fmf{plain}{v1,v2}
  \fmf{plain}{v2,v3}
  \fmf{plain}{v2,v4}
  \fmfdot{v1}
  \fmflabel{$x$}{v1}
  \fmfblob{.5w}{v3}
  \fmfblob{.5w}{v4}
  \end{fmfgraph*}
  \end{center}}
\end{picture}
+ \hbar\ \frac{\lambda}{2}  
\begin{picture}(22,4)(-6,0)
  \parbox{0cm}{\begin{center}
  \begin{fmfgraph*}(6,2)
  \setval
  \fmfforce{0w,1h}{v1}
  \fmfforce{1w,1h}{v2}
  \fmfforce{2w,1h}{v3}
  \fmf{plain}{v1,v2}
  \fmf{plain,left=1}{v2,v3,v2}
  \fmfdot{v1}
  \fmflabel{$x$}{v1}
  \fmfblob{.5w}{v3}
  \end{fmfgraph*}
  \end{center}}
\end{picture}.  
\end{align}
Note that in the limit $\hbar\to 0$, we recover the Euler-Lagrange
equation (\ref{Euler-Lagrange-diagrams}) for the field. 

The Dyson-Schwinger equation for the 1-point connected
Green's function of an isolated field 
(cf. Eq.~(\ref{Dyson-Schwinger-equation-1})), is the following:  
\begin{align}
\label{Dyson-Schwinger-1}
\begin{picture}(16,4)(-6,0)
  \parbox{0cm}{\begin{center}
  \begin{fmfgraph*}(6,2)
  \setval
  \fmfforce{0w,1h}{v1}
  \fmfforce{1w,1h}{v2}
  \fmf{plain}{v1,v2}
  \fmfdot{v1}
  \fmflabel{$x$}{v1}
  \fmfblob{.5w}{v2}
  \end{fmfgraph*}
  \end{center}}
\end{picture} 
&= \frac{\lambda}{2}  
\begin{picture}(20,4)(-6,0)
  \parbox{0cm}{\begin{center}
  \begin{fmfgraph*}(6,2)
  \setval
  \fmfforce{0w,1h}{v1}
  \fmfforce{1w,1h}{v2}
  \fmfforce{2w,2.5h}{v3}
  \fmfforce{2w,-.5h}{v4}
  \fmf{plain}{v1,v2}
  \fmf{plain}{v2,v3}
  \fmf{plain}{v2,v4}
  \fmfdot{v1}
  \fmflabel{$x$}{v1}
  \fmfblob{.5w}{v3}
  \fmfblob{.5w}{v4}
  \end{fmfgraph*}
  \end{center}}
\end{picture}
+ \hbar\ \frac{\lambda}{2}  
\begin{picture}(22,4)(-6,0)
  \parbox{0cm}{\begin{center}
  \begin{fmfgraph*}(6,2)
  \setval
  \fmfforce{0w,1h}{v1}
  \fmfforce{1w,1h}{v2}
  \fmfforce{2w,1h}{v3}
  \fmf{plain}{v1,v2}
  \fmf{plain,left=1}{v2,v3,v2}
  \fmfdot{v1}
  \fmflabel{$x$}{v1}
  \fmfblob{.5w}{v3}
  \end{fmfgraph*}
  \end{center}}
\end{picture}.  
\end{align}
For the 2-points connected Green's function, the 
Dyson-Schwinger equation is (cf. Eq.~(\ref{Dyson-Schwinger-equation-2})): 
\begin{align}
\label{Dyson-Schwinger-2}
\begin{picture}(22,4)(-6,0)
  \parbox{0cm}{\begin{center}
  \begin{fmfgraph*}(6,2)
  \setval
  \fmfforce{0w,1h}{v1}
  \fmfforce{1w,1h}{v2}
  \fmfforce{2w,1h}{v3}
  \fmf{plain}{v1,v2,v3}
  \fmfdot{v1,v3}
  \fmflabel{$x$}{v1}
  \fmflabel{$y$}{v3}
  \fmfblob{.5w}{v2}
  \end{fmfgraph*}
  \end{center}}
\end{picture} 
&= 
\begin{picture}(16,4)(-6,0)
  \parbox{0cm}{\begin{center}
  \begin{fmfgraph*}(6,2)
  \setval
  \fmfforce{0w,1h}{v1}
  \fmfforce{1w,1h}{v2}
  \fmf{plain}{v1,v2}
  \fmfdot{v1,v2}
  \fmflabel{$x$}{v1}
  \fmflabel{$y$}{v2}
  \end{fmfgraph*}
  \end{center}}
\end{picture} 
+ \lambda  
\begin{picture}(28,4)(-6,0)
  \parbox{0cm}{\begin{center}
  \begin{fmfgraph*}(6,2)
  \setval
  \fmfforce{0w,1h}{v1}
  \fmfforce{1w,1h}{v2}
  \fmfforce{2w,1h}{v3}
  \fmfforce{3w,1h}{v4}
  \fmfforce{1w,3h}{v5}
  \fmf{plain}{v1,v2,v3,v4}
  \fmf{plain}{v2,v5}
  \fmfdot{v1,v4}
  \fmflabel{$x$}{v1}
  \fmflabel{$y$}{v4}
  \fmfblob{.5w}{v3}
  \fmfblob{.5w}{v5}
  \end{fmfgraph*}
  \end{center}}
\end{picture}
+ \hbar\ \frac{\lambda}{2}  
\begin{picture}(30,4)(-6,0)
  \parbox{0cm}{\begin{center}
  \begin{fmfgraph*}(6,2)
  \setval
  \fmfforce{0w,1h}{v1}
  \fmfforce{1w,1h}{v2}
  \fmfforce{2w,1h}{v3}
  \fmfforce{3w,1h}{v4}
  \fmf{plain}{v1,v2}
  \fmf{plain,left=1}{v2,v3,v2}
  \fmf{plain}{v3,v4}
  \fmfdot{v1,v4}
  \fmflabel{$x$}{v1}
  \fmflabel{$y$}{v4}
  \fmfblob{.5w}{v3}
  \end{fmfgraph*}
  \end{center}}
\end{picture}.  
\end{align}
For the 3-points Green's function: 
\begin{align}
\label{Dyson-Schwinger-3}
\begin{picture}(22,4)(-6,0)
  \parbox{0cm}{\begin{center}
  \begin{fmfgraph*}(6,2)
  \setval
  \fmfforce{0w,1h}{v1}
  \fmfforce{1w,1h}{v2}
  \fmfforce{2w,2.5h}{v3}
  \fmfforce{2w,-.5h}{v4}
  \fmf{plain}{v1,v2,v3}
  \fmf{plain}{v2,v4}
  \fmfdot{v1,v3,v4}
  \fmflabel{$x$}{v1}
  \fmflabel{$y$}{v3}
  \fmflabel{$z$}{v4}
  \fmfblob{.5w}{v2}
  \end{fmfgraph*}
  \end{center}}
\end{picture} 
&= \lambda  
\begin{picture}(28,4)(-6,0)
  \parbox{0cm}{\begin{center}
  \begin{fmfgraph*}(6,2)
  \setval
  \fmfforce{0w,1h}{v1}
  \fmfforce{1w,1h}{v2}
  \fmfforce{2w,2.5h}{v3}
  \fmfforce{2w,-.5h}{v4}
  \fmfforce{3w,4h}{v5}
  \fmfforce{3w,-2h}{v6}
  \fmf{plain}{v1,v2,v3,v5}
  \fmf{plain}{v2,v4,v6}
  \fmfdot{v1,v5,v6}
  \fmflabel{$x$}{v1}
  \fmflabel{$y$}{v5}
  \fmflabel{$z$}{v6}
  \fmfblob{.5w}{v3}
  \fmfblob{.5w}{v4}
  \end{fmfgraph*}
  \end{center}}
\end{picture} 
+ \lambda  
\begin{picture}(28,4)(-6,0)
  \parbox{0cm}{\begin{center}
  \begin{fmfgraph*}(6,2)
  \setval
  \fmfforce{0w,1h}{v1}
  \fmfforce{1w,1h}{v2}
  \fmfforce{1w,3h}{v3}
  \fmfforce{2w,1h}{v4}
  \fmfforce{3w,2.5h}{v5}
  \fmfforce{3w,-.5h}{v6}
  \fmf{plain}{v1,v2,v3}
  \fmf{plain}{v2,v4,v5}
  \fmf{plain}{v4,v6}
  \fmfdot{v1,v5,v6}
  \fmflabel{$x$}{v1}
  \fmflabel{$y$}{v5}
  \fmflabel{$z$}{v6}
  \fmfblob{.5w}{v3}
  \fmfblob{.5w}{v4}
  \end{fmfgraph*}
  \end{center}}
\end{picture}
+ \hbar\ \frac{\lambda}{2}  
\begin{picture}(30,4)(-6,0)
  \parbox{0cm}{\begin{center}
  \begin{fmfgraph*}(6,2)
  \setval
  \fmfforce{0w,1h}{v1}
  \fmfforce{1w,1h}{v2}
  \fmfforce{2w,1h}{v3}
  \fmfforce{3w,2.5h}{v4}
  \fmfforce{3w,-.5h}{v5}
  \fmf{plain}{v1,v2}
  \fmf{plain,left=1}{v2,v3,v2}
  \fmf{plain}{v3,v4}
  \fmf{plain}{v3,v5}
  \fmfdot{v1,v4,v5}
  \fmflabel{$x$}{v1}
  \fmflabel{$y$}{v4}
  \fmflabel{$z$}{v5}
  \fmfblob{.5w}{v3}
  \end{fmfgraph*}
  \end{center}}
\end{picture}.  
\\ \nonumber &
\end{align}
\end{point}

\begin{point}{Perturbative expansion on graphs}
Then the perturbative solution of the Dyson-Schwinger equation is given
by a formal series expanded on Feynman diagrams, which are graphs in
the space. For the 1-point Green's function, the solution of 
(\ref{Dyson-Schwinger-W}) is ($J\neq 0$): 
\begin{align}
\label{Dyson-Schwinger-W-solution}
\begin{picture}(16,4)(-6,0)
  \parbox{0cm}{\begin{center}
  \begin{fmfgraph*}(6,2)
  \setval
  \fmfforce{0w,1h}{v1}
  \fmfforce{1w,1h}{v2}
  \fmf{plain}{v1,v2}
  \fmfdot{v1}
  \fmfblob{.5w}{v2}
  \end{fmfgraph*}
  \end{center}}
\end{picture} 
&= 
\begin{picture}(12,4)(-2,0)
  \parbox{0cm}{\begin{center}
  \begin{fmfgraph*}(6,2)
  \setval
  \fmfforce{0w,1h}{v1}
  \fmfforce{1w,1h}{v2}
  \fmf{plain}{v1,v2}
  \fmfdot{v1}
  \fmfv{decor.shape=cross,decor.size=.4w}{v2}
  \end{fmfgraph*}
  \end{center}}
\end{picture} 
+ \frac{\lambda}{2}  
\begin{picture}(16,4)(-2,0)
  \parbox{0cm}{\begin{center}
  \begin{fmfgraph*}(6,2)
  \setval
  \fmfforce{0w,1h}{v1}
  \fmfforce{1w,1h}{v2}
  \fmfforce{2w,2.5h}{v3}
  \fmfforce{2w,-.5h}{v4}
  \fmf{plain}{v1,v2}
  \fmf{plain}{v2,v3}
  \fmf{plain}{v2,v4}
  \fmfdot{v1}
  \fmfv{decor.shape=cross,decor.size=.4w}{v3,v4}
  \end{fmfgraph*}
  \end{center}}
\end{picture}
+ \hbar\ \frac{\lambda}{2}  
\begin{picture}(12,4)(-2,0)
  \parbox{0cm}{\begin{center}
  \begin{fmfgraph*}(6,2)
  \setval
  \fmfforce{0w,1h}{v1}
  \fmfforce{1w,1h}{v2}
  \fmfforce{1.5w,1h}{v3}
  \fmf{plain}{v1,v2}
  \fmf{plain,left=1}{v2,v3,v2}
  \fmfdot{v1}
  \end{fmfgraph*}
  \end{center}}
\end{picture} 
\\ \nonumber & \\ \nonumber 
&
+ \frac{\lambda^2}{2} 
\begin{picture}(22,6)(-2,0)
  \parbox{0cm}{\begin{center}
  \begin{fmfgraph*}(6,2)
  \setval
  \fmfforce{0w,1h}{v1}
  \fmfforce{1w,1h}{v2}
  \fmfforce{2w,2.5h}{v3}
  \fmfforce{2w,-.5h}{v4}
  \fmfforce{3w,4h}{v5}
  \fmfforce{3w,1h}{v6}
  \fmf{plain}{v1,v2,v3,v5}
  \fmf{plain}{v2,v4}
  \fmf{plain}{v3,v6}
  \fmfdot{v1}
  \fmfv{decor.shape=cross,decor.size=.4w}{v4,v5,v6}
  \end{fmfgraph*}
  \end{center}}
\end{picture}
+ \hbar\ \frac{\lambda^2}{2}  
\begin{picture}(24,6)(-2,0)
  \parbox{0cm}{\begin{center}
  \begin{fmfgraph*}(6,2)
  \setval
  \fmfforce{0w,1h}{v1}
  \fmfforce{1w,1h}{v2}
  \fmfforce{2w,2.5h}{v3}
  \fmfforce{2w,-.5h}{v4}
  \fmfforce{2.5w,3.25h}{v5}
  \fmf{plain}{v1,v2,v3}
  \fmf{plain}{v2,v4}
  \fmf{plain,left=1}{v3,v5,v3}
  \fmfdot{v1}
  \fmfv{decor.shape=cross,decor.size=.4w}{v4}
  \end{fmfgraph*}
  \end{center}}
\end{picture}
+ \hbar\ \frac{\lambda^2}{2}  
\begin{picture}(22,4)(-2,0)
  \parbox{0cm}{\begin{center}
  \begin{fmfgraph*}(6,2)
  \setval
  \fmfforce{0w,1h}{v1}
  \fmfforce{1w,1h}{v2}
  \fmfforce{2w,1h}{v3}
  \fmfforce{3w,1h}{v4}
  \fmf{plain}{v1,v2}
  \fmf{plain,left=1}{v2,v3,v2}
  \fmf{plain}{v3,v4}
  \fmfdot{v1}
  \fmfv{decor.shape=cross,decor.size=.4w}{v4}
  \end{fmfgraph*}
  \end{center}}
\end{picture} 
\\ \nonumber & \\ \nonumber 
&
+ \frac{\lambda^3}{8} 
\begin{picture}(26,8)(-2,0)
  \parbox{0cm}{\begin{center}
  \begin{fmfgraph*}(6,2)
  \setval
  \fmfforce{0w,1h}{v1}
  \fmfforce{1w,1h}{v2}
  \fmfforce{2w,3h}{v3}
  \fmfforce{2w,-1h}{v4}
  \fmfforce{3w,5h}{v5}
  \fmfforce{3w,2h}{v6}
  \fmfforce{3w,0h}{v7}
  \fmfforce{3w,-3h}{v8}
  \fmf{plain}{v1,v2,v3,v5}
  \fmf{plain}{v3,v6}
  \fmf{plain}{v2,v4,v7}
  \fmf{plain}{v4,v8}
  \fmfdot{v1}
  \fmfv{decor.shape=cross,decor.size=.4w}{v5,v6,v7,v8}
  \end{fmfgraph*}
  \end{center}}
\end{picture}
+ \frac{\lambda^3}{2} 
\begin{picture}(28,9)(-2,0)
  \parbox{0cm}{\begin{center}
  \begin{fmfgraph*}(6,2)
  \setval
  \fmfforce{0w,1h}{v1}
  \fmfforce{1w,1h}{v2}
  \fmfforce{2w,2.5h}{v3}
  \fmfforce{2w,-.5h}{v4}
  \fmfforce{3w,4h}{v5}
  \fmfforce{3w,1h}{v6}
  \fmfforce{4w,5.5h}{v7}
  \fmfforce{4w,2.5h}{v8}
  \fmf{plain}{v1,v2,v3,v5,v7}
  \fmf{plain}{v2,v4}
  \fmf{plain}{v3,v6}
  \fmf{plain}{v5,v8}
  \fmfdot{v1}
  \fmfv{decor.shape=cross,decor.size=.4w}{v4,v6,v7,v8}
  \end{fmfgraph*}
  \end{center}}
\end{picture}
+ \hbar\ \frac{\lambda^3}{2} 
\begin{picture}(24,6)(-2,0)
  \parbox{0cm}{\begin{center}
  \begin{fmfgraph*}(6,2)
  \setval
  \fmfforce{0w,1h}{v1}
  \fmfforce{1w,1h}{v2}
  \fmfforce{2w,2.5h}{v3}
  \fmfforce{2w,-.5h}{v4}
  \fmfforce{3w,4h}{v5}
  \fmfforce{3w,1h}{v6}
  \fmfforce{3.5w,.25h}{v7}
  \fmf{plain}{v1,v2,v3,v5}
  \fmf{plain}{v2,v4}
  \fmf{plain}{v3,v6}
  \fmf{plain,left=1}{v6,v7,v6}
  \fmfdot{v1}
  \fmfv{decor.shape=cross,decor.size=.4w}{v4,v5}
  \end{fmfgraph*}
  \end{center}}
\end{picture}
+ \hbar\ \frac{\lambda^3}{4} 
\begin{picture}(24,6)(-2,0)
  \parbox{0cm}{\begin{center}
  \begin{fmfgraph*}(6,2)
  \setval
  \fmfforce{0w,1h}{v1}
  \fmfforce{1w,1h}{v2}
  \fmfforce{2w,2.5h}{v3}
  \fmfforce{2w,-.5h}{v4}
  \fmfforce{3w,4h}{v5}
  \fmfforce{3w,1h}{v6}
  \fmfforce{2.5w,-1.25h}{v7}
  \fmf{plain}{v1,v2,v3,v5}
  \fmf{plain}{v2,v4}
  \fmf{plain}{v3,v6}
  \fmf{plain,left=1}{v4,v7,v4}
  \fmfdot{v1}
  \fmfv{decor.shape=cross,decor.size=.4w}{v5,v6}
  \end{fmfgraph*}
  \end{center}}
\end{picture}
\\ \nonumber & \\ \nonumber 
&
+ \hbar\ \frac{\lambda^3}{4}  
\begin{picture}(28,4)(-2,0)
  \parbox{0cm}{\begin{center}
  \begin{fmfgraph*}(6,2)
  \setval
  \fmfforce{0w,1h}{v1}
  \fmfforce{1w,1h}{v2}
  \fmfforce{2w,1h}{v3}
  \fmfforce{3w,1h}{v4}
  \fmfforce{4w,2.5h}{v5}
  \fmfforce{4w,-.5h}{v6}
  \fmf{plain}{v1,v2}
  \fmf{plain,left=1}{v2,v3,v2}
  \fmf{plain}{v3,v4,v5}
  \fmf{plain}{v4,v6}
  \fmfdot{v1}
  \fmfv{decor.shape=cross,decor.size=.4w}{v5,v6}
  \end{fmfgraph*}
  \end{center}}
\end{picture} 
+ \hbar\ \frac{\lambda^3}{2}  
\begin{picture}(28,4)(-2,0)
  \parbox{0cm}{\begin{center}
  \begin{fmfgraph*}(6,2)
  \setval
  \fmfforce{0w,1h}{v1}
  \fmfforce{1w,1h}{v2}
  \fmfforce{2w,1h}{v3}
  \fmfforce{3w,1h}{v4}
  \fmfforce{4w,1h}{v5}
  \fmfforce{1w,3h}{v6}
  \fmf{plain}{v1,v2,v3}
  \fmf{plain,left=1}{v3,v4,v3}
  \fmf{plain}{v4,v5}
  \fmf{plain}{v2,v6}
  \fmfdot{v1}
  \fmfv{decor.shape=cross,decor.size=.4w}{v5,v6}
  \end{fmfgraph*}
  \end{center}}
\end{picture} 
+ \hbar^2\ \frac{\lambda^3}{4}  
\begin{picture}(25,4)(-2,0)
  \parbox{0cm}{\begin{center}
  \begin{fmfgraph*}(6,2)
  \setval
  \fmfforce{0w,1h}{v1}
  \fmfforce{1w,1h}{v2}
  \fmfforce{2w,1h}{v3}
  \fmfforce{3w,1h}{v4}
  \fmfforce{3.5w,1h}{v5}
  \fmf{plain}{v1,v2}
  \fmf{plain,left=1}{v2,v3,v2}
  \fmf{plain}{v3,v4}
  \fmf{plain,left=1}{v4,v5,v4}
  \fmfdot{v1}
  \end{fmfgraph*}
  \end{center}}
\end{picture} 
\\ \nonumber 
&
+ \hbar\ \frac{\lambda^3}{4}  
\begin{picture}(24,6)(-2,0)
  \parbox{0cm}{\begin{center}
  \begin{fmfgraph*}(6,2)
  \setval
  \fmfforce{0w,1h}{v1}
  \fmfforce{1w,1h}{v2}
  \fmfforce{2w,2.5h}{v3}
  \fmfforce{2w,-.5h}{v4}
  \fmfforce{2.5w,3.25h}{v5}
  \fmfforce{2.5w,-1.25h}{v6}
  \fmf{plain}{v1,v2,v3}
  \fmf{plain}{v2,v4}
  \fmf{plain,left=1}{v3,v5,v3}
  \fmf{plain,left=1}{v4,v6,v4}
  \fmfdot{v1}
  \end{fmfgraph*}
  \end{center}}
\end{picture}
+ \hbar\ \frac{\lambda^3}{2}  
\begin{picture}(20,8)(-2,0)
  \parbox{0cm}{\begin{center}
  \begin{fmfgraph*}(6,2)
  \setval
  \fmfforce{0w,1h}{v1}
  \fmfforce{1w,1h}{v2}
  \fmfforce{1.7w,2.2h}{v3}
  \fmfforce{1.7w,-.2h}{v4}
  \fmfforce{2.7w,3.7h}{v5}
  \fmfforce{2.7w,-1.7h}{v6}
  \fmf{plain}{v1,v2}
  \fmf{plain,left=.6}{v2,v3,v4,v2}
  \fmf{plain}{v3,v5}
  \fmf{plain}{v4,v6}
  \fmfdot{v1}
  \fmfv{decor.shape=cross,decor.size=.4w}{v5,v6}
  \end{fmfgraph*}
  \end{center}}
\end{picture} 
+ \hbar^2\ \frac{\lambda^3}{4}  
\begin{picture}(16,4)(-2,0)
  \parbox{0cm}{\begin{center}
  \begin{fmfgraph*}(6,2)
  \setval
  \fmfforce{0w,1h}{v1}
  \fmfforce{1w,1h}{v2}
  \fmfforce{1.5w,2.45h}{v3}
  \fmfforce{1.5w,-.45h}{v4}
  \fmfforce{2w,1h}{v5}
  \fmf{plain}{v1,v2}
  \fmf{plain,left=1}{v2,v5,v2}
  \fmf{plain}{v3,v4}
  \fmfdot{v1}
  \end{fmfgraph*}
  \end{center}}
\end{picture} 
+ \ \O(\lambda^4). \\ \nonumber &
\end{align}
The coefficient of each graph $\G$ contains a factor $\lambda^V(\G)$ 
where $V(\G)$ is the number of internal vertices of the graph, 
and at the denominator the symmetry factor $\Sym(\G)$ of the graph, 
that is the number of permutations of the external crosses (the sources) 
and of the internal edges (joint to the same internal vertices) 
which leave the graph invariant, multiplied by a factor $2$ for each 
bubble (an internal edge connected to a single vertex). 

Of course, the solution of Eq.~(\ref{Dyson-Schwinger-1}) is ($J=0$): 
\begin{align}
\label{Dyson-Schwinger-1-solution}
\begin{picture}(12,4)(-2,0)
  \parbox{0cm}{\begin{center}
  \begin{fmfgraph*}(6,2)
  \setval
  \fmfforce{0w,1h}{v1}
  \fmfforce{1w,1h}{v2}
  \fmf{plain}{v1,v2}
  \fmfdot{v1}
  \fmfblob{.5w}{v2}
  \end{fmfgraph*}
  \end{center}}
\end{picture} 
&= \hbar\ \frac{\lambda}{2}  
\begin{picture}(12,4)(-2,0)
  \parbox{0cm}{\begin{center}
  \begin{fmfgraph*}(6,2)
  \setval
  \fmfforce{0w,1h}{v1}
  \fmfforce{1w,1h}{v2}
  \fmfforce{1.5w,1h}{v3}
  \fmf{plain}{v1,v2}
  \fmf{plain,left=1}{v2,v3,v2}
  \fmfdot{v1}
  \end{fmfgraph*}
  \end{center}}
\end{picture} 
+ \hbar^2\ \frac{\lambda^3}{4}  
\begin{picture}(24,4)(-2,0)
  \parbox{0cm}{\begin{center}
  \begin{fmfgraph*}(6,2)
  \setval
  \fmfforce{0w,1h}{v1}
  \fmfforce{1w,1h}{v2}
  \fmfforce{2w,1h}{v3}
  \fmfforce{3w,1h}{v4}
  \fmfforce{3.5w,1h}{v5}
  \fmf{plain}{v1,v2}
  \fmf{plain,left=1}{v2,v3,v2}
  \fmf{plain}{v3,v4}
  \fmf{plain,left=1}{v4,v5,v4}
  \fmfdot{v1}
  \end{fmfgraph*}
  \end{center}}
\end{picture} 
+ \hbar^2\ \frac{\lambda^3}{4}  
\begin{picture}(16,4)(-2,0)
  \parbox{0cm}{\begin{center}
  \begin{fmfgraph*}(6,2)
  \setval
  \fmfforce{0w,1h}{v1}
  \fmfforce{1w,1h}{v2}
  \fmfforce{1.5w,2.45h}{v3}
  \fmfforce{1.5w,-.45h}{v4}
  \fmfforce{2w,1h}{v5}
  \fmf{plain}{v1,v2}
  \fmf{plain,left=1}{v2,v5,v2}
  \fmf{plain}{v3,v4}
  \fmfdot{v1}
  \end{fmfgraph*}
  \end{center}}
\end{picture} 
+ \ \O(\lambda^4). 
\end{align}
For the 2-points Green's function, the solution of 
Eq.~(\ref{Dyson-Schwinger-2}) is 
\begin{align}
\label{Dyson-Schwinger-2-solution}
\begin{picture}(16,4)(-2,0)
  \parbox{0cm}{\begin{center}
  \begin{fmfgraph*}(6,2)
  \setval
  \fmfforce{0w,1h}{v1}
  \fmfforce{1w,1h}{v2}
  \fmfforce{2w,1h}{v3}
  \fmf{plain}{v1,v2,v3}
  \fmfdot{v1,v3}
  \fmfblob{.5w}{v2}
  \end{fmfgraph*}
  \end{center}}
\end{picture} 
&= 
\begin{picture}(10,4)(-2,0)
  \parbox{0cm}{\begin{center}
  \begin{fmfgraph*}(6,2)
  \setval
  \fmfforce{0w,1h}{v1}
  \fmfforce{1w,1h}{v2}
  \fmf{plain}{v1,v2}
  \fmfdot{v1,v2}
  \end{fmfgraph*}
  \end{center}}
\end{picture} 
+ \hbar\ \frac{\lambda^2}{2}  
\begin{picture}(16,8)(-2,0)
  \parbox{0cm}{\begin{center}
  \begin{fmfgraph*}(6,2)
  \setval
  \fmfforce{0w,1h}{v1}
  \fmfforce{1w,1h}{v2}
  \fmfforce{1w,3h}{v3}
  \fmfforce{1w,4.5h}{v4}
  \fmfforce{2w,1h}{v5}
  \fmf{plain}{v1,v2,v5}
  \fmf{plain}{v2,v3}
  \fmf{plain,left=1}{v3,v4,v3}
  \fmfdot{v1,v5}
  \end{fmfgraph*}
  \end{center}}
\end{picture}
+ \hbar\ \frac{\lambda^2}{2}  
\begin{picture}(22,4)(-2,0)
  \parbox{0cm}{\begin{center}
  \begin{fmfgraph*}(6,2)
  \setval
  \fmfforce{0w,1h}{v1}
  \fmfforce{1w,1h}{v2}
  \fmfforce{2w,1h}{v3}
  \fmfforce{3w,1h}{v4}
  \fmf{plain}{v1,v2}
  \fmf{plain,left=1}{v2,v3,v2}
  \fmf{plain}{v3,v4}
  \fmfdot{v1,v4}
  \end{fmfgraph*}
  \end{center}}
\end{picture}
\\ \nonumber 
&+ \hbar^2\ \frac{\lambda^4}{4}  
\begin{picture}(34,6)(-2,0)
  \parbox{0cm}{\begin{center}
  \begin{fmfgraph*}(6,2)
  \setval
  \fmfforce{0w,1h}{v1}
  \fmfforce{1w,1h}{v2}
  \fmfforce{2w,1h}{v3}
  \fmfforce{3w,1h}{v4}
  \fmfforce{4w,1h}{v5}
  \fmfforce{5w,1h}{v6}
  \fmf{plain}{v1,v2}
  \fmf{plain,left=1}{v2,v3,v2}
  \fmf{plain}{v3,v4}
  \fmf{plain,left=1}{v4,v5,v4}
  \fmf{plain}{v5,v6}
  \fmfdot{v1,v6}
  \end{fmfgraph*}
  \end{center}}
\end{picture}
+ \hbar^2\ \frac{\lambda^4}{2}  
\begin{picture}(22,6)(-2,0)
  \parbox{0cm}{\begin{center}
  \begin{fmfgraph*}(6,2)
  \setval
  \fmfforce{0w,1h}{v1}
  \fmfforce{1w,1h}{v2}
  \fmfforce{2w,1h}{v3}
  \fmfforce{3w,1h}{v4}
  \fmfforce{1.15w,2.1h}{v5}
  \fmfforce{1.85w,2.1h}{v6}
  \fmf{plain}{v1,v2}
  \fmf{plain,left=1}{v2,v3,v2}
  \fmf{plain}{v3,v4}
  \fmf{plain,left=.4}{v5,v6,v5}
  \fmfdot{v1,v4}
  \end{fmfgraph*}
  \end{center}}
\end{picture}
+ \hbar^2\ \frac{\lambda^4}{2}  
\begin{picture}(22,6)(-2,0)
  \parbox{0cm}{\begin{center}
  \begin{fmfgraph*}(6,2)
  \setval
  \fmfforce{0w,1h}{v1}
  \fmfforce{1w,1h}{v2}
  \fmfforce{1.5w,2.45h}{v3}
  \fmfforce{1.5w,-.45h}{v4}
  \fmfforce{2w,1h}{v5}
  \fmfforce{3w,1h}{v6}
  \fmf{plain}{v1,v2}
  \fmf{plain,left=1}{v2,v5,v2}
  \fmf{plain}{v3,v4}
  \fmf{plain}{v5,v6}
  \fmfdot{v1,v6}
  \end{fmfgraph*}
  \end{center}}
\end{picture} 
 \\ \nonumber 
&+ \hbar^2\ \frac{\lambda^4}{4}  
\begin{picture}(22,10)(-2,0)
  \parbox{0cm}{\begin{center}
  \begin{fmfgraph*}(6,2)
  \setval
  \fmfforce{0w,1h}{v1}
  \fmfforce{1w,1h}{v2}
  \fmfforce{1w,3h}{v3}
  \fmfforce{1w,4.5h}{v4}
  \fmfforce{2w,1h}{v5}
  \fmfforce{2w,3h}{v6}
  \fmfforce{2w,4.5h}{v7}
  \fmfforce{3w,1h}{v8}
  \fmf{plain}{v1,v2,v5,v8}
  \fmf{plain}{v2,v3}
  \fmf{plain,left=1}{v3,v4,v3}
  \fmf{plain}{v5,v6}
  \fmf{plain,left=1}{v6,v7,v6}
  \fmfdot{v1,v8}
  \end{fmfgraph*}
  \end{center}}
\end{picture}
+ \hbar^2\ \frac{\lambda^4}{4}  
\begin{picture}(28,10)(-2,0)
  \parbox{0cm}{\begin{center}
  \begin{fmfgraph*}(6,2)
  \setval
  \fmfforce{0w,1h}{v1}
  \fmfforce{1w,1h}{v2}
  \fmfforce{1w,3h}{v3}
  \fmfforce{1w,4.5h}{v4}
  \fmfforce{2w,1h}{v5}
  \fmfforce{3w,1h}{v6}
  \fmfforce{4w,1h}{v7}
  \fmf{plain}{v1,v2,v5}
  \fmf{plain}{v2,v3}
  \fmf{plain,left=1}{v3,v4,v3}
  \fmf{plain,left=1}{v5,v6,v5}
  \fmf{plain}{v6,v7}
  \fmfdot{v1,v7}
  \end{fmfgraph*}
  \end{center}}
\end{picture}
+ \hbar^2\ \frac{\lambda^4}{4}  
\begin{picture}(28,10)(-2,0)
  \parbox{0cm}{\begin{center}
  \begin{fmfgraph*}(6,2)
  \setval
  \fmfforce{0w,1h}{v1}
  \fmfforce{1w,1h}{v2}
  \fmfforce{2w,1h}{v3}
  \fmfforce{3w,1h}{v4}
  \fmfforce{3w,3h}{v5}
  \fmfforce{3w,4.5h}{v6}
  \fmfforce{4w,1h}{v7}
  \fmf{plain}{v1,v2}
  \fmf{plain,left=1}{v2,v3,v2}
  \fmf{plain}{v3,v4,v7}
  \fmf{plain}{v4,v5}
  \fmf{plain,left=1}{v5,v6,v5}
  \fmfdot{v1,v7}
  \end{fmfgraph*}
  \end{center}}
\end{picture}
\\ \nonumber 
&+ \hbar^2\ \frac{\lambda^4}{4}  
\begin{picture}(16,18)(-2,0)
  \parbox{0cm}{\begin{center}
  \begin{fmfgraph*}(6,2)
  \setval
  \fmfforce{0w,1h}{v1}
  \fmfforce{1w,1h}{v2}
  \fmfforce{1w,3h}{v3}
  \fmfforce{1w,5.5h}{v4}
  \fmfforce{1w,7.5h}{v5}
  \fmfforce{1w,9h}{v6}
  \fmfforce{2w,1h}{v7}
  \fmf{plain}{v1,v2,v7}
  \fmf{plain}{v2,v3}
  \fmf{plain,left=1}{v3,v4,v3}
  \fmf{plain}{v4,v5}
  \fmf{plain,left=1}{v5,v6,v5}
  \fmfdot{v1,v7}
  \end{fmfgraph*}
  \end{center}}
\end{picture}
+ \hbar^2\ \frac{\lambda^4}{4}  
\begin{picture}(16,14)(-2,0)
  \parbox{0cm}{\begin{center}
  \begin{fmfgraph*}(6,2)
  \setval
  \fmfforce{0w,1h}{v1}
  \fmfforce{1w,1h}{v2}
  \fmfforce{1w,3h}{v3}
  \fmfforce{.5w,5h}{v4}
  \fmfforce{.25w,6h}{v5}
  \fmfforce{1.5w,5h}{v6}
  \fmfforce{1.75w,6h}{v7}
  \fmfforce{2w,1h}{v8}
  \fmf{plain}{v1,v2,v8}
  \fmf{plain}{v2,v3,v4}
  \fmf{plain,left=1}{v4,v5,v4}
  \fmf{plain}{v3,v6}
  \fmf{plain,left=1}{v6,v7,v6}
  \fmfdot{v1,v8}
  \end{fmfgraph*}
  \end{center}}
\end{picture}
+ \hbar^2\ \frac{\lambda^4}{4}  
\begin{picture}(16,14)(-2,0)
  \parbox{0cm}{\begin{center}
  \begin{fmfgraph*}(6,2)
  \setval
  \fmfforce{0w,1h}{v1}
  \fmfforce{1w,1h}{v2}
  \fmfforce{1w,3h}{v3}
  \fmfforce{1w,6h}{v4}
  \fmfforce{.5w,4.5h}{v5}
  \fmfforce{1.5w,4.5h}{v6}
  \fmfforce{2w,1h}{v7}
  \fmf{plain}{v1,v2,v7}
  \fmf{plain}{v2,v3}
  \fmf{plain,left=1}{v3,v4,v3}
  \fmf{plain}{v5,v6}
  \fmfdot{v1,v7}
  \end{fmfgraph*}
  \end{center}}
\end{picture}
+ \hbar^2\ \frac{\lambda^4}{4}  
\begin{picture}(22,14)(-2,0)
  \parbox{0cm}{\begin{center}
  \begin{fmfgraph*}(6,2)
  \setval
  \fmfforce{0w,1h}{v1}
  \fmfforce{1w,1h}{v2}
  \fmfforce{2w,1h}{v3}
  \fmfforce{3w,1h}{v4}
  \fmfforce{1.5w,2.5h}{v5}
  \fmfforce{1.5w,4.5h}{v6}
  \fmfforce{1.5w,6h}{v7}
  \fmf{plain}{v1,v2}
  \fmf{plain,left=1}{v2,v3,v2}
  \fmf{plain}{v3,v4}
  \fmf{plain}{v5,v6}
  \fmf{plain,left=1}{v6,v7,v6}
  \fmfdot{v1,v4}
  \end{fmfgraph*}
  \end{center}}
\end{picture}
 + \O(\lambda^6). 
\end{align}

Note that the grey boxes contain all the connected graphs. This motivates 
the name of the connected Green's functions. 
\end{point}

\begin{point}{Exercise: 3-points connected Green's function}
Write the diagrammatic expansion of the 3-points connected Green's
function, that is the solution of Eq.~(\ref{Dyson-Schwinger-3}). 
\end{point}

\begin{point}{Feynman rules}
We can therefore conclude that each connected Green's function  
$G(x_1,...,x_k)=\sum_n \lambda^n\ G_n(x_1,...,x_k)$ 
has perturbative coefficients $G_n(x_1,...,x_k)$ given by the finite sum 
of the {\em amplitude\/} $A(\G;x_1,...,x_k)$ of all the Feynman graphs 
with $n$ internal vertices, constructed according to the following 
{\em Feynman's rules\/} (valid for $J=0$): 
\begin{itemize}
\item
consider all the graphs with internal vertices of valence 3, 
and $k$ external vertices of valence 1;  
\item 
label the external vertices by $x_1,...,x_k$; 
\item
label the internal vertices by free variables $y,z,u,v,...$;
\item 
assign a weigth $G_0(y-z)$ to each edge joining the vertices $y$ and
$z$;
\item 
assign a weigth $\lambda$ to each internal vertex 
\begin{picture}(12,4)(-2,0)
  \parbox{0cm}{\begin{center}
  \begin{fmfgraph*}(4,2)
  \setval
  \fmfforce{0w,1h}{v1}
  \fmfforce{1w,1h}{v2}
  \fmfforce{2w,2.5h}{v3}
  \fmfforce{2w,-.5h}{v4}
  \fmf{plain}{v1,v2}
  \fmf{plain}{v2,v3}
  \fmf{plain}{v2,v4}
  \end{fmfgraph*}
  \end{center}}
\end{picture};

\item 
assign a weigth $\hbar$ to each loop 
\begin{picture}(6,4)(-2,0)
  \parbox{0cm}{\begin{center}
  \begin{fmfgraph*}(4,2)
  \setval
  \fmfforce{0w,1h}{v1}
  \fmfforce{1w,1h}{v2}
  \fmf{plain,left=1}{v1,v2,v1}
  \end{fmfgraph*}
  \end{center}}
\end{picture};
\item 
to obtain $A(\G;x_1,...x_k)$ for a given graph $\G$, 
multiply all the weigths and integrate over the free variables;
\item
divide by the symmetry factor $\Sym(\G)$ of the graph.
\end{itemize}
\end{point}

\begin{point}{Exercise: Feynman's rules in presence of an external source}
Modify the Feynman's rules given above so that they are valid 
when $J\neq 0$. 
\end{point}

\begin{point}{Exercise: compute some amplitudes}
Compute the amplitudes of the first Feynman graphs appearing 
in the expansions of the 2-points Green's function given above, 
using the Feynman's rules, and compare them with the results in  
Exercise~\ref{G(x,y)-value}. 
\end{point}

\begin{point}{Conclusion}
For a typical quantum field $\phi$ with Lagrangian density of the form 
\begin{align*}
\L(\phi) &= \frac{1}{2}\ \phi^t A \phi-J(x)\ \phi(x)
-\frac{\lambda}{3!}\ \phi(x)^3, 
\end{align*}
the connected $k$-points Green's function can be described 
as a formal series 
\begin{align*}
G(x_1,...,x_k) &= \sum_{n=0}^\infty \lambda^n G_n(x_1,...,x_k),  
\end{align*}
where each coefficient $G_n(x_1,...,x_k)$ is a finite sum 
\begin{align*}
G_n(x_1,...,x_k) 
&= \sum_{V(\G)=n} \frac{\hbar^{L(\G)}}{\Sym(\G)}\ A(\G;x_1,...,x_k)
\end{align*}
of amplitudes $A(\G;x_1,...,x_k)$ associated to each connected Feynman 
diagram $\G$ with $n$ internal vertices of valence 3. 
Note that, in these lectures, the amplitude of a graph is considered 
modulo the factor $\frac{\hbar^{L(\G)}}{\Sym(\G)}$. 
\end{point}


\subsection{Field theory on the momentum space}

\begin{point}{Momentum coordinates}
In relativistic quantum mechanics, the {\em four-momentum\/} $p$, that we 
call simply {\em momentum\/} here, is the conjugate variable of the 
four-position $x$, seen as an operator of multiplication on the wave function. 
Therefore the momentum is the Fourier transform of the operator of derivation 
by the position, and belongs to the Fourier space $\widehat{\R^D}$. 

To express the field theory on the momentum variables, we 
Fourier transform all the components of the equation of motion: 
\begin{align*}
\widehat{\phi}(p) & = \int_{\R^D} d^Dx\ \phi(x)\ e^{i p\cdot x}, \\
\widehat{J}(p) & = \int_{\R^D} d^Dx\ J(x)\ e^{i p\cdot x}, \\ 
\widehat{G_0}(p) & = \int_{\R^D} d^Dx\ G_0(x-y)\ e^{i p\cdot (x-y)} , 
\end{align*}
for instance, for the Klein-Gordon field, 
$\widehat{G_0}(p)=\dfrac{1}{p^2+m^2}$ is the Fourier transform of 
the free propagator~(\ref{KG-propagator}). 
The classical Euler-Lagrange equation~(\ref{Euler-Lagrange-equation}) 
is then transformed into 
\begin{align*}
\widehat{\phi}(p) & = \widehat{G_0}(p)\ \widehat{J}(p)
+\frac{\lambda}{2}\ \widehat{G_0}(p)\ \int \frac{d^D q}{(2\pi)^D}\ 
\widehat{\phi}(q)\ \widehat{\phi}(p-q).  
\end{align*}
The Fourier transform of the Green's functions is
\begin{align*}
\widehat{G}^{(k)}(p_1,...,p_k) &= \int_{(\R^D)^k} d^Dx_1... d^Dx_k\ 
G(x_1,...,x_k)\ e^{i p_1\cdot (x_1-x_k)}\cdots e^{i p_k\cdot (x_{k-1}-x_k)}, 
\end{align*}
where the translation invariance of $G(x_1,...,x_k)$ implies that 
$\sum_{i=1,...,k} p_i =0$, and the Dyson-Schwinger equations 
(\ref{Dyson-Schwinger-equation-1}), (\ref{Dyson-Schwinger-equation-2}), etc, 
can easily be expressed in terms of external momenta: 
\begin{align*}
\widehat{G}^{(1)}(0) 
&= \frac{\lambda}{2}\ \widehat{G_0}(0)\ 
\int\frac{d^D q}{(2\pi)^D}\ \widehat{G}^{(1)}(q) \widehat{G}^{(1)}(-q)
+ \hbar\ \frac{\lambda}{2}\ \widehat{G_0}(0)\ 
\int\frac{d^D q}{(2\pi)^D}\ \widehat{G}^{(2)}(q), \\
\widehat{G}^{(2)}(p) 
&= \widehat{G_0}(p) 
+ \lambda\ \widehat{G_0}(p)\ \widehat{G}^{(1)}(0)\ \widehat{G}^{(2)}(p) 
+ \hbar\ \frac{\lambda}{2}\ \widehat{G_0}(p) \int \frac{d^D q}{(2\pi)^D}\ 
\widehat{G}^{(3)}(q,p-q,-p), 
\end{align*}
and so on. 
\end{point}

\begin{point}{Feynman graphs on the momentum space}
\label{Feynman-graphs-momentum}
The Feynman graphs on the momentum space look exactely like those on the 
space-time coordinates, except that the external legs are not fixed in the 
{\em dotted\/} positions $x_1,...,x_k$, but have {\em oriented\/} edges, 
and in particular oriented external legs labeled by momenta $p_1,...,p_k$, 
where the arrows tell what is the direction of the propagation. 
The Feynman notations are:  
\begin{itemize}
\item
field $\widehat{\phi}(p)$ = 
\begin{picture}(10,4)(-2,0)
  \parbox{0cm}{\begin{center}
  \begin{fmfgraph*}(6,2)
  \setval
  \fmfforce{0w,1h}{v1}
  \fmfforce{1w,1h}{v2}
  \fmf{plain_arrow,label=$p$}{v1,v2}
  \fmfblob{.5w}{v2}
  \end{fmfgraph*}
  \end{center}}
\end{picture}, 
or $k$-points connected Green's function $\widehat{G}^{(k)}(p_1,...,p_k)$ = 
\begin{picture}(16,8)(-6,0)
  \parbox{0cm}{\begin{center}
  \begin{fmfgraph*}(8,8)
  \setval
  \fmfsurroundn{v}{7}
  \fmfforce{c}{c}
  \fmfv{decor.shape=circle,decor.size=.33w,decor.filled=shaded}{c}
  \fmf{plain_arrow}{v1,c}
  \fmf{plain_arrow}{v2,c}
  \fmf{plain_arrow}{v3,c}
  \fmf{plain_arrow}{v4,c}
  \fmf{plain_arrow}{v5,c}
  \fmf{plain_arrow}{v6,c}
  \fmf{plain_arrow}{v7,c}
  \fmflabel{$p_1$}{v4}
  \fmflabel{$p_2$}{v3}
  \fmflabel{$p_k$}{v5}
  \end{fmfgraph*}
  \end{center}}
\end{picture};

\item
propagator $\widehat{G_0}(p)$ = 
\begin{picture}(10,4)(-2,0)
  \parbox{0cm}{\begin{center}
  \begin{fmfgraph*}(6,2)
  \setval
  \fmfforce{0w,1h}{v1}
  \fmfforce{1w,1h}{v2}
  \fmf{plain_arrow,label=$p$}{v1,v2}
  \end{fmfgraph*}
  \end{center}}
\end{picture}; 

\item
source $\widehat{J}(p)$ = 
\begin{picture}(10,4)(0,0)
  \parbox{0cm}{\begin{center}
  \begin{fmfgraph*}(6,2)
  \setval
  \fmfforce{.5w,1h}{v1}
  \fmfforce{1w,1h}{v2}
  \fmf{plain,label=$p$}{v2,v1}
  \fmfv{decor.shape=cross,decor.size=.4w}{v2}
  \end{fmfgraph*}
  \end{center}}
\end{picture} 
(short leg labelled by $p$), 
such that $\widehat{G_0}(p) \widehat{J}(p)$ has the same dimension as 
$\widehat{\phi}(p)$. 
\end{itemize}
The Feynman graphs with short external legs are sometimes called 
{\em truncated\/} or {\em amputated\/}. 
Modulo these few differences, the Euler-Lagrange equation, the 
Dyson-Schwinger equations, and their perturbative solutions, are 
the same as those already given on the space-time coordinates. 

To simplify the notations, from now on we denote by $G_0(p)$ the free 
propagator also in the momentum space, instead of $\widehat{G_0}(p)$, 
and in general we omit the hat symbol. Similarly, we omit the orientation 
of the propagators unless necessary. 
\end{point}

\begin{point}{One-particle irreducible graphs}
The Feynman rules, which allow us to write the amplitude of a Feynman graph, 
implicetely state that the amplitude of a non-connected graph is the product 
of the amplitudes of all its connected components
(cf. Eqs.~(\ref{full-connected-green-functions}) and 
Exercise~\ref{full-connected-diagrams}). 
If we work in the momentum space, then from the Feynman rules it also follows 
that if a graph $\G$ is the junction of two subgraphs $\G_1$ and $\G_2$, 
through a simple edge, that is 
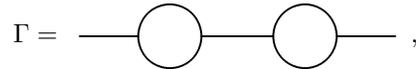
\begin{align*}
\G &= 
\begin{picture}(46,6)(-2,0)
  \parbox{0cm}{\begin{center}
  \begin{fmfgraph*}(12,2)
  \setval
  \fmfforce{0w,1h}{v1}
  \fmfforce{1w,1h}{v2}
  \fmfforce{2.5w,1h}{v3}
  \fmfforce{3.5w,1h}{v4}
  \fmf{plain,label=$p$}{v2,v1}
  \fmf{plain,label=$p$}{v3,v2}
  \fmf{plain,label=$p$}{v4,v3}
  \fmfv{decor.shape=circle,decor.size=.7w,decor.filled=empty,%
label=$\Gamma_1$,label.dist=0h}{v2}
  \fmfv{decor.shape=circle,decor.size=.7w,decor.filled=empty,%
label=$\Gamma_2$,label.dist=0h}{v3}
  \end{fmfgraph*}
  \end{center}}
\end{picture}, 
\\ & 
\end{align*}
then the amplitude of $\G$ is the product of the amplitude of the 
single graphs, that is 
\begin{align*}
A(\G;p) &= G_0(p)\ A(\G_1;p)\ G_0(p)\ A(\G_2;p)\ G_0(p),  
\end{align*}
where $\G_1$ and $\G_2$ here are truncated on both sides. 
(Note that the internal edge must have momentum $p$ because of the
conservation of total momentum at each vertex.) 

We say that a connected Feynman graph $\G$ is {\em one-particle
irreducible\/}, in short 1PI, if it remains connected when 
we cut one of its edges. In particular, the free propagator 
$\begin{picture}(10,4)(-2,0)
  \parbox{0cm}{\begin{center}
  \begin{fmfgraph*}(6,2)
  \setval
  \fmfforce{0w,1h}{v1}
  \fmfforce{1w,1h}{v2}
  \fmf{plain}{v2,v1}
  \end{fmfgraph*}
  \end{center}}
\end{picture}$ 
is not 1PI, therefore the 1PI graphs in the momentum space are truncated. 
For instance, the graphs
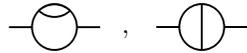
\begin{align*}
\begin{picture}(18,4)(0,0)
  \parbox{0cm}{\begin{center}
  \begin{fmfgraph*}(6,2)
  \setval
  \fmfforce{.5w,1h}{v1}
  \fmfforce{1w,1h}{v2}
  \fmfforce{2w,1h}{v3}
  \fmfforce{2.5w,1h}{v4}
  \fmfforce{1.15w,2.1h}{v5}
  \fmfforce{1.85w,2.1h}{v6}
  \fmf{plain}{v1,v2}
  \fmf{plain,left=1}{v2,v3,v2}
  \fmf{plain}{v3,v4}
  \fmf{plain,left=.4}{v5,v6,v5}
  \end{fmfgraph*}
  \end{center}}
\end{picture},  
\begin{picture}(18,4)(0,0)
  \parbox{0cm}{\begin{center}
  \begin{fmfgraph*}(6,2)
  \setval
  \fmfforce{.5w,1h}{v1}
  \fmfforce{1w,1h}{v2}
  \fmfforce{1.5w,2.45h}{v3}
  \fmfforce{1.5w,-.45h}{v4}
  \fmfforce{2w,1h}{v5}
  \fmfforce{2.5w,1h}{v6}
  \fmf{plain}{v1,v2}
  \fmf{plain,left=1}{v2,v5,v2}
  \fmf{plain}{v3,v4}
  \fmf{plain}{v5,v6}
  \end{fmfgraph*}
  \end{center}}
\end{picture} 
\end{align*}
are 1PI, while the graphs 
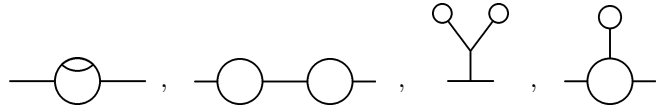
\begin{align*}
\begin{picture}(22,4)(-2,0)
  \parbox{0cm}{\begin{center}
  \begin{fmfgraph*}(6,2)
  \setval
  \fmfforce{0w,1h}{v1}
  \fmfforce{1w,1h}{v2}
  \fmfforce{2w,1h}{v3}
  \fmfforce{3w,1h}{v4}
  \fmfforce{1.15w,2.1h}{v5}
  \fmfforce{1.85w,2.1h}{v6}
  \fmf{plain}{v1,v2}
  \fmf{plain,left=1}{v2,v3,v2}
  \fmf{plain}{v3,v4}
  \fmf{plain,left=.4}{v5,v6,v5}
  \end{fmfgraph*}
  \end{center}}
\end{picture},  
\begin{picture}(30,4)(0,0)
  \parbox{0cm}{\begin{center}
  \begin{fmfgraph*}(6,2)
  \setval
  \fmfforce{.5w,1h}{v1}
  \fmfforce{1w,1h}{v2}
  \fmfforce{2w,1h}{v3}
  \fmfforce{3w,1h}{v4}
  \fmfforce{4w,1h}{v5}
  \fmfforce{4.5w,1h}{v6}
  \fmf{plain}{v1,v2}
  \fmf{plain,left=1}{v2,v3,v2}
  \fmf{plain}{v3,v4}
  \fmf{plain,left=1}{v4,v5,v4}
  \fmf{plain}{v5,v6}
  \end{fmfgraph*}
  \end{center}}
\end{picture},  
\begin{picture}(16,8)(-2,0)
  \parbox{0cm}{\begin{center}
  \begin{fmfgraph*}(6,2)
  \setval
  \fmfforce{.5w,1h}{v1}
  \fmfforce{1w,1h}{v2}
  \fmfforce{1w,3h}{v3}
  \fmfforce{.5w,5h}{v4}
  \fmfforce{.25w,6h}{v5}
  \fmfforce{1.5w,5h}{v6}
  \fmfforce{1.75w,6h}{v7}
  \fmfforce{1.5w,1h}{v8}
  \fmf{plain}{v1,v2,v8}
  \fmf{plain}{v2,v3,v4}
  \fmf{plain,left=1}{v4,v5,v4}
  \fmf{plain}{v3,v6}
  \fmf{plain,left=1}{v6,v7,v6}
  \end{fmfgraph*}
  \end{center}}
\end{picture},  
\begin{picture}(20,8)(0,0)
  \parbox{0cm}{\begin{center}
  \begin{fmfgraph*}(6,2)
  \setval
  \fmfforce{.5w,1h}{v1}
  \fmfforce{1w,1h}{v2}
  \fmfforce{2w,1h}{v3}
  \fmfforce{2.5w,1h}{v4}
  \fmfforce{1.5w,2.5h}{v5}
  \fmfforce{1.5w,4.5h}{v6}
  \fmfforce{1.5w,6h}{v7}
  \fmf{plain}{v1,v2}
  \fmf{plain,left=1}{v2,v3,v2}
  \fmf{plain}{v3,v4}
  \fmf{plain}{v5,v6}
  \fmf{plain,left=1}{v6,v7,v6}
  \end{fmfgraph*}
  \end{center}}
\end{picture}
\end{align*}  
are not 1PI. 
If we denote the junction of graphs through one of their external legs by 
the concatenation, for instance
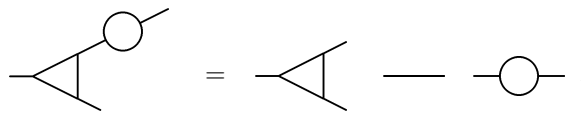
\begin{align*}
\begin{picture}(28,8)(0,0)
  \parbox{0cm}{\begin{center}
  \begin{fmfgraph*}(6,2)
  \setval
  \fmfforce{.5w,1h}{v1}
  \fmfforce{1w,1h}{v2}
  \fmfforce{2w,2.5h}{v3}
  \fmfforce{2w,-.5h}{v4}
  \fmfforce{3w,4h}{v5}
  \fmfforce{2.5w,-1.25h}{v6}
  \fmfforce{4w,5.5h}{v7}
  \fmf{plain,label=$p_1$}{v2,v1}
  \fmf{plain}{v2,v3,v5}
  \fmf{plain,label=$p_2$}{v5,v7}
  \fmf{plain}{v2,v4}
  \fmf{plain}{v3,v4}
  \fmf{plain,label=$p_3$,label.dist=0.5h}{v4,v6}
  \fmfv{decor.shape=circle,decor.size=.85w,decor.filled=empty}{v5}
  \end{fmfgraph*}
  \end{center}}
\end{picture} 
&= 
\begin{picture}(20,8)(0,0)
  \parbox{0cm}{\begin{center}
  \begin{fmfgraph*}(6,2)
  \setval
  \fmfforce{.5w,1h}{v1}
  \fmfforce{1w,1h}{v2}
  \fmfforce{2w,2.5h}{v3}
  \fmfforce{2w,-.5h}{v4}
  \fmfforce{2.5w,3.25h}{v5}
  \fmfforce{2.5w,-1.25h}{v6}
  \fmf{plain,label=$p_1$}{v2,v1}
  \fmf{plain}{v2,v3}
  \fmf{plain,label=$p_2$,label.dist=0.5h}{v5,v3}
  \fmf{plain}{v2,v4}
  \fmf{plain}{v3,v4}
  \fmf{plain,label=$p_3$,label.dist=0.5h}{v4,v6}
  \end{fmfgraph*}
  \end{center}}
\end{picture} 
\begin{picture}(12,4)(0,0)
  \parbox{0cm}{\begin{center}
  \begin{fmfgraph*}(8,2)
  \setval
  \fmfforce{0w,1h}{v1}
  \fmfforce{1w,1h}{v2}
  \fmf{plain,label=$p_2$}{v2,v1}
  \end{fmfgraph*}
  \end{center}}
\end{picture}
\begin{picture}(14,4)(0,0)
  \parbox{0cm}{\begin{center}
  \begin{fmfgraph*}(6,2)
  \setval
  \fmfforce{0w,1h}{v1}
  \fmfforce{1w,1h}{v2}
  \fmfforce{2w,1h}{v3}
  \fmf{plain,label=$p_2$}{v2,v1}
  \fmf{plain,label=$p_2$}{v3,v2}
  \fmfv{decor.shape=circle,decor.size=.85w,decor.filled=empty}{v2}
  \end{fmfgraph*}
  \end{center}}
\end{picture}, 
\\ & 
\end{align*}
then any connected graph can then be seen as the concatenantion of its 
1PI components and the free propagators necessary to joint them. 
To avoid these free propagators popping out at any cut, we can consider 
graphs which are truncated only on some of their external legs, and allow 
to joint truncated legs with full ones, for instance
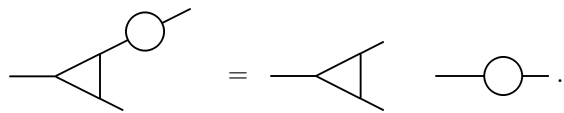
\begin{align*}
\begin{picture}(28,8)(0,0)
  \parbox{0cm}{\begin{center}
  \begin{fmfgraph*}(6,2)
  \setval
  \fmfforce{0w,1h}{v1}
  \fmfforce{1w,1h}{v2}
  \fmfforce{2w,2.5h}{v3}
  \fmfforce{2w,-.5h}{v4}
  \fmfforce{3w,4h}{v5}
  \fmfforce{2.5w,-1.25h}{v6}
  \fmfforce{4w,5.5h}{v7}
  \fmf{plain,label=$p_1$}{v2,v1}
  \fmf{plain}{v2,v3,v5}
  \fmf{plain,label=$p_2$}{v5,v7}
  \fmf{plain}{v2,v4}
  \fmf{plain}{v3,v4}
  \fmf{plain,label=$p_3$,label.dist=0.5h}{v4,v6}
  \fmfv{decor.shape=circle,decor.size=.85w,decor.filled=empty}{v5}
  \end{fmfgraph*}
  \end{center}}
\end{picture} 
&= 
\begin{picture}(22,8)(-2,0)
  \parbox{0cm}{\begin{center}
  \begin{fmfgraph*}(6,2)
  \setval
  \fmfforce{0w,1h}{v1}
  \fmfforce{1w,1h}{v2}
  \fmfforce{2w,2.5h}{v3}
  \fmfforce{2w,-.5h}{v4}
  \fmfforce{2.5w,3.25h}{v5}
  \fmfforce{2.5w,-1.25h}{v6}
  \fmf{plain,label=$p_1$}{v2,v1}
  \fmf{plain}{v2,v3}
  \fmf{plain,label=$p_2$,label.dist=0.5h}{v5,v3}
  \fmf{plain}{v2,v4}
  \fmf{plain}{v3,v4}
  \fmf{plain,label=$p_3$,label.dist=0.5h}{v4,v6}
  \end{fmfgraph*}
  \end{center}}
\end{picture} 
\begin{picture}(18,4)(-2,0)
  \parbox{0cm}{\begin{center}
  \begin{fmfgraph*}(6,2)
  \setval
  \fmfforce{0w,1h}{v1}
  \fmfforce{1.5w,1h}{v2}
  \fmfforce{2.5w,1h}{v3}
  \fmf{plain,label=$p_2$}{v2,v1}
  \fmf{plain,label=$p_2$}{v3,v2}
  \fmfv{decor.shape=circle,decor.size=.85w,decor.filled=empty}{v2}
  \end{fmfgraph*}
  \end{center}}
\end{picture}. 
\\ & 
\end{align*}
With this trick, any connected graph $\G$ can be seen as the junction 
$\G=\G_1\cdots\G_s$ of its 1PI components (modulo some free propagators). 
\end{point}

\begin{point}{Proper or 1PI Green's functions}
The fact that any connected Feynman graph can be reconstructed from its 
1PI components implies that the connected Green's function 
\begin{align*}
G^{(k)}(p_1,...,p_k) &= \sum_{E(\G)=k} \lambda^{V(\G)} 
\frac{\hbar^{L(\G)}}{\Sym(\G)}\ A(\G;p_1,...,p_k), 
\end{align*}
where the sum is over all connected graphs with $k$ external legs, can be 
reconstructed from the set of {\em proper\/} or {\em 1PI Green's functions\/}  
\begin{align*}
\GOPI^{(k)}(p_1,...,p_k) 
&= \underset{\mbox{\scriptsize 1PI\ } \G}{\sum_{E(\G)=k}} 
\lambda^{V(\G)} \frac{\hbar^{L(\G)}}{\Sym(\G)}\ A(\G;p_1,...,p_k), 
\end{align*}
where the sum is now over 1PI graphs suitably truncated. 
The precise relation between connected and proper Green's functions can be 
given easily only for the 2-point Green's functions: in this case we have 
\begin{align*}
G^{(2)}(p) &= G_0(p)\ \left[1-\GOPI^{(2)}(p)\ G_0(p)\right]^{-1}.
\end{align*}
The general case is much more involved, and was treated recently using 
algebraic tools by \^A.~Mestre and R.~Oeckl in \cite{MestreOeckl}. 
\end{point}

\begin{point}{Conclusion}
In summery, for a typical quantum field $\phi$ with Lagrangian density 
of the form 
\begin{align*}
\L(\phi) &= \frac{1}{2}\ \phi^t A \phi-J(x)\ \phi(x)
-\frac{\lambda}{3!}\ \phi(x)^3, 
\end{align*}
the connected $k$-points Green's function on the 
momentum space can be described as a formal series 
\begin{align*}
G(p_1,...,p_k) &= \sum_{n=0}^\infty \lambda^n G_n(p_1,...,p_k),  
\end{align*}
where each coefficient $G_n(p_1,...,p_k)$ is a finite sum of amplitudes 
associated to each (partially amputated) connected Feynman diagram with $n$ 
internal vertices of valence 3, and the amplitude of each graph $\G$ is the 
product of the amplitudes of its 1PI components $\G_i$, that is
\begin{align*}
G_n(p_1,...,p_k) 
&= \sum_{V(\G)=n} \frac{\hbar^{L(\G)}}{\Sym(\G)}\ A(\G;p_1,...,p_k) \\ 
&= \sum_{V(\G)=n}\ \prod_{\G=\G_1\cdots\G_s} \frac{\hbar^{L(\G_i)}}{\Sym(\G_i)}\ 
A(\G_i;p^{(i)}_1,...,p^{(i)}_{k_i}). 
\end{align*}
\end{point}


\section*{Lecture IV - Renormalization}
\addcontentsline{toc}{section}{\bf Lecture IV - Renormalization}
\label{lecture4}

In Lecture~II we computed the first terms of the
perturbative solution of the classical and the quantum interacting
fields. As we saw in Lecture~III, these terms can be
regarded as the amplitudes of some useful combinatorial objects, 
the rooted trees and the Feynman's graphs. These analitic expressions,
the amplitudes, are constructed as repeated integrals of products 
of the field propagator $G_0$ and eventually an external field $J$. 
The field propagator $G_0(x)$ is a distribution of the point $x$, 
and it is singular in $x=0$ if $n>1$. 
Then, the square $G_0(x)^2$ is a continuous function for $x\neq 0$, 
but it is not defined in $x=0$. 
On the momentum space, this problem is translated into the divergency 
of the integral containing powers of the free propagator. 

The powers of a free propagator never occur in the amplitude of the 
trees labelling the perturbative expansion of classical fields, cf. 
Eq.~(\ref{Euler-Lagrange-solution}). Similarly, they do not
occur in the classical part of the perturbative expansion of
Green's functions for a quantum field (that is, those terms which
are not factors of $\hbar$). 
Instead, such terms occur in the quantum corrections, that is, the
terms which are factors of $\hbar$. For instance, the last two terms in 
Eq.~(\ref{Dyson-Schwinger-solution-W}) contain $G_0(y-y)=G_0(0)$ and 
the square $G_0(y-z)^2$ which is meaningless for $y=z$. 

In this lecture we explain some tools developped to give a meaning
to the ill-defined terms appearing in the perturbative expansions of
the Green's functions. This technique is known as the theory of 
renormalization. 


\subsection{Renormalization of Feynman amplitudes}
\label{Renormalization-graphs}

The renormalization of the ill-defined amplitudes can be done for graphs 
on the momentum variables as well as on the space-time variables. 
On the space-time variables, the renormalization program has been described 
by H.~Epstein and V.~J.~Glaser in \cite{EpsteinGlaser}, in the context 
of the {\em causal perturbation theory\/}. 
However, to describe renormalization, it is convenient to work on the 
momentum space and to consider 1PI graphs. 
\bigskip 

\begin{point}{Problem of divergent integrals: ultraviolet and infrared
divergencies}
In dimension $D=1$, all the integrals appearing in the perturbative
expansion of the Green's functions are convergent. 
For example, if we consider the Klein-Gordon field $\phi$, 
the free propagator 
\begin{align*}
G_0(x-y) &= \int_{\R} \frac{d p}{2\pi} \frac{e^{-i p (x-y)}}{p^2+m^2}
\end{align*}
is a continuous function. Therefore all the products of propagators 
are also continuous functions, and the integrals are well defined. 

In dimension $D>1$, the free propagator $G_0(x-y)$ is a singular
distribution on the diagonal $x=y$, and the product with other
distributions which are singular at the same points, such as its 
powers $G_0(x-y)^m$, makes no sense. 
For the Klein-Gordon field, for example, this happens already 
in the simple loop 
\begin{align*}
\G &=  
\begin{picture}(30,4)(-6,0)
  \parbox{0cm}{\begin{center}
  \begin{fmfgraph*}(6,2)
  \setval
  \fmfforce{0w,1h}{v1}
  \fmfforce{1w,1h}{v2}
  \fmfforce{2w,1h}{v3}
  \fmfforce{3w,1h}{v4}
  \fmf{plain}{v1,v2}
  \fmf{plain,left=1}{v2,v3,v2}
  \fmf{plain}{v3,v4}
  \fmfdot{v1,v4}
  \fmflabel{$x$}{v1}
  \fmflabel{$y$}{v4}
  \end{fmfgraph*}
  \end{center}}
\end{picture}, 
\end{align*}
whose amplitude 
\begin{align*}
A(\G;x,y)  
&= \int d^D u\ d^D v\ G_0(x-u)\ G_0(u-v)^2\ G_0(v-z) 
\end{align*}
contains the square $G_0(u-v)^2$. 
To understand how the integral is affected by the singularity, we better 
write the simple loop on the momentum space. 
The Fourier transform of $\G$ gives the (truncated) simple loop 
\begin{align*}
\begin{picture}(18,4)(0,0)
  \parbox{0cm}{\begin{center}
  \begin{fmfgraph*}(6,2)
  \setval
  \fmfforce{.5w,1h}{v1}
  \fmfforce{1w,1h}{v2}
  \fmfforce{2w,1h}{v3}
  \fmfforce{2.5w,1h}{v4}
  \fmf{plain,label=$p$}{v2,v1}
  \fmf{plain,right=1}{v3,v2,v3}
  \fmf{plain,label=$p$}{v4,v3}
  \end{fmfgraph*}
  \end{center}}
\end{picture}. 
\end{align*}
To compute its amplitude, we write the integrated momentum $q$ in spherical 
coordinates, with $|q|$ denoting the module. Then we see that for 
$|q|\rightarrow \infty$ the integral 
\begin{align*}
\int \frac{d^D q}{(2\pi)^D}\ \frac{1}{q^2+m^2}\ \frac{1}{(p-q)^2+m^2}
\end{align*}
roughly behaves like 
\begin{align*}
\int_{|q|_{min}}^\infty d |q|^D \frac{1}{|q|^4} &\simeq 
\int_{|q|_{min}}^\infty d |q|\ \frac{1}{|q|^{4-(D-1)}} .
\end{align*}
This integral converges if and only if $4-(D-1)>1$, that is $D<4$. 
Therefore $A(\G;x,y)$ diverges when the dimension of the
base-space is $D\geq 4$. 

The divergency of an amplitude $A(\G;p)$ which occurs when 
an integrated variable $q$ has module $|q|\rightarrow\infty$ 
is called {\em ultraviolet\/}. The divergency which occurs when 
$|q|\rightarrow |q|_{min}$ is called {\em infrared\/}. 
The infrared divergencies appear typically when the mass $m$ is zero 
and $|q|_{min}=0$ (for instance, for photons). 
In this lecture we only deal with ultraviolet divergencies. 

To simplify the notations, if $\G$ is a graph with $k$ external legs, 
we denote its amplitudes $A(\G;x_1,...,x_k)$ or $A(\G;p_1,...,p_k)$ 
simply by $A(\G)$, when the dependence on the external parameters 
$x_1,...,x_k$ or $p_1,...,p_k$ is not relevant. 
\end{point}

\begin{point}{Renormalized amplitudes, normalization conditions and 
renormalisable theories}
There is a general procedure to estimate which integrals are
divergent, and then to extract from each infinite value a finite 
contribution which has a physical meaning. This program is called 
the {\em renormalization\/} of the amplitude of Feynman graphs. 

Given a graph $\G$ with divergent amplitude $A(\G)$, the aim of the 
renormalization program is to find a finite contribution $\Ar(\G)$, 
called {\em renormalized amplitude\/}, which satisfies some physical 
requirements. In contraposition to the renormalized amplitude, the 
original divergent amplitude is often called {\em bare\/} or {\em nude\/}. 

The physical conditions required, called {\em normalization conditions\/},  
are those which guarantee that the connected Green's function and 
its derivatives have a precise value at a given point. 
The theory is called {\em renormalisable\/} if the number of conditions 
that we have to impose to determine the amplitude of all Feynaman graphs 
is finite.  
For instance, the $\phi^3$ theory is renormalizable in dimension 
$D\leq 6$. 
\end{point}

\begin{point}{Power counting: classification of one loop divergencies}
\label{power counting}
The {\em superficial degree of divergency\/} of a 1PI graph $\G$ 
measures the degree of singularity $\w(\G)$ of the integral in
$A(\G)$ with respect to the integrated variables $q_1,q_2,...$. 
By definition, $\w(\G)$ is the integer such that, under the
transformation of momentum $q_i \rightarrow t q_i$, with $t\in\R$, 
the amplitude is transformed as 
\begin{align*}
A(\G) \quad\longrightarrow\quad t^{\w(\G)}\ A(\G). 
\end{align*}
The superficial degree of divergency detects the ``real'' divergency only for
diagrams with one single loop: in this case $A(\G)$ converges if and only if 
$\w(\G)$ is negative. The divergencies for single-loop graphs are then 
classified according to $\w(\G)$: 
\begin{itemize}
\item 
a graph $\G$ has a {\em logarithmic\/} divergency if
$\w(\G)=0$; 
\item 
it has a {\em polynomial\/} divergency of degree $\w(\G)$ if 
$\w(\G)>0$. That is, the divergency is {\em linear\/} if 
$\w(\G)=1$, it is {\em quadratic\/} if $\w(\G)=2$, 
and so on. 
\end{itemize}
If the graph contains many loops, instead, it can have a negative value of 
$\w(\G)$ and at the same time contain some divergent subgraphs. 
Therefore $\w(\G)$ can not be used to estimate the real (not superficial) 
divergency of a graph $\G$ with many loops. In this case, we first have 
to compute $\w(\gamma)$ for each single 1PI subgraph $\gamma$ of $\G$, 
starting from the subgraphs with a simple loop and proceding by enlarging 
the subgraphs until we reach $\G$ itself. 
This recursive procedure on the subgraphs will be discussed in details
for the renormalization of the graph with many loops. 

The superficial degree of divergency can be computed easily knowing only 
the combinatorial datas of each graph. If we denote by 
\begin{itemize}
\item 
$I$ the number of internal edges of a given graph, 
\item 
$E$ the number of external edges,
\item 
$V$ the number of vertices,  
\item 
$L$ the number of loops ($L=I-V+1$ because of conservation of
momentum at each vertex),
\end{itemize}
then for the Klein-Gordon field we have 
\begin{align}
\label{degree-divergency}
\w(\G) &= D\ L-2\ I = D+(D-2)\ I-D\ V,
\end{align}
where $D$ is the dimension of the base-space. 
In fact, the transformation $q \rightarrow t q$ gives
\begin{align*}
\frac{d^D\ q}{(2\pi)^D} &\quad\longrightarrow\quad 
t^D\ \frac{d^D\ q}{(2\pi)^D}, \\ 
\frac{1}{q^2+m^2} &\quad\longrightarrow\quad 
t^{-2}\ \frac{1}{q^2+m^2},  
\end{align*}
therefore to compute $\w(\G)$ we have to add a term $D$ for each loop, 
and a term $-2$ for each internal edge.

In particular, for the $\phi^3$-theory (the field $\phi$ with
interacting Lagrangian proportional to $\phi^3$), we have an
additional relation $3V=E+2I$, and therefore  
\begin{align*}
\w(\G) &= D+\frac{D-6}{2}\ V -\frac{D-2}{2}\ E.
\end{align*}
\end{point}

\begin{point}{Regularization: yes or not}
Let $\G$ be a divergent graph, that is, we suppose that the amplitude 
$A(\G)$ presents an ultraviolet divergency. 
In order to extract the renormalized amplitude $\Ar(\G)$, we can not work 
directly on $A(\G)$, which is infinite. Instead, there are the following 
two main possibilities. 
\bigskip  

\noindent{\bf Regularization:} 
We can modify $A(\G)$ into a new integral $A_{\rho}(\G)$, called 
{\em regularized amplitude\/}, by introducing a {\em regularization 
parameter\/} $\rho$ such that 
\begin{itemize}
\item
$A_{\rho}(\G)$ converges, 
\item 
$A_{\rho}(\G)$ reproduces the divergency of $A(\G)$ in a certein limit 
$\rho\rightarrow\rho_0$. 
\end{itemize}
The regularized amplitude $A_{\rho}(\G)$ is then a well-defined function of 
the external momenta with values which depends on the parameter $\rho$. 
Let us denote by $\Rr$ the ring of such values. 

Then we can modify the function $A_{\rho}(\G)$ into a 
new function $\Ar_{\rho}(\G)$ such that the limit 
\begin{align*}
\Ar(\G) &= \lim_{\rho\rightarrow\rho_0} \Ar_{\rho}(\G)
\end{align*}
is finite and compatible with the normalization conditions. 

Since we are dealing here with ultraviolet divergencies, it suffices
to choose as regularization parameter a {\em cut-off\/}
$\Lambda\in\R^+$ which bounds the integrated variables by above. 
If we denote by $I(\G;q_1,...,q_\ell)$ the integrand of $A(\G)$, 
that is 
\begin{align*}
A(\G) &= \int \frac{d^D\ q_1}{(2\pi)^D}\cdots \frac{d^D\ q_\ell}{(2\pi)^D}\ 
I(\G;q_1,...,q_\ell), 
\end{align*}
the regularized amplitude can be choosen as 
\begin{align*}
A_{\Lambda}(\G) &= \int_{|q_i|\leq\Lambda} 
\frac{d^D\ q_1}{(2\pi)^D}\cdots \frac{d^D\ q_\ell}{(2\pi)^D}\ I(\G;q_1,...,q_\ell),
\end{align*}
which reproduces the divergency of $A(\G)$ for $\Lambda\rightarrow\infty$. 
Alternatively, the regularized amplitude $A_{\Lambda}(\G)$ 
can also be described as 
\begin{align*}
A_{\Lambda}(\G) 
&= \int \frac{d^D\ q_1}{(2\pi)^D}\cdots \frac{d^D\ q_\ell}{(2\pi)^D}\ 
\chi_{\Lambda}(|q_1|,...,|q_\ell|)\ I(\G;q_1,...,q_\ell), 
\end{align*}
where $\chi_{\Lambda}(|q_1|,...,|q_\ell|)$ is the step function with 
value $1$ for $|q_1|,...,|q_\ell|\leq\Lambda$ and value $0$ for 
$|q_1|,...,|q_\ell|>\Lambda$. 

Beside the cut-off, there exist other possible regularizations. One of the most
frequently used is the {\em dimensional regularization\/}, which
modifies the real dimension $D$ by a complex parameter $\varepsilon$ 
such that $A_{\varepsilon}(\G)$ reproduces the divergency of $A(\G)$ for 
$\varepsilon\rightarrow 0$. Since this regularization demands many
explanations, and we are not going to use it here, we omit the
details which can be found in \cite{'tHooftVeltman} or \cite{ItzyksonZuber}. 
\bigskip  

\noindent{\bf Integrand functions:} 
The integrand $I(\G;q_1,...,q_\ell)$ of $A(\G)$ is a well defined (rational) 
function of the variables $q_1,...,q_\ell$. Therefore we can work directly 
with the integrand in order to modify it into a new function 
$\Ir(\G;q_1,...,q_\ell)$, called {\em renormalized integrand\/}, such that 
\begin{align*}
\Ar(\G) &= \int \frac{d^D\ q_1}{(2\pi)^D}\cdots \frac{d^D\ q_\ell}{(2\pi)^D}\ 
\Ir(\G;q_1,...,q_\ell)
\end{align*}
is finite. This method was used by Bogoliubov in his first
formulation of the renormalization, and by Zimmermann in the final 
prove of the so-called BPHZ formula (cf.~\ref{BPHZ}).  
Its main advantage is that it is independent of the choice of a 
regularization. For these reasons we adopt it here. 
\end{point}

\begin{point}{Renormalization of a simple loop: Bogoliubov's
subtraction scheme}
Let $\G$ be a 1PI graph with one loop and superficial degree 
of divergency $\w(\G)\geq 0$. We suppose that $\G$ has 
$k$ external legs with external momentum $\p=(p_1,...,p_k)$, then 
the bare amplitude of the graph is
\begin{align*}
A(\G;\p)&= \int \frac{d^D\ q}{(2\pi)^D}\ I(\G;\p;q). 
\end{align*}
Let $T^{\w(\G)}$ denote the operator which computes the Taylor expansion 
in the external momentum variables $\p$ around 
the point $\p=0$, up to the degree $\w(\G)$. 
Then Bogoliubov and Parasiuk proved in \cite{BogoliubovParasiuk,Parasiuk} 
(see also \cite{BogoliubovShirkov}) that the integral 
\begin{align*}
\Ar(\G;\p)&=\int \frac{d^D\ q}{(2\pi)^D} 
\left(I(\G;\p;q)- T^{\w(\G)}[I(\G;\p;q)]\right)
\end{align*}
is finite. Changing the value of $\p=0$ to another value $\p=\p_0$ amounts 
to change $\Ar(\G)$ by a finite value. Eventually, the parameter $\p_0$ 
can then be chosen according to the normalization conditions. 
\end{point}

\begin{point}{Local counterterms}
If we fix some regularization $\rho$, the renormalized (finite)
amplitude can be expressed as a sum 
\begin{align}
\label{Ar-1loop}
\Ar_{\rho}(\G;\p)
&= A_{\rho}(\G;\p)- T^{\w(\G)}\big[A_{\rho}(\G;\p)\big], 
\end{align}
where the removed divergency is contained in a polynomial of the 
external momenta $\p$,  
\begin{align*}
-T^{\w(\G)}\big[A_{\rho}(\G;\p)\big] 
&=-\int \frac{d^Dq}{(2\pi)^D} I_{\rho}(\G)\Big|_{\p=0} 
- \sum_{i,\mu}\ p_i^\mu \int \frac{d^Dq}{(2\pi)^D} 
\frac{\partial I_{\rho}(\G)}{\partial p_i^\mu}\Big|_{\p=0} 
- \frac{1}{2} \underset{\mu,\nu}{\sum_{i,j}}\ p_i^\mu p_j^\nu 
\int \frac{d^Dq}{(2\pi)^D} 
\frac{\partial^2 I_{\rho}(\G)}{\partial p_i^\mu\partial p_j^\nu}\Big|_{\p=0} 
-... 
\end{align*}
In matrix notations, with $\p=(p_1,...,p_k)$, we can write 
\begin{align*}
-T^{\w(\G)}\big[A_{\rho}(\G;\p)\big] 
&=C_0^{\rho}(\Gamma) + C_1^{\rho}(\Gamma)\ \p 
+ \cdots + C_{\w(\Gamma)}^{\rho}(\Gamma)\ \p^{\w(\Gamma)}, 
\end{align*}
where the coefficients
\begin{align}
\label{C-1loop}
C_r^{\rho}(\G) &= - \frac{1}{r!} 
\int \frac{d^Dq}{(2\pi)^D}\ \partial^r_{\p} I_{\rho}(\G)\Big|_{\p=0}
\end{align}
are called the {\em counterterms\/} of the graph $\G$. 
If $\w(\Gamma)=0$, we denote by $C(\Gamma)$ the unique counterterm
in degree $0$. 

The counterterms are usually directly related to the normalization conditions, 
therefore having a finite number of countertems is equivalent to 
the renormalisability of the theory. 

From now on, any time we mention the counterterms we suppose that 
a regularization has been fixed a priori, and we omit the
regularization parameter $\rho$ in the notation. 
\end{point}

\begin{point}{Examples: renormalization of a simple loop}
\label{example-ren-1loop}

\noindent{\bf a)}\ Let us consider the graph 
$\G = 
\begin{picture}(16,6)(-2,0)
  \parbox{0cm}{\begin{center}
  \begin{fmfgraph*}(6,3)
  \setval
  \fmfforce{0w,.5h}{v1}
  \fmfforce{.5w,.5h}{v2}
  \fmfforce{1.5w,.5h}{v3}
  \fmfforce{2w,.5h}{v4}
  \fmf{plain,label=$p$}{v2,v1}
  \fmf{plain,left=1}{v2,v3,v2}
  \fmf{plain,label=$p$}{v4,v3}
  \end{fmfgraph*}
  \end{center}}
\end{picture}
$,  
in dimension $D=4$. Its amplitude (that we suppose regularized) is 
\begin{align*}
A(\G;p) &= \int \frac{d^4 q}{(2\pi)^4}\ 
\frac{1}{q^2+m^2} \frac{1}{(p-q)^2+m^2}.  
\end{align*}
Since $E=2$ and $V=2$ we have $\w(\G)= 0$, therefore the graph $\G$ 
has a logarithmic divergency.
According to the subtraction scheme, its renormalized amplitude is 
$\Ar(\G;p)= A(\G;p)+C(\G)$ where the counterterm is  
\begin{align*}
C(\G) &= -\int \frac{d^4 q}{(2\pi)^4}\ I(\G)\Big|_{p=0}
= -\int \frac{d^4 q}{(2\pi)^4}\ \frac{1}{(q^2+m^2)^2}. 
\end{align*}
The integral $\Ar(\G;p)$ is indeed finite, because 
\begin{align*}
I(\G)-I(\G)\Big|_{p=0} &= \frac{1}{(q^2+m^2)^2} \frac{2pq-p^2}{(p-q)^2+m^2} 
\end{align*}
behaves like $\dfrac{1}{|q|^5}$ for $|q|\rightarrow\infty$, and 
therefore $\ds \Ar(\G;p) = \int \frac{d^4 q}{(2\pi)^4}\ 
\Big(I(\G)-I(\G)\Big|_{p=0}\Big)$
behaves like 
\begin{align*}
\int_{|q|_{min}}^\infty \frac{d^4 |q|}{|q|^5} 
&\simeq \int_{|q|_{min}}^\infty \frac{d |q|}{|q|^{5-3}} 
= \left[-\frac{1}{|q|}\right]_{|q|_{min}}^\infty = \frac{1}{|q|_{min}}. 
\end{align*}
\bigskip 

\noindent{\bf b)}\ Let us consider the same graph 
$\G = 
\begin{picture}(16,4)(-2,0)
  \parbox{0cm}{\begin{center}
  \begin{fmfgraph*}(6,3)
  \setval
  \fmfforce{0w,.5h}{v1}
  \fmfforce{.5w,.5h}{v2}
  \fmfforce{1.5w,.5h}{v3}
  \fmfforce{2w,.5h}{v4}
  \fmf{plain,label=$p$}{v2,v1}
  \fmf{plain,left=1}{v2,v3,v2}
  \fmf{plain,label=$p$}{v4,v3}
  \end{fmfgraph*}
  \end{center}}
\end{picture}
$,  
but in dimension $D=6$. Its amplitude is 
\begin{align*}
A(\G;p) &= \int \frac{d^6 q}{(2\pi)^6}\ 
\frac{1}{q^2+m^2} \frac{1}{(p-q)^2+m^2}.  
\end{align*}
Since $E=2$ and $V=2$ we have $\w(\G)=2$, therefore the graph $\G$ has 
a quadratic divergency.
Then $\Ar(\G;p)= A(\G;p)-T^2\big[A(\G;p)\big]$ with 
\begin{align*}
-T^2\big[A(\G;p)\big] &= C_0(\G) + p\ C_1(\G) + p^2\ C_2(\G),   
\end{align*}
and the local counterterms of $\Gamma$ are 
\begin{align*}
C_0(\G) &= -\int \frac{d^6 q}{(2\pi)^6}\ \frac{1}{(q^2+m^2)^2}, \\  
C_1(\G) &= -\int \frac{d^6 q}{(2\pi)^6}\ \frac{2q}{(q^2+m^2)^3} = 0 
\quad\mbox{(because the integrand is an odd function)}, \\ 
C_2(\G) &= -\int \frac{d^6 q}{(2\pi)^6}\ \frac{3q^2-m^2}{(q^2+m^2)^4}. 
\end{align*}
Since the function 
\begin{align*}
I(\G)-T^2[I(\G)] 
&= \frac{4p^3q^3-3p^4q^2-4m^2p^3q+m^2p^4}{(q^2+m^2)^4 [(p-q)^2+m^2]}. 
\end{align*}
has leading term of order 
$\dfrac{|q|^3}{|q|^{10}}=\dfrac{1}{|q|^7}$ for $|q|\rightarrow\infty$,  
its integral $\Ar(\G;p)$ behaves like 
\begin{align*} 
\int_{|q|_{min}}^\infty \frac{d^6 |q|}{|q|^7}
&\simeq \int_{|q|_{min}}^\infty \frac{d |q|}{|q|^{7-5}} 
= \left[-\frac{1}{|q|}\right]_{|q|_{min}}^\infty = \frac{1}{|q|_{min}}, 
\end{align*}
and therefore it converges. 
\bigskip 

\noindent{\bf Exercice:} Check that the counterterm $C_0(\G)$ alone is not 
sufficient to make the amplitude converging. 
\bigskip

\noindent{\bf c)}\ Let us consider the graph 
$\G = 
\begin{picture}(22,6)(-6,0)
  \parbox{0cm}{\begin{center}
  \begin{fmfgraph*}(6,3)
  \setval
  \fmfforce{0w,.5h}{v1}
  \fmfforce{1w,.5h}{v2}
  \fmfforce{1.7w,1.7h}{v3}
  \fmfforce{1.7w,-.7h}{v4}
  \fmf{plain_arrow}{v1,v2}
  \fmf{plain_arrow}{v2,v3}
  \fmf{plain_arrow}{v2,v4}
  \fmfv{decor.shape=circle,decor.filled=empty,decor.size=.8w}{v2}
  \fmflabel{$p_1$}{v1}
  \fmflabel{$p_2$}{v3}
  \fmfforce{.8w,-.5h}{v5}
  \fmflabel{$p_1-p_2$}{v5}
  \end{fmfgraph*}
  \end{center}}
\end{picture}$ 
in dimension $D=6$. Its amplitude is 
\begin{align*}
& \\ 
A(\G;p_1,p_2) &= \int \frac{d^6 q}{(2\pi)^6}\ 
\frac{1}{q^2+m^2} \frac{1}{(q+p_2)^2+m^2} \frac{1}{(q-p_1)^2+m^2}. 
\end{align*}
Since $E=3$ and $V=3$ we have $\w(\G)=0$, therefore $\G$ has a 
logarithmic divergency.
Then the renormalized amplitude is 
$\Ar(\G;p_1,p_2) = A(\G;p_1,p_2)+C(\G)$, 
where the counterterm is 
\begin{align*} 
C(\G) &= - \int \frac{d^6 q}{(2\pi)^6}\ I(\G;p_1,p_2;q)\Big|_{p_i=0} 
= - \int \frac{d^6 q}{(2\pi)^6}\ \frac{1}{(q^2+m^2)^3}. 
\end{align*} 
In fact, the function 
\begin{align*}
I(\G)-I(\G)\Big|_{p_i=0} &= \frac{1}{q^2+m^2} 
\left(\frac{1}{[(q-p_1)^2+m^2]\ [(q+p_2)^2+m^2]}-\frac{1}{(q^2+m^2)^2}\right)\\
&=\frac{2(p_1-p_2)q^3 - (p_1^2-4p_1p_2+p_2^2)q^2 
- 2[p_1p_2(p_1-p_2)+m^2(p_1+p_2)]q -(p_1^2p_2^2+m^2p_1^2+m^2p_2^2)}
{(q^2+m^2)^3\ [(q-p_1)^2+m^2]\ [(q+p_2)^2+m^2]}
\end{align*}
has leading term $\dfrac{|q|^3}{|q|^{10}}=\dfrac{1}{|q|^7}$, and therefore 
its integral in dimension $6$ converges, as in example b).  
\end{point}

\begin{point}{Divergent subgraphs} 
The subtraction scheme employed for graphs with one loop does not work 
for graphs with many loops, because of the possible presence of 
divergent subgraphs. 

For instance, consider the graph
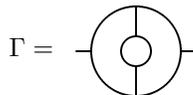
\begin{align*}
\G &= 
\begin{picture}(22,4)(0,0)
  \parbox{0cm}{\begin{center}
  \begin{fmfgraph*}(4,4)
  \setval
  \fmfforce{.5w,.5h}{v1}
  \fmfforce{1w,.5h}{v2}
  \fmfforce{2.5w,2h}{v3}
  \fmfforce{2.5w,1h}{v4}
  \fmfforce{2.5w,0h}{v5}
  \fmfforce{2.5w,-1h}{v6}
  \fmfforce{4w,.5h}{v7}
  \fmfforce{4.5w,.5h}{v8}
  \fmf{plain}{v1,v2}
  \fmf{plain,left=1}{v2,v7,v2}
  \fmf{plain,left=1}{v4,v5,v4}
  \fmf{plain}{v3,v4}
  \fmf{plain}{v5,v6}
  \fmf{plain}{v7,v8}
  \end{fmfgraph*}
  \end{center}}
\end{picture} 
\\ &
\end{align*}
in dimension $D=4$. Since $E=2$ and $V=6$, the graph has negative superficial 
degree of divergency $\w(\G)=-4$. According to the subtraction's scheme, then, 
it should have a zero counterterm $C(\G)$. 
However, the graph $\Gamma$ contains the 1PI subgraph 
$\gamma = 
\begin{picture}(12,4)(0,0)
  \parbox{0cm}{\begin{center}
  \begin{fmfgraph*}(4,2)
  \setval
  \fmfforce{.5w,1h}{v1}
  \fmfforce{1w,1h}{v2}
  \fmfforce{2w,1h}{v3}
  \fmfforce{2.5w,1h}{v4}
  \fmf{plain}{v1,v2}
  \fmf{plain,left=1}{v2,v3,v2}
  \fmf{plain}{v3,v4}
  \end{fmfgraph*}
  \end{center}}
\end{picture}
$ 
which has $\w(\gamma)=0$ in dimension $D=4$ (as we computed in the first 
of Examples~\ref{example-ren-1loop}). Since $\gamma$ diverges, the graph 
$\G$ diverges too, even if $\w(\G)$ is strictly negative.  
\end{point}

\begin{point}{Renormalization of many loops: BPHZ algorithm}
\label{BPHZ}
Let $\G$ be a 1PI graph with many loops and superficial degree 
of divergency $\w(\G)\geq 0$, and/or containing some divergent 
subgraphs. 
Let $A(\G)$ be its amplitude (we omit the external momenta $\p$)
and $I(\G)$ or $I(\G;\mathbf{q})$ its integrand, where 
$\mathbf{q}=(q_1,...,q_\ell)$ are the integrated momenta and $\ell$ 
is the number of loops of $\G$. 

Then, the {\em BPHZ Formula\/} states that the renormalized (i.e. finite) 
amplitude of $\G$ is given by 
\begin{align}
\label{Ar-manyloops}
\Ar(\G) 
&=\int \frac{d^D\ q_1}{(2\pi)^D}\cdots \frac{d^D\ q_\ell}{(2\pi)^D}\ 
\left(\Ip(\G;\mathbf{q})
-T^{\w(\G)}\Big[\Ip(\G;\mathbf{q})\Big]\right), 
\end{align}
where $\Ip(\G)$ denotes a {\em prepared term\/} where all the 
divergent subgraphs have been renormalized. 
The prepared term is defined recursively on the 1PI divergent subgraphs 
of $\G$, by the formula 
\begin{align*}
\Ip(\G;\mathbf{q}) &= I(\G;\mathbf{q}) 
+ \sum_{\gamma_i} \prod_i \left(-T^{\w(\gamma_i)} 
\Big[\Ip(\gamma_i;\mathbf{q}_i)\Big]\right)\ 
\frac{I(\Gamma;\mathbf{q})}
{\prod_i I(\gamma_i;\mathbf{q}_i)}, 
\end{align*}
where the sum is over all 1PI divergent proper subgraphs $\gamma_i$ of $\G$ 
(that is, the subgraphs different from $\G$ itself), such that 
$\gamma_i\cap\gamma_j=\emptyset$ (that is, they are disjoint). 
The proof was first partially given by Bogoliubov and Parasiuk in 1957 
\cite{BogoliubovParasiuk}, then ameliorated by Hepp in 1966 \cite{Hepp} 
and finally established by Zimmermann in 1969 \cite{Zimmermann}, 
who gave a non-recursive formulation in terms of {\em forests\/} of 
divergent subgraphs. 
\medskip 

Formula (\ref{Ar-manyloops}) is usually expressed in a more uniform way.  
Suppose that in the quotient $\frac{I(\Gamma;\mathbf{q})}
{\prod_i I(\gamma_i;\mathbf{q}_i)}$ there remain the first $\ell'$ 
momenta $\mathbf{q}'=(q_1,\ldots ,q_{\ell'})$ appearing explicitely.  
If we set  
\begin{align*}
I(\Gamma/\{\gamma_i\};\mathbf{q}') 
&:= I(\Gamma;\mathbf{q})/\prod_i I(\gamma_i;\mathbf{q}_i), 
\end{align*}
and we integrate over the momenta $\mathbf{q}'$, we define a new graph
$\Gamma/ \{\gamma_i\}$ through its amplitude 
\begin{align*}
A(\Gamma/\{\gamma_i\}) 
&=\int \frac{d^D\ q_1}{(2\pi)^D}\cdots \frac{d^D\ q_{\ell'}}{(2\pi)^D}\ 
I(\Gamma/\{\gamma_i\};\mathbf{q}').   
\end{align*}
This graph can be defined graphically by sqeezing each vertex subgraph 
$\gamma_i$ of $\Gamma$ to the corresponding usual vertex point, and each
propagator subgraph $\gamma_j$ (with 2 external legs) to a new kind of 
vertex point
\begin{align*}
\begin{picture}(12,4)(0,0)
  \parbox{0cm}{\begin{center}
  \begin{fmfgraph*}(6,2)
  \setval
  \fmfforce{.5w,1h}{v1}
  \fmfforce{2w,1h}{v2}
  \fmfforce{3.5w,1h}{v3}
  \fmf{plain}{v3,v1}
  \fmfv{decor.shape=cross,decor.size=.6w}{v2}
  \end{fmfgraph*}
  \end{center}}
\end{picture} 
\end{align*}
which separates two distinguished free propagators (and therefore it
is not considered to be 1PI). 
Then the prepared term can be written 
\begin{align}
\label{Ip-manyloops}
\Ip(\G;\mathbf{q}) &= I(\G;\mathbf{q}) + \sum_{\gamma_i} \prod_i 
\left(-T^{\w(\gamma_i)}\Big[\Ip(\gamma_i;\mathbf{q}_i)\Big]\right)\ 
I(\Gamma/\{\gamma_i\};\mathbf{q}') , 
\end{align}
and the integrand of the renormalized amplitude can be given in a
recursive manner,  
\begin{align}
\label{Ir-manyloops}
\Ir(\G) &= I(\G) 
+\sum_{\gamma_i} 
\left\{ \prod_i \left(-T^{\w(\gamma_i)}\Big[\Ip(\gamma_i)\Big]\right)\
I(\Gamma/\{\gamma_i\}) \right\}
- T^{\w(\Gamma)}\Big[\Ip(\Gamma)\Big].  
\end{align}
\end{point}

\begin{point}{Recursive definition of the counterterms}
The definition of the counterterms given for one-loop graphs by
(\ref{C-1loop}) can then be naturally extended to graphs with many loops 
by applying the Taylor expansion to the prepared integrand  
$\Ip(\G)$ instead of the bare integrand $I(\G)$, that is, by considering
\begin{align*}
- T^{\w(\G)}\left[ 
\int \frac{d^D\ \q}{(2\pi)^D}\ 
\Ip(\G;\mathbf{q}) \right] &= 
C_0(\G) + C_1(\G)\ \p + \cdots + C_{\w(\G)}(\G)\ \p^{\w(\G)},   
\end{align*} 
where we symbolically denote by $\frac{d^D\ \q}{(2\pi)^D}$ the full
expression $\frac{d^D\ q_1}{(2\pi)^D}\cdots \frac{d^D\
q_{\ell}}{(2\pi)^D}$. 

To express the counterterms $C_r(\G)$ in a recursive way, we must 
separate the integrals of each component appearing in the prepared 
term $\Ip(\G)$. 
Consider the complete integral 
\begin{align*}
\int \frac{d^D\ \q}{(2\pi)^D}\ \Ip(\G;\q) 
&= \int \frac{d^D\ \q}{(2\pi)^D}\ I(\G;\mathbf{q}) 
+ \sum_{\gamma_i} \prod_i \int \frac{d^D\ \q'}{(2\pi)^D}\ 
\int \frac{d^D\ \q_i}{(2\pi)^D}\ 
\left(-T^{\w(\gamma_i)}\Big[\Ip(\gamma_i;\mathbf{q}_i)\Big]\right)\ 
I(\G/\{\gamma_i\};\mathbf{q'}). 
\end{align*}
If we denote by $\p_i$ the external momenta of the subgraph
$\gamma_i$, we have 
\begin{align*}
\int \frac{d^D\ \q_i}{(2\pi)^D}\ 
\left(-T^{\w(\gamma_i)}\Big[\Ip(\gamma_i;\mathbf{q}_i)\Big]\right)
&= C_0(\gamma_i) + C_1(\gamma_i)\ \mathbf{p}_i +\cdots + 
C_{\w(\gamma_i)}(\gamma_i)\ \mathbf{p}_i^{\w(\gamma_i)}.  
\end{align*} 
Of course the momenta $\p_i$ are integrated over $\q'$, because they
are internal in $\Gamma$. 
To separate the integrals, it suffices to modify the amplitude of the 
graph $\Gamma/\{\gamma_i\}$ by multiplying it by each remaining momenta
$\p_i^r$. 
In practice, it suffices to label each new crossed vertex obtained by
squeezing $\gamma_i$ by a label $(r)$, with
$r=0,1,...,\w(\gamma_i)$ and to define its amplitude by 
\begin{align}
\label{A_(r)}
A(\Gamma/\{{\gamma_i}_{(r)}\}) 
&=\int \frac{d^D\ \q'}{(2\pi)^D}\ 
\prod_i\ \mathbf{q}_i^r\ I(\Gamma/\{\gamma_i\};\mathbf{q}').   
\end{align}

Finally, if we label each scratched subgraph $\gamma_i$ by the same label $(r)$
used in its associated crossed vertex, and we define its counterterm by 
\begin{align}
\label{counterterm-map}
C({\gamma_i}_{(r)}) &= C_r(\gamma_i), 
\end{align}
we can describe the counterterms in a recursive way as 
\begin{align}
\label{C-manyloops-recursive}
C(\G_{(r)}) &= -\frac{1}{r!}\ \partial_{\p}^r\Big|_{\p=0} 
\left[A(\G;\mathbf{p}) 
+\sum_{\gamma_i} \prod_i \sum_{r_i=0}^{\w(\gamma_i)} 
C({\gamma_i}_{(r_i)})\ A(\Gamma/\{{\gamma_i}_{(r_i)}\};\mathbf{p})\right]. 
\end{align}
The labels $(r)$ are useful only for graphs with positive superficial degree of
divergency. If $\w(\Gamma)=0$, the subfix $(0)$ is systematically
omitted. 

As a consequence, the extention of formula (\ref{Ar-1loop}) to graphs
with many loops is given by 
\begin{align}
\label{Ar-recursive}
\Ar(\G;\mathbf{p})&= A(\G;\mathbf{p}) 
+\sum_{\gamma_i} \prod_i \sum_{r_i=0}^{\w(\gamma_i)} 
C({\gamma_i}_{(r_i)})\ A(\Gamma/\{{\gamma_i}_{(r_i)}\};\mathbf{p}) 
+ C(\G_{(0)}) + \cdots +
\p^{\w(\G)}\ C(\G_{(\w(\G))}). 
\end{align}
\end{point}

\begin{point}{Examples: renormalization of many loops}
\label{example-ren-many-loops}

\noindent{\bf a)}\ Let us consider the graph 
$\G = 
\begin{picture}(34,8)(-6,0)
  \parbox{0cm}{\begin{center}
  \begin{fmfgraph*}(6,3)
  \setval
  \fmfforce{0w,.5h}{v1}
  \fmfforce{.5w,.5h}{v2}
  \fmfforce{1.5w,1h}{v3}
  \fmfforce{1.5w,0h}{v4}
  \fmfforce{2.5w,1.5h}{v5}
  \fmfforce{2.5w,-.5h}{v6}
  \fmfforce{3w,1.75h}{v7}
  \fmfforce{3w,-.75h}{v8}
  \fmf{plain_arrow}{v1,v2}
  \fmf{plain_arrow, label=$q_1$}{v2,v3}
  \fmf{plain_arrow, label=$q_2$}{v3,v5}
  \fmf{plain_arrow}{v5,v7}
  \fmf{plain}{v2,v4,v6}
  \fmf{plain_arrow}{v6,v8}
  \fmf{plain}{v3,v4}
  \fmf{plain}{v5,v6}
  \fmflabel{$p_1$}{v1}
  \fmflabel{$p_2$}{v7}
  \fmflabel{$p_1-p_2$}{v8}
  \end{fmfgraph*}
  \end{center}}
\end{picture} 
$ 
in dimension $D=6$. Its amplitude is 

\begin{align*}
A(\G;p_1,p_2) &= 
\int \frac{d^6 q_1}{(2\pi)^6} \frac{d^6 q_2}{(2\pi)^6}\ 
\frac{1}{q_1^2+m^2}\ \frac{1}{(p_1-q_1)^2+m^2}\ 
\frac{1}{(q_1-q_2)^2+m^2}\ 
\\ & \hspace{5cm} \times 
\frac{1}{q_2^2+m^2}\ \frac{1}{(p_1-q_2)^2+m^2}\ \frac{1}{(q_2-p_2)^2+m^2}. 
\end{align*}
Since $E=3$ and $V=5$ we have $\w(\G)=0$,  
therefore the graph $\G$ has a logarithmic superficial divergency. 
Beside this, the graph $\Gamma$ has two 1PI subgraphs: 
\begin{itemize}
\item
the graph 
$\gamma = 
\begin{picture}(26,6)(-6,0)
  \parbox{0cm}{\begin{center}
  \begin{fmfgraph*}(6,3)
  \setval
  \fmfforce{0w,.5h}{v1}
  \fmfforce{.5w,.5h}{v2}
  \fmfforce{1.5w,1h}{v3}
  \fmfforce{1.5w,0h}{v4}
  \fmfforce{2w,1.25h}{v5}
  \fmfforce{2w,-.25h}{v6}
  \fmf{plain_arrow}{v1,v2}
  \fmf{plain}{v2,v3}
  \fmf{plain_arrow}{v3,v5}
  \fmf{plain}{v2,v4,v6}
  \fmf{plain}{v3,v4}
  \fmflabel{$p_1$}{v1}
  \fmflabel{$q_2$}{v5}
  \end{fmfgraph*}
  \end{center}}
\end{picture}$ 
has a logarithmic divergency; 

\item
the graph 
$\gamma' =  
\begin{picture}(26,6)(-6,0)
  \parbox{0cm}{\begin{center}
  \begin{fmfgraph*}(6,3)
  \setval
  \fmfforce{0w,1.5h}{v1}
  \fmfforce{0w,-.5h}{v2}
  \fmfforce{.5w,1h}{v3}
  \fmfforce{.5w,0h}{v4}
  \fmfforce{1.5w,1h}{v5}
  \fmfforce{1.5w,0h}{v6}
  \fmfforce{2w,1.5h}{v7}
  \fmfforce{2w,-.5h}{v8}
  \fmf{plain_arrow}{v1,v3}
  \fmf{plain}{v2,v4}
  \fmf{plain}{v3,v4,v6,v5,v3}
  \fmf{plain_arrow}{v5,v7}
  \fmf{plain}{v6,v8}
  \fmfforce{0.2w,1h}{v0}
  \fmflabel{$q_1$}{v0}
  \fmflabel{$p_2$}{v7}
  \end{fmfgraph*}
  \end{center}}
\end{picture}$ 
has $E=4$ and $V=4$, therefore $\w(\gamma')=-2$: it converges.  
\bigskip
\end{itemize}
In conclusion, $\Gamma$ has one divergent 1PI subgraph, $\gamma$. 
According to the BPHZ formula~(\ref{Ip-manyloops}), the prepared amplitude 
of $\Gamma$ is 
\begin{align*}
\Ip(\Gamma;p_1,p_2;q_1,q_2) &= 
I(\Gamma) -T^0[I(\gamma)]\ I(\Gamma/\gamma) 
\end{align*} 
where for the graph $\gamma$ we have 
\begin{align*}
-T^0[I(\gamma)] 
&= - I(\gamma;p_1,q_2;q_1)\Big|_{p_1,q_2=0} 
= -\frac{1}{(q_1^2+m^2)^3}, 
\end{align*}
and for the graph 
$\Gamma/\gamma = 
\begin{picture}(26,6)(-6,0)
  \parbox{0cm}{\begin{center}
  \begin{fmfgraph*}(6,3)
  \setval
  \fmfforce{0w,.5h}{v1}
  \fmfforce{.5w,.5h}{v2}
  \fmfforce{1.5w,1h}{v3}
  \fmfforce{1.5w,0h}{v4}
  \fmfforce{2w,1.25h}{v5}
  \fmfforce{2w,-.25h}{v6}
  \fmf{plain_arrow}{v1,v2}
  \fmf{plain_arrow,label=$q_2$}{v2,v3}
  \fmf{plain_arrow}{v3,v5}
  \fmf{plain}{v2,v4,v6}
  \fmf{plain}{v3,v4}
  \fmflabel{$p_1$}{v1}
  \fmflabel{$p_2$}{v5}
  \end{fmfgraph*}
  \end{center}}
\end{picture}$
we have 
\begin{align*}
I(\Gamma/\gamma;p_1,p_2;q_2) 
&= \frac{1}{q_2^2+m^2}\ \frac{1}{(p_1-q_2)^2+m^2}\ \frac{1}{(q_2-p_2)^2+m^2}. 
\end{align*}
Therefore 
\begin{align*}
\Ap(\Gamma;p_1,p_2) 
&= \int \frac{d^6\ q_1}{(2\pi)^6} \frac{d^6\ q_2}{(2\pi)^6}\ 
\left(I(\Gamma)-I(\gamma)\Big|_{p_1,q_2=0}\ I(\Gamma/\gamma)\right) 
\end{align*}
and the overall counterterm $C(\Gamma) = -T^0\Big[\Ap(\Gamma;p_1,p_2)\Big]$ 
of $\Gamma$ is then 
\begin{align*}
C(\Gamma) 
&= -\int \frac{d^6\ q_1}{(2\pi)^6} \frac{d^6\ q_2}{(2\pi)^6}\ 
\left(\frac{1}{(q_1^2+m^2)^2}\ \frac{1}{(q_1-q_2)^2+m^2}\ 
\frac{1}{(q_2^2+m^2)^3}-\frac{1}{(q_1^2+m^2)^3}\ 
\frac{1}{(q_2^2+m^2)^3}\right).
\end{align*}
\bigskip

\noindent{\bf b)}\ Let us consider the graph 
$\G = 
\begin{picture}(24,6)(-4,0)
  \parbox{0cm}{\begin{center}
  \begin{fmfgraph*}(4,4)
  \setval
  \fmfforce{.5w,.5h}{v1}
  \fmfforce{1w,.5h}{v2}
  \fmfforce{2w,1.5h}{v3}
  \fmfforce{2w,-.5h}{v4}
  \fmfforce{3w,.5h}{v5}
  \fmfforce{3.5w,.5h}{v6}
  \fmf{plain_arrow}{v1,v2}
  \fmf{plain,left=1}{v2,v5,v2}
  \fmf{plain_arrow}{v5,v6}
  \fmfv{decor.shape=circle,decor.size=.9w,decor.filled=empty}{v3}
  \fmfv{decor.shape=circle,decor.size=.9w,decor.filled=empty}{v4}
  \fmfforce{.49w,.5h}{v0}
  \fmflabel{$p$}{v0}
  \fmflabel{$p$}{v6}
  \end{fmfgraph*}
  \end{center}}
\end{picture}$ 
in dimension $D=6$. The integrand of its amplitude is 
\begin{align*}
&\\
I(\G;p;q_1,q_2,q_3) &= 
\frac{1}{(q_1^2+m^2)^2}\ \frac{1}{q_2^2+m^2}\ \frac{1}{(q_1-q_2)^2+m^2}\ \\ 
& \hspace{5cm} \times 
\frac{1}{\big((p-q_1)^2+m^2\big)^2}\ \frac{1}{q_3^2+m^2}\ 
\frac{1}{(p-q_1-q_3)^2+m^2}. 
\end{align*}
Since $E=2$ and $V=6$ we have $\w(\G)=2$, 
therefore the graph $\G$ has a quadratic superficial divergency. 
Moreover, the graph $\Gamma$ has two 1PI subgraphs, 
$\gamma_1 = \gamma_2 =  
\begin{picture}(12,4)(0,0)
  \parbox{0cm}{\begin{center}
  \begin{fmfgraph*}(4,2)
  \setval
  \fmfforce{.5w,1h}{v1}
  \fmfforce{1w,1h}{v2}
  \fmfforce{2w,1h}{v3}
  \fmfforce{2.5w,1h}{v4}
  \fmf{plain}{v1,v2}
  \fmf{plain,left=1}{v2,v3,v2}
  \fmf{plain}{v3,v4}
  \end{fmfgraph*}
  \end{center}}
\end{picture}$, 
which have a quadratic divergency. 
Let us compute the counterterms of $\Gamma$ using the BPHZ formula. 
We have
\begin{align*}
\Ip(\G) &= I(\G) - T^2\Big[ I(\gamma_1)\ I(\G/\gamma_1)
+ I(\gamma_2)\ I(\G/\gamma_2) 
+ I(\gamma_1)\ I(\gamma_2)\ I(\G/\gamma_1\gamma_2) \Big], 
\end{align*} 
where 
\begin{align*}
\G/\gamma_1=
\begin{picture}(16,6)(0,0)
  \parbox{0cm}{\begin{center}
  \begin{fmfgraph*}(4,4)
  \setval
  \fmfforce{.5w,.5h}{v1}
  \fmfforce{1w,.5h}{v2}
  \fmfforce{2w,1.5h}{v3}
  \fmfforce{2w,-.5h}{v4}
  \fmfforce{3w,.5h}{v5}
  \fmfforce{3.5w,.5h}{v6}
  \fmf{plain}{v1,v2}
  \fmf{plain,left=1}{v2,v5,v2}
  \fmf{plain}{v5,v6}
  \fmfv{decor.shape=circle,decor.size=.9w,decor.filled=empty}{v3}
  \fmfv{decor.shape=cross,decor.size=.4w}{v4}
  \end{fmfgraph*}
  \end{center}}
\end{picture}
, \qquad
\G/\gamma_2=
\begin{picture}(16,6)(0,0)
  \parbox{0cm}{\begin{center}
  \begin{fmfgraph*}(4,4)
  \setval
  \fmfforce{.5w,.5h}{v1}
  \fmfforce{1w,.5h}{v2}
  \fmfforce{2w,1.5h}{v3}
  \fmfforce{2w,-.5h}{v4}
  \fmfforce{3w,.5h}{v5}
  \fmfforce{3.5w,.5h}{v6}
  \fmf{plain}{v1,v2}
  \fmf{plain,left=1}{v2,v5,v2}
  \fmf{plain}{v5,v6}
  \fmfv{decor.shape=circle,decor.size=.9w,decor.filled=empty}{v4}
  \fmfv{decor.shape=cross,decor.size=.4w}{v3}
  \end{fmfgraph*}
  \end{center}}
\end{picture}
, 
\qquad\mbox{and}\qquad 
\G/\gamma_1\gamma_2= 
\begin{picture}(16,6)(0,0)
  \parbox{0cm}{\begin{center}
  \begin{fmfgraph*}(4,4)
  \setval
  \fmfforce{.5w,.5h}{v1}
  \fmfforce{1w,.5h}{v2}
  \fmfforce{2w,1.5h}{v3}
  \fmfforce{2w,-.5h}{v4}
  \fmfforce{3w,.5h}{v5}
  \fmfforce{3.5w,.5h}{v6}
  \fmf{plain}{v1,v2}
  \fmf{plain,left=1}{v2,v5,v2}
  \fmf{plain}{v5,v6}
  \fmfv{decor.shape=cross,decor.size=.4w}{v3}
  \fmfv{decor.shape=cross,decor.size=.4w}{v4}
  \end{fmfgraph*}
  \end{center}}
\end{picture}. 
\\ 
\end{align*} 
Since the graphs $\gamma_1$ and $\gamma_2$ give the same contribution 
when integrated, we can call them both $\gamma$ and sum them up. 
The counterterms 
$C(\Gamma_{(r)}) = -\frac{1}{r!} \partial_{\p}^r\Big|_{\p=0}\ 
\big[\Ap(\Gamma)\big]$, for $r=0,2$, are then given explicitely as follows: 
\begin{align*}
C(\Gamma_{(r)}) 
&= -\frac{1}{r!} \partial_{\p}^r\Big|_{\p=0}\ 
\Big[A(\Gamma) + 2 C(\gamma_{(0)})\ A(\Gamma/\gamma_{(0)}) 
+ 2 C(\gamma_{(2)})\ A(\Gamma/\gamma_{(2)}) \\ 
& \hspace{1cm}
+ C(\gamma_{(0)})^2 \ A(\Gamma/(\gamma_{(0)})^2) 
+ 2 C(\gamma_{(0)})\ C(\gamma_{(2)})\ A(\Gamma/\gamma_{(0)}\gamma_{(2)})
+ C(\gamma_{(2)})^2\ A(\Gamma/(\gamma_{(2)})^2) \Big]. 
\end{align*}
\end{point} 


\subsection{Dyson's renormalization formulas for Green's functions}
\label{Renormalization-Dyson}

As we fixed in Section~\ref{ReviewQFT}, the aim of quantum field theory 
is to compute the full Green's functions 
$\la\phi(x_1)\cdots\phi(x_k)\ra$. To do this, we need to compute the 
connected Green's functions $G(x_1,...,x_k)$, which can only be found 
perturbatively, as formal series in the powers of the coupling constant 
$\lambda$. 
In Lecture~III we showed that the coefficients of these 
series can be labelled by suitable Feynman graphs. 
Therefore the connected Green's functions can be written as 
\begin{align*}
G(x_1,...,x_k) &= \sum_{n=0}^\infty \lambda^n\ 
\sum_{V(\G)=n} \frac{\hbar^{L(\G)}}{\Sym(\G)}\ A(\G;x_1,...,x_k), 
\end{align*}
where the sum is over all the connected Feynman graphs with $k$ external legs. 
In Section~\ref{Renormalization-graphs}, then, we pointed out the problem 
of divergencies, which affects some graphs with loops, and showed how 
to extract a finite contribution for each graph, the renormalized amplitude. 
Summing up all the renormalized amplitudes, we obtain the renormalized 
connected Green's functions 
\begin{align*}
\Gr(x_1,...,x_k) &= \sum_{n=0}^\infty \lambda^n\ 
\sum_{V(\G)=n} \frac{\hbar^{L(\G)}}{\Sym(\G)}\ \Ar(\G;x_1,...,x_k), 
\end{align*}
and finally the searched renormalized full Green's functions 
$\la\phi(x_1)\cdots\phi(x_k)\ra^{ren}$. 

In this section, we discuss the direct way from the bare Green's 
functions $G(x_1,...,x_k)$ to the renormalized ones, $\Gr(x_1,...,x_k)$, 
without making use of Feynman graphs.  

\begin{point}{Bare and renormalized Lagrangian}
From the BPHZ formula~(\ref{Ar-recursive}), it is clear that the passage from 
the bare to the renormalized amplitudes amounts to adding many terms 
which contain the counterterms of the divergent subgraphs, 
\begin{align*}
\Ar(\G) &= A(\G) + \mbox{terms}. 
\end{align*}
Inserting these terms in the connected Green's functions, then, 
amounts to adding a series in $\lambda$, 
\begin{align*}
\Gr(x_1,...,x_k;\lambda) &= G(x_1,...,x_k;\lambda) + \mbox{series}(\lambda). 
\end{align*}
Since the connected Green's functions are completely determined from 
the Lagrangian $\L(\phi)$, as we saw in Section~\ref{ReviewQFT}, the
new terms added to $G(x_1,...,x_k;\lambda)$ must correspond to 
new terms added to $\L(\phi)$, 
\begin{align*}
\Lr(\phi,\lambda) &= \L(\phi,\lambda) + \D\L(\phi,\lambda). 
\end{align*}
This Lagrangian is called {\em renormalized\/}, in contraposition with 
the original Lagrangian $\L(\phi,\lambda)$ called {\em bare\/}. 

Let us stress that, beside its name, the renormalized Lagrangian has no 
particular physical meaning: it is only a formal Lagrangian which gives 
rise to the renormalized (hence physically meaningful) Green's functions, 
through the standard procedure described in Section~\ref{ReviewQFT}. 

The number of terms appearing in $\D\L(\phi)$ tells us if the theory is 
renormalizable or not: the theory is not renormalizable if the number of 
terms to be added is infinite. 
\end{point}

\begin{point}{Renormalization factors}
If the theory is renormalizable, then $\D\L(\phi)$ contains exactly 
one term proportional to each term of $\L(\phi)$. 
The factors appearing in each term of the renormalized Lagrangian are 
called {\em renormalization factors\/}. 

To be precise, let us consider again the interacting Klein-Gordon Lagrangian 
\begin{align*}
\L(\phi,m,\lambda) &= 
\frac{1}{2} |\partial_\mu \phi(x)|^2 + \frac{m^2}{2} \phi(x)^2 
- \frac{\lambda}{3!} \phi(x)^3,  
\end{align*}
as a function of the field $\phi$ and of the physical parameters $m$, 
the mass, and $\lambda$, the coupling constant. 
Then the terms added by the renormalization can be organized as 
\begin{align*}
\D\L(\phi,m,\lambda) &= 
\frac{1}{2} \D_k(\lambda) |\partial_\mu \phi(x)|^2 
+ \frac{m^2}{2} \D_m(\lambda) \phi(x)^2 
- \frac{\lambda}{3!} \D_{\lambda}(\lambda) \phi(x)^3,  
\end{align*}
where $\D_k$, $\D_m$ and $\D_\lambda$ are series in $\lambda$ containing 
the counterterms of all Feynman graphs. 
Hence the renormalized Lagrangian is of the form 
\begin{align*}
\Lr = \L + \D\L &= 
\frac{1}{2} |\partial_\mu \phi(x)|^2 + \frac{m^2}{2} \phi(x)^2 
- \frac{\lambda}{3!}\phi(x)^3 
+ \D_k(\lambda) \frac{1}{2} |\partial_\mu \phi(x)|^2 
+ \D_m(\lambda) \frac{m^2}{2} \phi(x)^2 
- \D_{\lambda}(\lambda) \frac{\lambda}{3!}\phi(x)^3 \\ 
&= \frac{1}{2}\ Z_3(\lambda)\ |\partial_\mu \phi(x)|^2 
+ \frac{m^2}{2}\ Z_m(\lambda)\ \phi(x)^2 
- \frac{\lambda}{3!}\ Z_1(\lambda)\ \phi(x)^3 , 
\end{align*}
where $Z_3(\lambda)= 1+\D_k(\lambda)$, $Z_m(\lambda) = 1+\D_m(\lambda)$ 
and $Z_1(\lambda) = 1+\D_{\lambda}(\lambda)$ are the renormalization factors. 

The renormalization factors are completely determined by the counterterms 
of the divergent graphs. 
For the $\phi^3$ theory in dimension $D=6$, for instance, a graph $\Gamma$ 
with $E=2$ is quadratically divergent (as we saw in 
Example~\ref{example-ren-1loop} b) and its counterterms are of the form 
$C_0(\Gamma)+p^2\ C_2(\Gamma)$.
According to our previous notations, and up to the scalar factor $m^2$, 
this can also be written as $m^2\ C(\Gamma_{(0)})+p^2\ C(\Gamma_{(2)})$.
Instead, a graph $\Gamma$ with $E=3$ is logarithmically divergent (as we saw 
in Example~\ref{example-ren-1loop} c) and has a single counterterm 
$C(\Gamma)$. 
It turns out that in this case the renormalization factors are organized 
as follows:
\begin{align}
\label{Z-C}
\nonumber
Z_3(\lambda) &= 1 -  
\sum_{E(\Gamma)=2} \frac{C(\Gamma_{(2)})}{\Sym(\Gamma)}\ \lambda^{V(\Gamma)}, \\ 
Z_m(\lambda) &= 1 -  
\sum_{E(\Gamma)=2} \frac{C(\Gamma_{(0)})}{\Sym(\Gamma)}\ \lambda^{V(\Gamma)}, \\ 
\nonumber
\lambda\ Z_1(\lambda) &= \lambda + 
\sum_{E(\Gamma)=3} \frac{C(\Gamma)}{\Sym(\Gamma)}\ \lambda^{V(\Gamma)}, 
\end{align}
\end{point}

\begin{point}{Bare and effective parameters}
If we call $\phi_b = Z_3(\lambda)^{\frac{1}{2}} \phi$, then we have 
\begin{align*}
\Lr(\phi,m,\lambda) &= 
\frac{1}{2}\ |\partial_\mu \phi_b(x)|^2 
+ \frac{m^2}{2}\ Z_m(\lambda)\ Z_3(\lambda)^{-1} \phi_b(x)^2 
- \frac{\lambda}{3!}\ Z_1(\lambda)\ Z(\lambda)^{-\frac{3}{2}} \phi_b(x)^3 , 
\end{align*}
and if we set also 
\begin{align}
\label{bare-mass}
m_b &= m\ Z_m(\lambda)^{\frac{1}{2}} Z_3(\lambda)^{-\frac{1}{2}}, \\  
\label{bare-lambda}
\lambda_b &= \lambda\ Z_1(\lambda)\ Z_3(\lambda)^{-\frac{3}{2}}, 
\end{align}
we finally obtain 
\begin{align}
\label{Lren}
\Lr(\phi,m,\lambda) &= 
\frac{1}{2}\ |\partial_\mu \phi_b(x)|^2 + \frac{1}{2}\ m_b^2\ \phi_b(x)^2 
- \frac{\lambda_b}{3!}\ \phi_b(x)^3 
= \L(\phi_b,m_b,\lambda_b). 
\end{align}
In other words, the ``formal'' Lagrangian in $\phi,m,\lambda$ which 
produces the ``real'' (renormalized) Green's functions, 
is exactely the original Lagrangian, but on ``unreal'' values 
of the field, $\phi_b$, of the mass, $m_b$, and of the coupling constant, 
$\lambda_b$. By definition, the parameters $\phi_b, m_b, \lambda_b$ 
are formal series in $\lambda$ with coefficients given by the counterterms 
of the graphs. The are called {\em bare\/}, in contraposition with the 
physical ones, $\phi, m, \lambda$, which are called {\em effective\/} 
because they are the measured ones. 
\end{point}

\begin{point}{Dyson's formulas}
According to Eq.~(\ref{Lren}), the renormalized Lagrangian in the effective 
parametrs, $\Lr(\phi,m,\lambda)$, is equal to the bare Lagrangian in the 
bare parameters, $\L(\phi_b,m_b,\lambda_b)$. 
Therefore, the renormalized Green's functions in the effective parameters, 
$\Gr(x_1,...,x_k;m,\lambda)$, must be related to the bare Green's functions 
in the bare parameters, $G(x_1,...,x_k;m_b,\lambda_b)$. 
This relation is given by the following formula 
\begin{align}
\label{Dyson-formula}
\Gr(p_1,...,p_k;m,\lambda) &= 
Z_3^{-\frac{k}{2}}(\lambda)\ G(p_1,...,p_k;m_b,\lambda_b), 
\end{align}
where $m_b=m_b(m,\lambda)$ and $\lambda_b=\lambda_b(\lambda)$ are the 
formal series in the powers of $\lambda$ given by 
Eqs.~(\ref{bare-mass}) and (\ref{bare-lambda}). 

In this lecture, this equality is called {\em Dyson's formula\/}, because 
it was firstly introduced by F.~Dyson for quantum electrodynamics in 1949, 
cf.~\cite{Dyson}. 
\end{point}

\begin{point}{Renormalization and semidirect product of series}
Dyson's formula~(\ref{Dyson-formula}), together with the 
formulae~(\ref{bare-mass}) and (\ref{bare-lambda}), answers to the question 
that we posed at the beginning of this section. 
Combining all of them, in fact, we get the explicit expression of the 
renormalized Green's functions from the bare ones, by means of a 
product and a substitution by suitable formal series in $\lambda$. 
The transformation from bare to renormalized Green's functions 
is a {\em semidirect product\/} law. 

To show this, let us rewrite Dyson's formula by pointing out only the 
dependence of the formal series on the parameters $m$ and $\lambda$: 
\begin{align}
\label{Dyson-semidirect-product}
\Gr(m,\lambda) &= 
Z_3^{-\frac{k}{2}}(\lambda)\ G(m_b(m,\lambda),\lambda_b(\lambda)). 
\end{align}
In this formula, the quantities 
\begin{align*} 
\Gr(m,\lambda) &= G_0 + \O(\lambda), \\ 
G(m_b,\lambda_b) &= G_0 + \O(\lambda_b), \\ 
Z_3^{-\frac{k}{2}}(\lambda) &= (1+\O(\lambda))^{-\frac{k}{2}} 
= 1+\O(\lambda)
\end{align*}
are invertible series in $\lambda$ (with respect to the multiplication, 
cf.~Example~\ref{invertible-series}), and the two bare parameters 
\begin{align*}
m_b &= m + \O(\lambda), \\  
\lambda_b &= \lambda + \O(\lambda^2) 
\end{align*}
are formal diffeomorphisms in $\lambda$ (with respect to the substitution 
or composition, cf.~Example~\ref{diffeomorphisms}). 
Therefore Eq.~(\ref{Dyson-semidirect-product}) tells us 
that the renormalized Green's function can be found as a semidirect 
product of suitable series in $\lambda$. 

The relationship between the renormalization of the Green's functions 
and the renormalization of each single graph appearing in the perturbative 
expansions is the main topic of these lectures. 
It is described in details in the next section. 
\end{point}


\section*{Lecture V - Hopf algebra of Feynman graphs 
and combinatorial groups of renormalization}
\addcontentsline{toc}{section}{\bf Lecture V - Hopf algebra of Feynman graphs 
and combinatorial groups of renormalization}
\label{lecture5}

In Lecture~I we described the Hopf algebra canonically associated 
to an algebraic or to a proalgebraic group, and gave some examples, 
for the most common groups.
In this lecture, we start from a Hopf algebra on graphs related to the 
renormalization, and discuss what is the physical meaning of its associated 
proalgebraic group. 

\subsection{Connes-Kreimer Hopf algebra of Feynaman graphs 
and diffeographisms}

In the context of renormalization, a Hopf algebra is suitable to describe 
the combinatorics of the BPHZ formula, and can be given for any quantum 
field theory which is renormalizable by loal counterterms. 
Its aim is precisely to describe the recursive definition of the 
counterterms. 

Following the works \cite{ConnesKreimerI,ConnesKreimerII} of A.~Connes 
and D.~Kreimer, we choose as a toy model the $\phi^3$ theory in 
dimension $D=6$, for a scalar field $\phi$. 
In this theory the superficial divergent graphs are those with a 
number of exterior legs $E\leq 3$. 
Among these, the tadpole graphs, which have $E=1$, are not considered 
because we assume that the 1-point Green function $\la\phi(x)\ra$ 
vanishes. 

\begin{point}{Graded algebra of Feynaman graphs}
Let $\HCK$ be the polynomial algebra over $\C$ generated by the Feynman 
graphs which describe the local counterterms of the $\phi^3$ theory. 
These are the 1PI graphs with 2 or 3 external legs, constructed on three 
types of vertices: 
\begin{align*}
\begin{picture}(12,4)(-2,0)
  \parbox{0cm}{\begin{center}
  \begin{fmfgraph*}(4,2)
  \setval
  \fmfforce{0w,1h}{v1}
  \fmfforce{1w,1h}{v2}
  \fmfforce{2w,2.5h}{v3}
  \fmfforce{2w,-.5h}{v4}
  \fmf{plain}{v1,v2}
  \fmf{plain}{v2,v3}
  \fmf{plain}{v2,v4}
  \end{fmfgraph*}
  \end{center}}
\end{picture}
\quad , \quad 
\begin{picture}(20,4)(0,0)
  \parbox{0cm}{\begin{center}
  \begin{fmfgraph*}(6,2)
  \setval
  \fmfforce{.5w,1h}{v1}
  \fmfforce{2w,1h}{v2}
  \fmfforce{3.5w,1h}{v3}
  \fmf{plain}{v3,v1}
  \fmfv{decor.shape=cross,decor.size=.6w}{v2}
  \fmfforce{1.3w,1.8h}{v4}
  \fmflabel{$(0)$}{v4}
  \end{fmfgraph*}
  \end{center}}
\end{picture} 
\quad , \quad 
\begin{picture}(20,4)(0,0)
  \parbox{0cm}{\begin{center}
  \begin{fmfgraph*}(6,2)
  \setval
  \fmfforce{.5w,1h}{v1}
  \fmfforce{2w,1h}{v2}
  \fmfforce{3.5w,1h}{v3}
  \fmf{plain}{v3,v1}
  \fmfv{decor.shape=cross,decor.size=.6w}{v2}
  \fmfforce{1.3w,1.8h}{v4}
  \fmflabel{$(2)$}{v4}
  \end{fmfgraph*}
  \end{center}}
\end{picture} \quad . 
\end{align*}
The free commutative multiplication between graphs is denoted by the 
concatenation, and the formal unit is denoted by $1$.

On the algebra $\HCK$ we consider the grading induced by the number $L$ 
of loops of the Feynman graphs: the degree of a monomial 
$\Gamma_1\cdots\Gamma_s$ in $\HCK$ is given by 
$L(\Gamma_1)+\cdots+L(\Gamma_s)$. 
Then in degree $0$ we have only the scalars (multiples of the unit $1$), 
and therefore $\HCK$ is a connected graded algebra. 
In degree $1$ we have only linear combinations of the 1-loop graphs 
$\begin{picture}(12,4)(0,-1)
  \parbox{0cm}{\begin{center}
  \begin{fmfgraph*}(4,2)
  \setval
  \fmfforce{.5w,1h}{v1}
  \fmfforce{1w,1h}{v2}
  \fmfforce{2w,1h}{v3}
  \fmfforce{2.5w,1h}{v4}
  \fmf{plain}{v1,v2}
  \fmf{plain,left=1}{v2,v3,v2}
  \fmf{plain}{v3,v4}
  \end{fmfgraph*}
  \end{center}}
\end{picture}
$ 
and 
$\begin{picture}(16,4)(-2,-1)
  \parbox{0cm}{\begin{center}
  \begin{fmfgraph*}(6,3)
  \setval
  \fmfforce{0w,.5h}{v1}
  \fmfforce{.5w,.5h}{v2}
  \fmfforce{1.5w,1h}{v3}
  \fmfforce{1.5w,0h}{v4}
  \fmfforce{2w,1.25h}{v5}
  \fmfforce{2w,-.25h}{v6}
  \fmf{plain}{v1,v2,v3,v5}
  \fmf{plain}{v2,v4,v6}
  \fmf{plain}{v3,v4}
  \end{fmfgraph*}
  \end{center}}
\end{picture} 
$, 
eventually containing some crossed vertices. 
In degree $2$ we have linear combinations of products of two 1-loop graphs 
and graphs with 2 loops, and so on for all higher degrees. 

The number of non-crossed vertices $V$ of Feynman graphs can be used as an 
alternative grading of $\HCK$. Note, however, that it is not equivalent 
to the grading by $L$. In fact, according to paragraph~\ref{power counting}, 
if $E$ is the number of external legs of a $\phi^3$-graph in $D=6$, 
then the number of its vertices is $V=2L+E-2$. 
Then, at a given degree $L$ by loops, the degree by vertices is 
$V=2L$ for graphs with 2 external legs, and $V=2L+1$ for graphs with 
3 external legs. Therefore the grading induced by $V$ is finer then that 
induced by $L$. 
\end{point}

\begin{point}{Hopf algebra of Feynaman graphs}
On the graded algebra $\HCK$ we consider the coproduct 
$\D:\HCK\longrightarrow\HCK\otimes\HCK$ defined as the multiplicative 
and unital map given on a generator $\Gamma$ by 
\begin{align}
\label{HCK-coproduct}
\D (\Gamma) &= \Gamma\otimes 1 + 1\otimes\Gamma 
+ \sum_{\gamma_i,r_i} 
\Gamma/\{{\gamma_i}_{(r_i)}\} \otimes \prod_i {\gamma_i}_{(r_i)} 
\end{align}
where the sum is over any possible choice of 1PI proper and disjoint 
divergent subgraphs $\gamma_i$ of $\Gamma$, and $r_i=0,...,\w(\gamma_i)$.  
The notations used here were fixed in Section~\ref{Renormalization-graphs}: 
\begin{itemize}
\item
The term $\Gamma/\{{\gamma_i}_{(r_i)}\}$ is the graph obtained 
from $\Gamma$ by replacing each subgraph $\gamma_i$ having 2 
external legs with a labeled crossed vertex 
$\begin{picture}(22,8)(0,0)
  \parbox{0cm}{\begin{center}
  \begin{fmfgraph*}(6,2)
  \setval
  \fmfforce{.5w,1h}{v1}
  \fmfforce{2w,1h}{v2}
  \fmfforce{3.5w,1h}{v3}
  \fmf{plain}{v3,v1}
  \fmfv{decor.shape=cross,decor.size=.6w}{v2}
  \fmfforce{1.3w,1.8h}{v4}
  \fmflabel{$(r_i)$}{v4}
  \end{fmfgraph*}
  \end{center}}
\end{picture}$, 
and each subgraph $\gamma_j$ having 3 external legs with a vertex graphs
$\begin{picture}(12,4)(-2,0)
  \parbox{0cm}{\begin{center}
  \begin{fmfgraph*}(4,2)
  \setval
  \fmfforce{0w,1h}{v1}
  \fmfforce{1w,1h}{v2}
  \fmfforce{2w,2.5h}{v3}
  \fmfforce{2w,-.5h}{v4}
  \fmf{plain}{v1,v2}
  \fmf{plain}{v2,v3}
  \fmf{plain}{v2,v4}
  \end{fmfgraph*}
  \end{center}}
\end{picture}$. 
\item 
Each graph ${\gamma_i}_{(r_i)}$ means in fact the graph ${\gamma_i}$ 
with a prescribed counterterm map given as the partial derivative of 
order $r_i$ (evaluated at zero external momenta). 
\item 
The term $\prod_i\ {\gamma_i}_{(r_i)}$ is a monomial in $\HCK$, 
that is a free product of graphs. 
\end{itemize}
On $\HCK$ we also consider the counit $\varepsilon:\HCK\longrightarrow\C$ 
defined as the multiplicative and unital map which annihilates the generators, 
that is such that $\varepsilon(1)=1$ and $\varepsilon(\Gamma) = 0$. 

The coproduct and the counit so defined are graded algebra maps. Since 
the algebra $\HCK$ is connected, we can use the 5-terms equality of 
paragraph~\ref{antipode-def} to define recursively the antipode 
$S:\HCK\longrightarrow\HCK$. Explicitely, it is the multiplicative 
and unital map defined on the generators as 
\begin{align*}
S(\Gamma) &= -\Gamma - \sum_{\gamma_i,r_i} 
\Gamma/\{{\gamma_i}_{(r_i)}\} \ \prod_i S({\gamma_i}_{(r_i)}) . 
\end{align*}

In \cite{ConnesKreimerI}, A.~Connes and D.~Kreimer showed that 
$\HCK$ is a commutative and connected graded Hopf algebra, that is, 
the coproduct, the counit and the antipode satisfy all the compatibility 
properties listed in Section~\ref{Hopf-def}. 
\end{point}

\begin{point}{\bf Group of diffeographisms and renormalization} 
The Hopf algebra $\HCK$ is commutative but of course it is not finitely 
generated. Then, according to the paragraph~\ref{proalgebraic group}, 
$\HCK$ defines a pro-algebraic group: for any associative and 
commutative algebra $A$, the set $\GCK(A)$ of $A$-valued characters on 
$\HCK$ is a group with the convolution product 
$\alpha\star\beta = m_{A}\circ(\alpha\otimes\beta)\circ\D$. 

Connes and Kreimer showed in \cite{ConnesKreimerI} that if $\Arho$ is 
the algebra of regularized amplitudes for the $\phi^3$ theory 
in dimension $D=6$, then the BPHZ renormalization recursion takes place 
in the so-called {\em diffeographisms group\/}  
\begin{align}
\label{GCK}
\GCK(\Arho) & = \Hom_{Alg}(\HCK,\Arho).  
\end{align}
More precisely, this means that the bare amplitude map $A$, the regularized 
amplitude map $\Ar$ and the counterterm map $C$ are characters 
$\HCK \longrightarrow \Arho$, and moreover that the BPHZ renormalization 
Formula~(\ref{Ar-recursive}) is equivalent to  
\begin{align}
\label{Ar=A*C}
\Ar &= A\star C.
\end{align}

In fact, for a given 1PI $\phi^3$-graph $\Gamma$, Eq.~(\ref{Ar=A*C}) 
means that 
\begin{align*}
(A\star C)(\Gamma) &=  A(\Gamma) C(1) + A(1) C(\Gamma) 
+ \sum_{\gamma_i,r_i} 
A(\Gamma/\{{\gamma_i}_{(r_i)}\}) \ \prod_i C({\gamma_i}_{(r_i)})
= \Ar(\Gamma),  
\end{align*}
then, comparing the BPHZ Formula~(\ref{Ar-recursive}) with the expression 
(\ref{HCK-coproduct}) of the coproduct in $\HCK$, we see that 
Eq.~(\ref{Ar=A*C}) is trivially verifyed provided that the counterterm 
map $C$ is indeed an algebra homomorphism, and therefore 
$C(\gamma_1\cdots\gamma_s)= C(\gamma_1)\cdots C(\gamma_s)$. 
This fact is due to a peculiar property of the truncated Taylor operator 
$T^{\w(\G)}$ which appears in the counterterm of any graph $\Gamma$. 
Namely, if we denote by $T$ the truncated Taylor expansion, then 
for any functions $f$ and $g$ of the external momenta we have 
\begin{align*}
T[f g] + T[f] T[g] &= T\big[T[f]g + fT[g]\big]. 
\end{align*} 
An operator having this property is called a {\em Rota-Baxter operator\/}. 
The relationship between Rota-Baxter operators and renormalization 
has been largely investigated by K.~Ebrahimi-Fard and L.~Guo, see for 
instance \cite{Ebrahimi-FardGuo}. 
\end{point}

\begin{point}{\bf Diffeographisms and diffeomorphisms} 
In \cite{ConnesKreimerII}, Connes and Kreimer showed that the 
renormalization of the coupling constant, that is the formula 
(\ref{bare-lambda}) 
\begin{align*}
\lambda_b (\lambda) &= \lambda\ Z_1(\lambda)\ Z_3(\lambda)^{-\frac{3}{2}}, 
\end{align*}
defines an inclusion of the coordinate ring of the group of formal 
diffeomorphisms into the Hopf algebra $\HCK$. 

Let us denote by $\Hd$ the complex coordinate ring of the proalgebraic 
group $\Gd$ of formal diffeomorphisms in one variable, 
as illustrated in paragraph \ref{diffeomorphisms}. 
Recall that $\Hd=\C[x_1,x_2,...]$ is an infinitely generated commutative Hopf 
algebra with coproduct 
\begin{align*}
\D x_n &= x_n\otimes 1+1\otimes x_n
+\sum_{m=1}^{n-1} x_m\otimes 
\underset{p_0,...,p_m\geq 0}{\sum_{p_0+p_1+\cdots +p_m=n-m}} 
x_{p_0}x_{p_1}\cdots x_{p_m}
\end{align*}
and counit $\varepsilon(x_n)=0$. 
Then, the inclusion $\Hd \hookrightarrow\HCK$ is defined as follows: 
consider the expansion (\ref{Z-C}) of the renormalization factors 
in terms of the counterterms of the divergent graphs, namely 
\begin{align*}
Z_1(\lambda) &= 1 + 
\sum_{E(\Gamma)=3} \frac{C(\Gamma_{(0)})}{\Sym(\Gamma)}\ \lambda^{V(\Gamma)}, \\ 
Z_3(\lambda) &= 1 -  
\sum_{E(\Gamma)=2} \frac{C(\Gamma_{(2)})}{\Sym(\Gamma)}\ \lambda^{V(\Gamma)}, 
\end{align*}
and assign to a generator $x_n$ of $\Hd$ the combination of Feynman 
graphs appearing in the coefficient of $\lambda^{n+1}$ in the series 
$\lambda_b = \lambda\ Z_1(\lambda)\ Z_3(\lambda)^{-\frac{3}{2}}$. 
In \cite{ConnesKreimerII}, Connes and Kreimer proved that this map 
preserves the coproduct, and therefore it is a morphism of Hopf algebras. 
\end{point}

\begin{point}{\bf Diffeographisms as generalized series}
Connes and Kreimer's result summerized above means in particular that 
the group of diffeographisms $\GCK(\Arho)$ is projected onto the group 
of formal diffeomorphisms $\Gd(\Arho)$ in one variable, with coefficients 
in the algebra of regularized amplitudes. 
In this context, formal diffeomorphisms are formal series in the powers 
of the coupling constant $\lambda$, that is, series of the form 
\begin{align*}
f(\lambda) &= \sum_{n=0}^\infty f_n\ \lambda^{n+1}, 
\end{align*}
endowed with the composition law. 

A useful way to understand the map $\GCK(\Arho)\longrightarrow\Gd(\Arho)$ 
is to represent the diffeographisms as a generalization of usual series 
of the form 
\begin{align}
\label{diffeographism}
f(\lambda) &= \sum_{\G} f_{\G}\ \lambda^{\G}, 
\end{align}
where the sum is over suitable Feynman diagrams $\G$, the coefficients 
$f_{\G}$ are taken in the algebra $\Arho$, and the powers $\lambda^{\G}$ 
are not monomials in a possibly complex variable $\lambda$, but just 
formal symbols. 
The projection $\pi:\GCK(\Arho)\longrightarrow\Gd(\Arho)$ is simply the 
dual map of the inclusion $\Hd\longrightarrow\HCK$, and sends a diffeographism 
of the form (\ref{diffeographism}) into the formal diffeomorphism
\begin{align}
\label{physical-diffeomorphism}
\pi(f)(\lambda) 
&= \sum_{n=0}^\infty \left(\sum_{V(\G)=n+1} f_{\G}\right)\ \lambda^{n+1}. 
\end{align}
In other words, the projection is induced on the series by the map 
which sends a graph $\G$ to the number $V(\G)$ of its internal vertices. 

Series of the form (\ref{diffeographism}) are unreal, and of course 
have no physical meaning. Instead, their images (\ref{physical-diffeomorphism}) 
are usual series, and have a physical meaning in the context of 
perturbative quantum field theory: the coupling constants are exactely 
series of this form, summed up over suitable sets of Feynman diagrams. 
Moreover, the Green's functions and the renormalization factors are series 
of this form modulo a constant term which makes them being invertible series 
instead of formal diffeomorphisms. 
In conclusion, the meaning of Connes and Kreimer's results is that the 
renormalization procedure takes place in the group $\GCK(\Arho)$, even if 
the physical results are read in the group $\Gd(\Arho)$. 
\end{point}

\begin{point}{\bf Diffeographisms and Dyson's formulas}
According to Section~\ref{Renormalization-Dyson}, the result of renormalization 
is described by Dyson's formulas (\ref{Dyson-formula}) directly on 
usual series in the powers of the coupling constant $\lambda$. 
As we said, this happens in the semidirect product 
$\Gd(\Arho)\ltimes\Gi(\Arho)$ of the groups of formal diffeomorphisms 
by that of invertible series. 

However, these formulas require the knowledge of the renormalization factors. 
According to (\ref{Z-C}), these are known through the computations of 
the counterterms of all Feynman graphs. In other words, the physical 
results given by Dyson's formulas seem to be the projection of computations 
which take place in the semidirect product 
$\GCK(\Arho)\ltimes\Gi_{graphs}(\Arho)$, where $\GCK(\Arho)$ is the 
diffeographisms group dual to the Connes-Kreimer Hopf algebra, and 
$\Gi_{graphs}(\Arho)$ is a suitable lifting of the group of invertible 
series whose coordinate ring is spanned by Feynman graphs. 

This conjecture has been proved for quantum electrodynamics in the sequel of 
works \cite{BFqedren}, \cite{BFqedtree} and \cite{BFK}. In those works, 
the Green's functions are expanded over {\em planar binary trees\/}, 
that is, planar trees with internal vertices of valence 3, which were 
used by C.~Brouder in \cite{Brouder} as intermediate summation terms 
between integer numbers and Feynman graphs. 
It has also been proved by W.~van Suijlekom in \cite{VanSuijlekom} 
for any gauge theory. For the $\phi^3$-theory the work is in progress. 
\end{point}

\begin{point}{\bf Groups of ``combinatorial'' series}
If the diffeographisms are represented as generalized series of the form 
(\ref{diffeographism}), the group law dual to the coproduct in $\HCK$ 
should be represented as a ``composition'' among them. 
This operation has been defined in principle by P.~van der Laan in 
\cite{VanDerLaan}, using operads. An {\em operad\/} is the set of all possible 
operations of a given type that one can do on any algebra of that type. 
A particular algebra is then a representation of the corresponding operad. 
For instance, there exists the operad of associative algebras, that of 
Lie algebras, and many other examples of operads giving rise to 
corresponding types of algebras. 
By assumption, operads are endowed with an intrinsic {\em operadic 
composition\/} which allows to perform the operations one after another one 
in the corresponding algebras, and still get the result of an operation. 
The group $\Gd$ of formal diffeomorphisms is deeply related to the operad 
$\As$ of associative algebras, and in particular the composition of formal 
series in one variable can be directly related to the operadic composition 
in $\As$. Based on this observation, Van der Laan had the idea to realize 
the ``composition'' among diffeographisms as the operadic composition of 
a suitable operad constructed on Feynman graphs. 
In \cite{VanDerLaan}, he indeed defined an {\em operad of all Feynman 
graphs\/}, but didn't describe explicitely how to restrict the general 
construction to the particular case of Feynman graphs for a given theory. 
In particular, the explicit form of the group $\GCK(\Arho)$ related to the 
renormalization of the $\phi^3$-theory is not achieved. 

A complete description of the generalized series and their composition law 
is given in \cite{Frabetti} for the renormalization of quantum 
electrodynamics, on the intermediate coordinate rings spanned by 
planar binary trees. However, trees are combinatorial objets much simplier 
to handle then Feynman graphs, and the generalization of this construction 
to diffeographisms is still uncomplete. 

Groups of series expanded over other ``combinatorial objects'', such as 
rooted (non-planar) trees, also appear in the context of renormalization. 
Such trees, in fact, can be used to describe the perturbative expansion 
of Green's functions, and were used by D.~Kreimer in \cite{Kreimer} 
to describe the first Hopf algebra of renormalization appearing in the 
literature. The dual group of tree-expanded series was then used by 
F.~Girelli, T.~Krajewski and P.~Martinetti in 
\cite{GirelliKrajewskiMartinetti}, in their study of Wilson's continuous 
renormalization group. 

Furthermore, the series expanded over various ``combinatorial objects''
make sense not only in the context of the renormalization of a quantum 
field theory, but already for classical interacting fields. In fact, 
as we pointed out in Section~\ref{series-classical}, these fields are 
described perturbatively as series expanded over trees. Then, any result 
on usual series which has a physical meaning should be the projection 
of computations which take place in the corresponding set of 
``combinatorial series''. 

Finally, all the Hopf algebras constructed on ``combinatorial objects'' 
which appear in physics share some properties which are investigated 
in various branches of mathematics. On one side, as we already mentioned, 
these Hopf algebras seem to be deeply related to operads or to some 
generalization of them, see for instance the works by J.-L.~Loday.  
On the other side they turn out to be related to the various 
generalizations of the algebras of symmetric functions, 
see for instance the several works by J.~Y.~Thibon and coll., or those 
by M.~Aguiar and F.~Sottile, and seem related to the so-called 
{\em combinatorial Hopf algebras\/}. 
\end{point}


\end{fmffile}
\end{document}